\documentclass[bibyear]{aa}  
\usepackage[T1]{fontenc}

\makeatletter
\renewcommand*\aa@pageof{, page \thepage{} of \pageref*{LastPage}}
\makeatother

\usepackage{graphicx}
\usepackage{txfonts}
\usepackage{natbib}
\bibpunct{(}{)}{;}{a}{}{,} 
\usepackage{booktabs}
\usepackage{multicol}
\usepackage{graphicx}
\usepackage{fancyhdr}
\usepackage[hidelinks]{hyperref}
\usepackage{amsmath}	
\usepackage{amssymb}	
\usepackage[inter-unit-product=\cdot]{siunitx}
\usepackage{enumerate}
\usepackage{multirow}
\usepackage{mathtools}
\usepackage{comment}
\usepackage{longtable}
\usepackage{tabularx}
\usepackage{xcolor} 
\usepackage{ulem} 
\usepackage{comment}
\usepackage{cuted}
\usepackage{color, soul}
\usepackage{array}

\newcommand{\orcidicon}[1]{\href{https://orcid.org/#1}{\includegraphics[width=11pt]{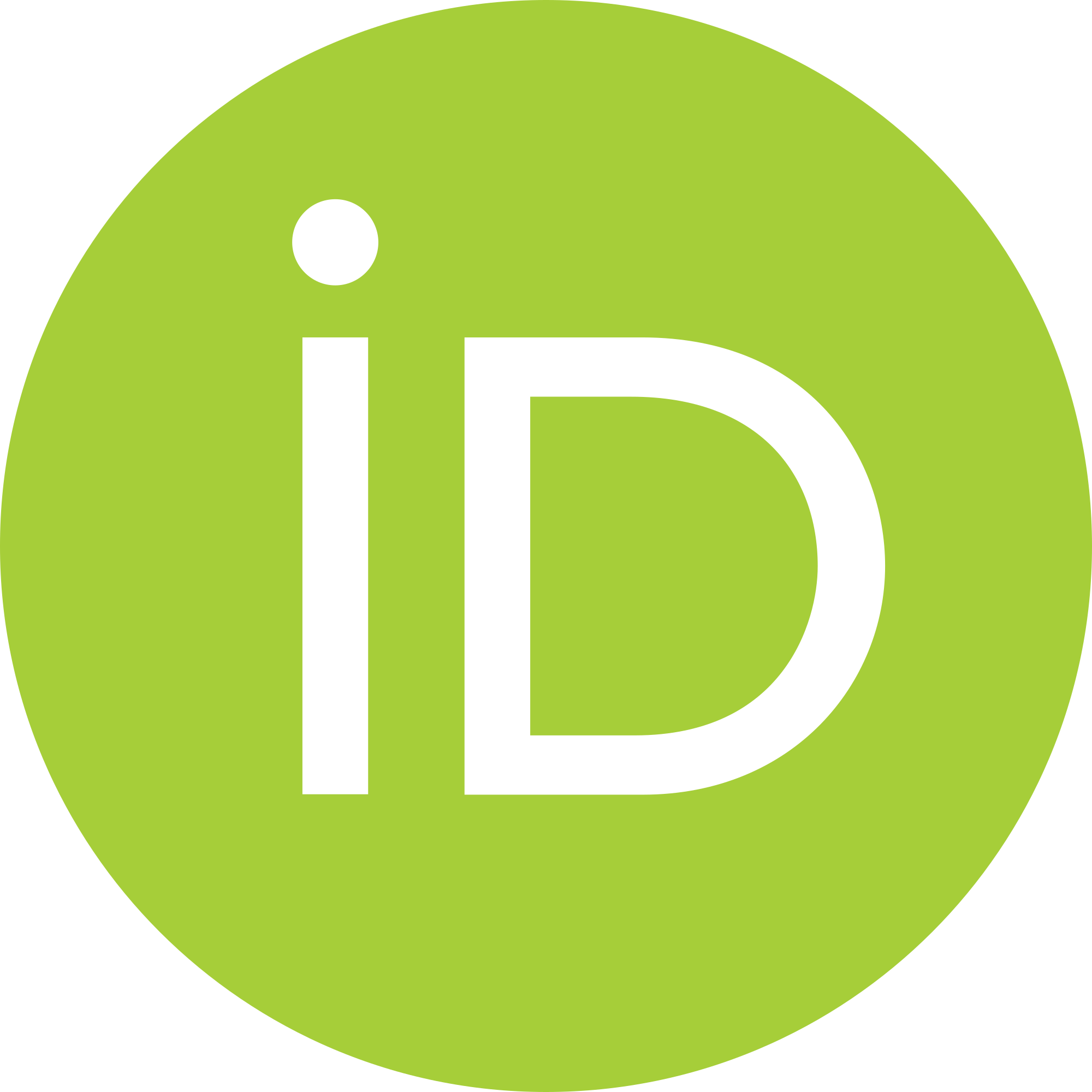}}}
\newcommand{\orcid}[1]{\href{https://orcid.org/#1}{\protect\orcidicon{#1}}}

\usepackage{graphicx}
\usepackage{txfonts}
\begin{document}

\title{Impact of accretion-induced chemically homogeneous evolution on stellar and compact binary populations}

\author{
Marco Dall'Amico,\inst{1,2,3}\orcid{0000-0003-0757-8334}\thanks{\href{mailto:marco.fromfriend@gmail.com}{marco.dallamico@pd.infn.it}}
Michela Mapelli,\inst{3,1,2,4}\orcid{0000-0001-8799-2548}\thanks{\href{mailto:mapelli@uni-heidelberg.de}{mapelli@uni-heidelberg.de}}
Giuliano Iorio\inst{5,1,2,4}\orcid{0000-0003-0293-503X},
Guglielmo Costa\inst{6}\orcid{0000-0002-6213-6988},
St\'ephane~Charlot\inst{7}\orcid{0000-0003-3458-2275},
Erika Korb\inst{1,2,3}\orcid{0009-0007-5949-9757},
Cecilia Sgalletta\inst{8,9,10}\orcid{0009-0003-7951-4820},
Marie Lecroq\inst{7}\orcid{0009-0008-2198-9651}
}
\authorrunning{M. Dall'Amico et al.}
\titlerunning{Impact of accretion-induced CHE on stellar and compact binary populations}
\institute{
$^{1}$ Physics and Astronomy Department Galileo Galilei, University of Padova, Vicolo dell’Osservatorio 3, I–35122, Padova, Italy\\
$^{2}$ INFN–Padova, Via Marzolo 8, I–35131 Padova, Italy\\
$^{3}$ Institut f{\"u}r Theoretische Astrophysik, ZAH, Universit{\"a}t Heidelberg, Albert-Ueberle-Stra{\ss}e 2, D-69120, Heidelberg, Germany\\
$^{4}$ INAF–Osservatorio Astronomico di Padova, Vicolo dell’Osservatorio 5, I–35122, Padova, Italy\\
$^{5}$ Departament de Física Quàntica i Astrofísica, Institut de Ciències del Cosmos, Universitat de Barcelona, Martí i Franquès 1, E-08028 Barcelona, Spain\\
$^{6}$ Univ Lyon, Univ Lyon1, ENS de Lyon, CNRS, Centre de Recherche Astrophysique de Lyon UMR5574, F-69230 Saint-Genis-Laval, France\\
$^{7}$ Sorbonne Universit\'e, CNRS, UMR 7095, Institut d’Astrophysique de Paris, 98 bis bd Arago, 75014 Paris, France\\
$^{8}$ SISSA, via Bonomea 365, I–34136 Trieste, Italy\\
$^{9}$ INFN, Sezione di Trieste, I–34127 Trieste, Italy\\
$^{10}$ INAF – Osservatorio Astronomico di Roma, Via Frascati 33, I–00040, Monteporzio Catone, Italy
}
\date{Received XX; accepted YY}

 
  \abstract{
 In binary star systems, mass transfer can spin up the accretor, possibly leading to efficient chemical mixing and chemically quasi-homogeneous evolution (CHE). Here, we explore the effects of accretion-induced CHE on both stellar populations and their compact binary remnants with the state-of-the-art population synthesis code {\sc sevn}. We find that CHE efficiently enhances the formation of Wolf-Rayet stars (WRs) from secondary stars, which are spun-up by accretion, while simultaneously preventing their evolution into red supergiant stars (RSGs). Including CHE in our models increases the fraction of WRs in our stellar sample by nearly a factor of $\approx3$ at low metallicity ($Z=0.001$). WRs formed through CHE are, on average, more massive and luminous than those formed without CHE. Most WRs formed via CHE end their life as black holes. As a direct consequence, the CHE mechanism enhances the formation of binary black holes (BBHs) and black hole-neutron star (BHNS) systems, while simultaneously quenching the production of binary neutron stars (BNSs). However, CHE  significantly quenches the merger rate of BBHs, BHNSs and BNSs at low metallicity ($Z\leq{}0.004$), because most binary compact objects formed via CHE have large orbital periods. For instance, the number of BBH and BHNS mergers decreases by  one order of magnitude  at $Z=0.004$ in the CHE model compared to the standard scenario. Finally, we find that secondary stars experiencing CHE frequently produce the most massive compact object in the binary system. In BHNSs, this implies that the black hole progenitor is the secondary star. Conversely, BBHs formed through accretion-induced CHE likely have asymmetric black hole components, but only a negligible fraction of these asymmetric systems ultimately merge within an Hubble time.
  
  }


   \maketitle
%

\section{Introduction}

Massive stars drive the evolution of the Universe. Their strong stellar winds and supernova events serve as a primary source of mechanical feedback for the interstellar medium, playing a key role in the genesis of new stars and planets \citep{Elmegreen,McKee}. Massive star radiative feedback, characterized by strong ultraviolet radiation, leads to the formation of HII regions and is commonly assumed to be the main source of the reionization epoch 
\citep{Haiman1997,Loeb2001,Schaerer2002,Maio,Bromm2004,Bromm2013,Klessen2023}. Moreover, massive stars are considered the main forge of alpha elements in the Universe, shaping the chemical evolution of galaxies \citep{Woosley1995,Woosley2002,Rauscher2002,Nomoto2013}. In addition, they are deemed progenitors of high-energy events like gamma-ray bursts, and in their final stages, they give birth to compact objects such as neutron stars and black holes \citep{Narayan1992,Heger2003,Woosley2006}. Lastly, massive stars can actively contribute to cluster dynamics, potentially triggering catastrophic events like stellar collisions \citep[e.g.][]{Zwart1999,Fregau2004,Mapelli2016,Zwart2016,DiCarlo2020,Torniamenti2022}.

Massive stars are rare and short-lived, with most of their crucial evolutionary phases lasting just a blink of an eye in the context of cosmological timescales. Despite their importance, several aspects of the evolution of massive stars remain uncertain. Mass loss episodes due to winds, pulsations, and outflows \citep[e.g.][]{Smith2014}, combined with rotation \citep[e.g.][]{Maeder2009}, magnetic fields \citep[e.g.][]{Donati2009}, and thermally induced mixing by overshooting, semiconvection, and dredge-up \citep[e.g.][]{Costa2021} are crucial events that govern the evolution of massive stars but still lack a complete understanding. To add an additional layer of complexity to the problem, it is now well-established that the vast majority of massive OB stars reside in binary systems, and that over $70\%$ of them are expected to interact with their companions \citep{Sana2012,Moe2017}. Binary interactions strongly affect the evolution of massive stars, and largely influence the production of compact binary mergers possibly detectable by ground-based interferometers \citep[e.g.][and references therein]{Marchant2023,Costa2023}.

One of the possible outcomes of the interplay between massive stellar evolution and binary processes is chemically quasi-homogeneous evolution (CHE). In massive metal-poor stars, large rotational velocities can generate strong mixing currents, causing the star to be partially or fully mixed before the depletion of hydrogen \citep{Eddington1925,Sweet1950}. Rotational mixing transports nuclear fusion byproducts from the core to the surface, simultaneously replenishing the core with hydrogen drawn from the outer envelope, beyond the limits of convection. These mixing currents prevent the formation of a strong chemical gradient between core and envelope, inducing the star to undergo CHE \citep{Maeder1987,Langer1992,Maeder2002,Heger2000,DeMink2009,Brott2011a,Kholer2015,Szecsi2015}.

CHE can be prevented if the massive star is a rapid, single rotator: angular momentum mass loss by stellar winds and magnetic braking can effectively spin down the star even before the mixing process becomes efficient \citep{Ivanova2003,Meynet2011}. In binary systems, on the other hand, various processes such as mass transfer episodes \citep[e.g.][]{Packet1981,Pols1991,Petrovic2005, Eldridge2011,Rensbergen2011,Shao2014,Renzo2021}, tidal interactions \citep[e.g.][]{Zahn1975,Hut1981,DeMink2009,Song2016}, and stellar mergers \citep[e.g.][]{Podsiadlowski1992,Tylendia2011,Vanbeveren2013} contribute as spin-up mechanisms, potentially sustaining the formation of rapidly rotating stars that undergo CHE.

CHE has a profound impact on both the observational properties of a massive stellar population \citep{Eldridge2011,Brott2011b,Martins2013,Eldridge2017,Schootemeijer2018,Cui2018,Ramachandran2019,Stanway2020,Massey2021a,Ghodla2022,Sharpe2024,Lecroq2024} and its efficiency in the production of compact binary mergers \citep{Eldridge2016,DeMink2016,Mandel2016,Marchant2016,Eldridge2019,DuBuisson2020,Riley2021}.
The overabundance of helium in a CHE star's envelope reduces opacity of the outer layers, leading to the formation of more compact and luminous stars \citep{DeMink2009}. Consequently, stars entering the CHE state evolve differently compared to their non-homogeneous counterparts. CHE stars become hotter and bluer as they evolve, likely appearing first as Oe/Of/slash-type stars, and later as Wolf-Rayet stars in their post-main sequence stage \citep{Brott2011a,DeMink2013}.
Due to the mixing currents that replenish the star's core with fresh envelope hydrogen, these stars can develop a large helium core and, at the end of their life, give birth to a massive compact object \citep{Mandel2016,DeMink2016}. Furthermore, rapidly rotating CHE stars are considered progenitors of hypernovae and long gamma-ray bursts \citep{Podsiadlowski2004,Yoon2005,Yoon2006,Woosley2006,Cantiello2007,Georgy2009,Eldridge2011}.

In this work, we explore the effects of accretion-induced CHE on a population of massive binary stars with primary mass in the range $[5,150]\,$M$_{\odot}$. First, we study the impact of CHE on the production of Wolf-Rayet stars (WRs) and red supergiant stars (RSGs) as a function of the binary fraction and the metallicity of the stellar population, assuming a constant star formation rate. We then explore the subsequent impact of CHE on the formation of binary black holes (BBHs), binary neutron stars (BNSs), and black hole-neutron star (BHNSs) binary systems. We discuss how the properties of binary compact mergers and their formation efficiency change as a function of our assumptions for CHE. We model our stellar population through the state-of-the-art population synthesis code {\sc sevn} \citep{Iorio2023}. 

\section{Methods}

\subsection{Population synthesis with {\sc sevn}}

We evolved our binary and single stellar populations with the stellar evolution for \textit{N}-body code {\sc sevn} \citep{spera2017,spera2019,mapelli2020,Iorio2023}\footnote{In this work, we adopt the version $2.7.5$ of {\sc sevn}, updated to commit hash \texttt{d9a351ce207eac6290b1d8a5ccd07f9eeada36b4}. {\sc sevn} is publicly available at \href{https://gitlab.com/sevncodes/sevn}{https://gitlab.com/sevncodes/sevn}.} {\sc sevn} incorporates single stellar evolution by interpolating pre-computed stellar tracks on the fly, while also modeling binary processes by means of analytic and semi-analytic prescriptions \citep{spera2017,spera2019,mapelli2020,Iorio2023}. Our simulations are based on stellar tracks\footnote{Here, we specifically make use of the tracks named \texttt{SEVNtracks\_parsec\_ov05\_AGB} for stars initialized in the hydrogen main sequence, and \texttt{SEVNtracks\_parsec\_pureHe36} for pure helium stars.} computed with the stellar evolution code {\sc parsec} \citep{Bressan2012,Chen2015,Costa2019,Costa2021}. Such tracks start from the pre-main sequence \citep{Bressan2012}, which can also be included in the population synthesis simulations. {\sc sevn} traces the evolution of individual stars by adaptively interpolating the stellar properties from the four closest stellar tracks to match the star's zero-age-main-sequence mass and metallicity. At the end of a star's lifetime, if a star is sufficiently massive, {\sc sevn} computes the
mass of the compact remnant as in \cite{Giacobbo2019} for electron-capture SNe, \cite{Fryer2012} for core-collapse SNe, and \cite{mapelli2020} for pulsational and pair-instability SNe. In our simulations, we specifically employed the \textit{delayed} core-collapse SN model by \cite{Fryer2012}. 
{\sc sevn} assumes that the compact remnant is a neutron star if its mass is less than $3\,$M$_{\odot}$, and a black hole otherwise. The supernova kicks applied to the compact remnants are modeled by {\sc sevn} following the approach outlined by \cite{Giacobbo2020}. In this framework, neutron star kick velocities are assigned based on the proper-motion distribution of young Galactic pulsars \citep{Hobbs2005}, while reduced-magnitude kicks are applied for black hole progenitors and stripped stars \citep{Tauris2017}. In addition, {\sc sevn} incorporates a wide range of binary evolution processes, including stable mass transfer by Roche-lobe overflow and winds, common envelope evolution, angular momentum dissipation by magnetic braking, tidal interactions, orbital decay by gravitational-wave emission, dynamical hardening, chemically homogeneous evolution, and stellar mergers. For a comprehensive description of the implementation of each of these processes, we refer to \cite{Iorio2023}. 

In the following, we discuss only the most significant assumptions implemented in our simulations for these binary-evolution processes. We adopted the same stability criterion for Roche-lobe overflow as the fiducial model presented by \cite{Iorio2023}. This model assumes a critical mass ratio, $q_{\rm c}$, between the donor and accretor star, above which the mass transfer becomes unstable on a dynamical timescale. Here, $q_{\rm c}$ is defined as in \cite{Hurley2002}, but for the following exception: the mass transfer is always considered stable if the donor has a radiative envelope, i.e. is a main sequence or a Hertzsprung gap star. For stable mass transfer, we assumed the standard prescription in {\sc sevn}, where the mass and angular momentum transfer is considered non-conservative, with an accretion efficiency on the secondary set to $f_{\rm MR}=0.5$.  {\sc sevn} describes the common-envelope phase with the $\alpha-\lambda$ formalism \citep{Webbink,Livio1988,Iben1993}. Here, $\alpha$ represents the fraction of orbital energy injected into the envelope, while $\lambda$ measures the concentration of the envelope and incorporates all the uncertainties related to the envelope structure. In our simulations, we adopted $\alpha=1$ and $\lambda$ that depends on the mass of the star, except for pure helium stars for which $\lambda=0.5$ \citep{Claeys2014}. 

{\sc sevn} manages tidal interactions by employing the analytical formalism presented by \cite{Hut1981} and further detailed by \cite{Hurley2002}. This includes the treatment of synchronization between stellar and orbital rotation, as well as the orbital circularization of the binary. Spin-down by magnetic braking is treated as in \cite{Hurley2002}, and it is active only if the star has an envelope and a core with, respectively, non-zero mass. Finally, in our work we considered only the evolution of isolated binaries, therefore we do not take into account the effects of dynamical hardening.

\subsection{Formation of CHE stars}\label{sec:CHE}

In this work, we exclusively explore the formation of CHE stars through accretion spin-up. This scenario occurs when the most evolved star in the binary system expands to its Roche radius, initiating a Roche-lobe overflow mass transfer. If the mass transfer remains dynamically stable, the companion star can efficiently accrete mass and angular momentum at a rate sufficient to effectively spin up. When the star is massive enough and sufficiently metal-poor, rotational mixing becomes extremely efficient, inducing the accretor star to evolve into the CHE state.  In principle, CHE can also be achieved in tight binaries due to tidal interactions. Strong tides tend to synchronize the rotational periods of the stars with the orbital periods of the binary and, as a result, both stars rapidly spin up, possibly triggering CHE. Here, we focus our study only on the former discussed CHE case. This choice and its impact on our results are further discussed in section~\ref{sec:discussion}.

{\sc sevn} allows for the formation of CHE stars through the accretion spin-up mechanism, adopting the prescription first introduced by \cite{Eldridge2011}. In this framework, a star becomes CHE if it satisfies three criteria. 

1) The star must accrete enough material and angular momentum during the Roche-lobe overflow phase to be considered rapidly spinning and fully mixed. In {\sc sevn}, this is assumed when a star accretes a fraction of its initial mass through Roche-lobe overflow. 
Following \cite{Eldridge2011}, we assume that a metal-poor accretor enters the CHE phase if it accretes more than $5\%$ of its initial mass. Note that if a non-rotating star accretes $\approx10\%$ of its initial mass, the star gathers enough angular momentum to reach its critical rotation velocity \citep{Eldridge2011}.

2) Rotational mixing is more efficient in more massive stars \citep{Heger2000,Maeder2001,Ekstrom2012,Georgy2013} and the minimum rotation rate required for CHE decreases with increasing mass of the star \citep{Yoon2006}. Moreover, massive stars are more likely to interact in binaries, meaning that the possibility of getting spun up by accretion is larger \citep{Sana2012}. {\sc sevn} enables the CHE only if a star is characterized by a post-accretion mass larger than a given threshold $M_{\rm min}$. In our fiducial model we set $M_{\rm min}=10\,$M$_{\odot}$ following the approach of \cite{Eldridge2011}. In Appendix~\ref{sec:app1} we present an alternative model with $M_{\rm min}=20\,$M$_{\odot}$, set as in \cite{Eldridge2017}.

3) Wind mass loss in massive stars strongly depends on metallicity \citep[e.g.][]{Vink2001}. Since mass loss removes angular momentum from the star, its metallicity must be low enough to weaken the winds and prevent the star from spinning down. We set the maximum metallicity below which CHE is possible as $Z_{\rm max}=0.004$ \citep{Yoon2006,Eldridge2011}.

If a star fulfills these three conditions, it is flagged as a CHE. In {\sc sevn}, its evolution proceeds with a fixed radius during the main sequence, at the end of which it directly evolves into a pure helium star. 



\subsection{Initial conditions and run setup}\label{sec:IC}

We simulated a total of $1.8\times10^{8}$ binary systems divided into six distinct sets, as summarized in Table~\ref{tab1:runs}. We further simulated $10^{8}$ single stars. We explored three different cases for CHE: one simulation set where CHE is never allowed, one with $M_{\rm min}=10\,$M$_{\odot}$, and one with $M_{\rm min}=20\,$M$_{\odot}$. Each of these simulation sets has been tested by initializing the stars at both their zero-age main sequence and pre-main sequence phases. Finally, we simulated our single and binary stars at ten different metallicities $Z=0.0001$, 0.0004, 0.0008, 0.001, 0.002, 0.004, 0.008, 0.014, 0.02, 0.04.
The main 
text and figures describe the results of our fiducial models, CHE10preMS and NoCHEpreMS. We present the other four models in Appendix~\ref{sec:app1}. In most Figures, we focus on metallicities $Z=0.001$ and $0.004$, because these are two cases in which considering accretion-induced CHE has maximal impact for binary compact objects.

We generate the mass of the primary and single stars from a Kroupa initial mass function \citep{Kroupa2001} in the range $[5,150]\,$M$_{\odot}$. The secondary mass in our binary systems is sampled assuming a mass ratio distribution $\mathcal{F}(q)\propto q^{-0.1}$ in the range $[0.1,1]$ \citep{Sana2012} and  imposing that the minimum possible secondary mass is  $2.2\,$M$_{\odot}$. 
The initial orbital period and eccentricity of our binaries are generated according to \cite{Sana2012}. The period follows a distribution $\mathcal{F}(\mathcal{P})\propto \mathcal{P}^{-0.55}$, with $\mathcal{P}=\log_{10}(P/{\rm days})\in[0.15,15]$. The eccentricities are extracted from the distribution $\mathcal{F}(e)\propto e^{-0.45}$ in the range $[10^{-5},e_{\rm max}(P)]$. Here, the upper limit of the eccentricity distribution is a function of the orbital period as $e_{\rm max}(P)=1-[P/(2\,{\rm days})]^{-2/3}$ \citep{Moe2017}. To date, {\sc sevn} does not incorporate evolutionary tracks for rotating stars. Therefore, we initialize our binary and single populations with stars having zero initial rotation\footnote{{\sc sevn} tracks the spin evolution of both stars, accounting for the effects of tidal interactions and mass transfer episodes.}. We stop our simulations either if the binary is disrupted by a supernova kick or when only compact objects remain.

\begin{table*}
\begin{center}
\caption{Models employed for the simulations.}
\label{tab1:runs}
\centering
\begin{tabular}{lccccc}
\hline\hline
Name & phase & CHE & $M_{\rm min}$ [M$_{\odot}$] &  $M_{\rm acc}$ & $Z_{\rm max}$\\
\hline
NoCHEzams  & ZAMS       & No CHE                & & & \\
\textbf{NoCHEpreMS} & \textbf{preMS} & \textbf{No CHE}                & & & \\
CHE10zams  & ZAMS       & \cite{Eldridge2011} & 10 & 5\% & 0.004 \\
\textbf{CHE10preMS} & \textbf{preMS} & \textbf{\cite{Eldridge2011}} & \textbf{10} & \textbf{5\%} & \textbf{0.004} \\
CHE20zams  & ZAMS       & \cite{Eldridge2017} & 20 & 5\% & 0.004 \\
CHE20preMS & preMS & \cite{Eldridge2017} & 20 & 5\% & 0.004 \\
\hline
\end{tabular}
\end{center}
\footnotesize{Column~1: name of the simulation set; column~2: initial phase of the stars at the beginning of the simulation (phase); column~3: the model includes (CHE) or does not include (No CHE)  chemically quasi-homogeneous evolution; column~4: minimum mass required for a star to enter a CHE phase ($M_{\rm min}$); column~5: minimum accreted mass relative to its ZAMS mass required for a star to enter a CHE phase ($M_{\rm acc}$); column~6: maximum metallicity required for a star to enter a CHE phase ($Z_{\rm max}$). Our fiducial models (CHEpreMS and NoCHEpreMS) are highlighted in boldface. In the main body of the paper, all Figures refer to the fiducial models. The other four models are described in Appendix~\ref{sec:app1}.}
\end{table*}


\subsection{Wolf-Rayet \& red supergiant stars}\label{sec:def}

\begin{figure*}
   \centering
   \includegraphics[width=0.34\textwidth]{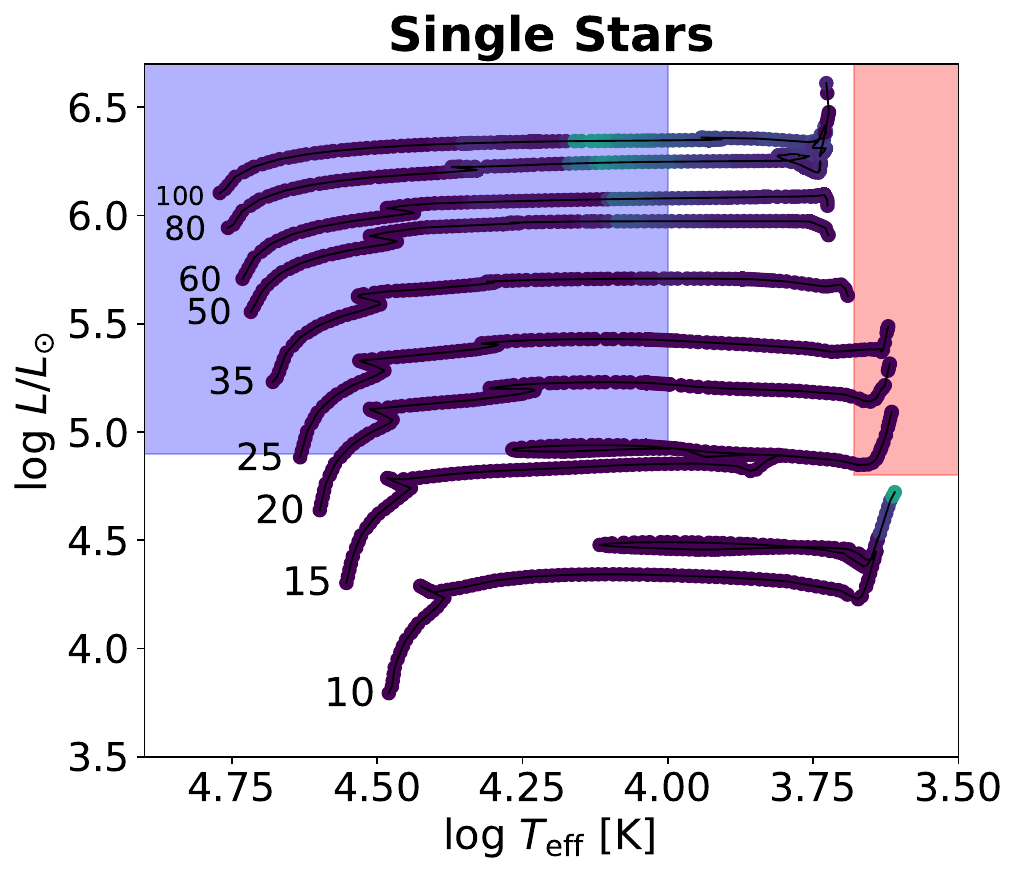}
   \includegraphics[width=0.65\textwidth]{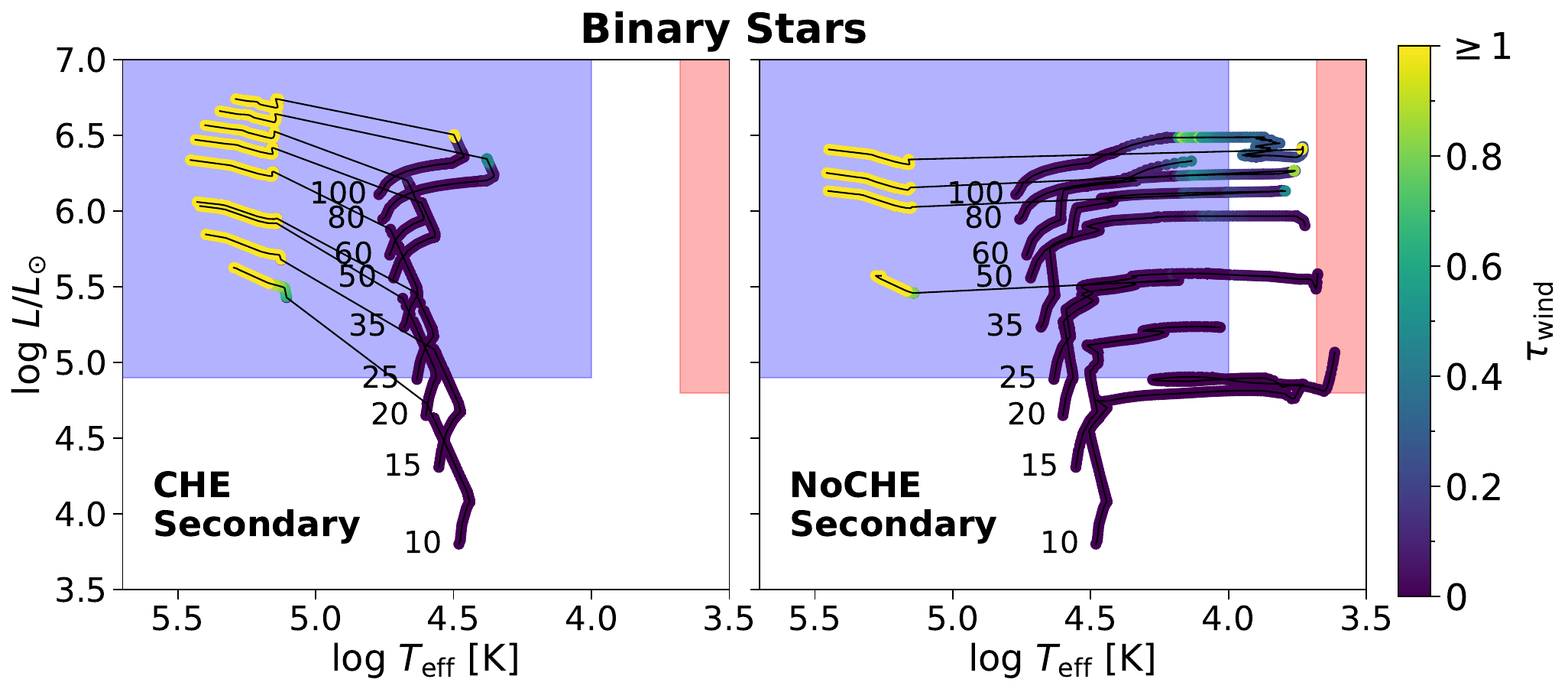}
      \caption{Evolutionary tracks starting from the zero-age-main-sequence at metallicity $Z=0.001$ of single stars (left) and some secondary stars in binary systems (right) with zero-age main sequence masses of 10, 15, 20, 25, 35, 50, 60, 80, and 100 M$_{\odot}$ and metallicity $Z=0.001$. The central and right-hand plots show the secondary stars in the CHE10preMS and NoCHEpreMS models. The colour-map shows the wind optical depth during the stars' evolution. The blue and red-filled regions indicate the threshold within which the luminosity and temperature are necessary to classify a star as a WR or a RSG.}
         \label{fig:HR}
   \end{figure*}

We explore the effects of CHE on a stellar population by studying the 
statistics of RSGs and WRs, as these are commonly associated with evolved stages of two distinct stellar sub-groups. RSGs are typically helium-burning stars with ages in the $\approx8-35\,$Myr range that arise from the evolution of stars with initial mass $\approx8-30\,$M$_{\odot}$. WRs descend instead from the evolution of stars with a mass larger than $\approx30\,$M$_{\odot}$, and are characterized by ages lower than $\approx8\,$Myr. Observationally, these two categories of stars lie in distinct regions on the Hertzsprung-Russell diagram, highlighted in blue and red in Figure~\ref{fig:HR}. We categorize RSGs as stars exhibiting temperatures typical of M and K spectral type stars, defined by $\log(T/K)\leq3.68$ \citep{Klencki2020}. We additionally set a minimum luminosity criterion of $\log(L/L_{\odot})\geq4.8$ for classifying a star as an RSG \citep{Massey2021a}. This threshold limits the contamination from stars in the asymptotic giant branch \citep{Brunish1986,Massey2003}. We further categorized stars as RSGs only if they have a hydrogen surface abundance of $X > 0.001$ \citep{Eldridge2008}. For WRs, we set luminosity and temperature thresholds of $\log(L/L_{\odot})\geq4.9$ \citep{Shenar2020} and $\log(T/K)\geq4$ to avoid the luminous blue variable region \citep{Eldridge2008}. Additionally, we require the condition of optically thick winds, computing the wind optical depth using the formulation presented by \cite{Langer1989} and commonly adopted in recent works \citep[e.g.,][]{Aguilera2022,Pauli2022,Sen2023,Sabhahit2023}:
\begin{equation}
\tau_{\rm wind} = \frac{\kappa_{\rm e}\, \lvert \dot{M} \rvert}{4\,\pi\,R(v_{\infty}-v_{\rm 0})}\,\ln{\Bigg[\frac{v_{\infty}}{v_{\rm 0}}\Bigg]}>1.
    \label{eq:1}
\end{equation}
Here, $R$ is the stellar radius, $\dot{M}$ is the wind mass loss, $v_{\rm 0}=20\,$km$\,$s$^{-1}$ is a typical value of the expansion velocity near the stellar surface, and $\kappa_{\rm e}\approx 0.20\,(1+X_{\rm s})\,$cm$^{2}\,$g$^{-1}$ is the Thomson scattering opacity with $X_{\rm s}$ the hydrogen surface abundance of the star. This is a conservative prescription as we assume that the main opacity mechanism is electron-scattering, while we ignore any line opacity contribution. Eq.~\ref{eq:1} assumes a $\beta=1$ velocity law for the velocity stratification \citep{Vink2001}, and includes a terminal wind velocity dependence $v_{\infty}$ on the escape velocity $v_{\rm esc}$ and the Eddington factor $\Gamma_{\rm e}$ as \citep{Lamers1995}:
\begin{equation}
\begin{cases}
v_{\infty}=2.6\,v_{\rm esc}\,(1-\Gamma_{\rm e})^{2}, & \text{if } T_{\rm eff} > 22.5\,\text{kK} \\
v_{\infty}=1.3\,v_{\rm esc}\,(1-\Gamma_{\rm e})^{2}, & \text{if } T_{\rm eff} < 22.5\,\text{kK} \\
\end{cases}
\end{equation}
with $T_{\rm eff}$ the effective temperature of the star, and 
\begin{equation}
    v_{\rm esc}=\sqrt{\frac{2\,G\,M}{R}}, \,\,\,\,\,\,\,\Gamma_{\rm e}=\frac{\kappa_{\rm e}\,L}{4\,\pi\,c\,G\,M}\approx10^{-4.813}\,(1+X_{\rm s})\frac{L}{M}.
\end{equation}
In the equations, $M$ is the total mass of the star, $L$ is its luminosity, $G$ is the gravitational constant, and $c$ is the speed of light.

\section{Results: Stellar populations}


\subsection{Single stellar evolution}

The left-hand panel in Figure~\ref{fig:HR} shows an example of evolutionary tracks for single stars at $Z=0.001\approx1/20\,Z_{\odot}$. In these models, our single stellar population does not yield any WR. The mass loss is nearly negligible, and the stellar winds lack the strength required to effectively strip the star of its hydrogen envelope, thereby preventing the formation of a WR, and appearing optically thin \citep{Conti1975,Chiosi1979,Chiosi1986}. 
The higher the metallicity is, the higher the mass loss and the lower the minimum mass for a star to become a WR. We find that at $Z=0.004$ only  stars with an initial zero-age-main-sequence mass $M_{\rm ZAMS}\geq59\,$M$_{\odot}$ become WRs, while at $Z=0.008$ WRs form  from the evolution of stars with mass $M_{\rm ZAMS}\geq 40\,$M$_{\odot}$. At solar ($Z=0.02$) metallicity, stars become WRs already for $M_{\rm ZAMS}\geq 29\,$M$_{\odot}$. This also affects the fraction of WRs produced in a population of single stars, which increases with $Z$, as shown in Table~\ref{tab:stars}. We further find that WR progenitors experience a relatively short RSG phase at high metallicity before completely expelling their envelopes. This is true for stars with $M_{\rm ZAMS}=40-51\,$M$_{\odot}$ at $Z=0.008$, and $M_{\rm ZAMS}=29-39\,$M$_{\odot}$ at $Z=0.02$. Finally, the least massive WRs in our single-star sample have masses $M_{\sc WR}^{\rm min}=45,\,22,\,10,\,$M$_{\odot}$ at metallicity $Z=0.002,\,0.008,\,0.02$, respectively.

The maximum mass for a star to become a RSG shows a less distinct metallicity dependence than the minimum mass required to form a WR. In our single stellar evolution models, the maximum zero-age-main-sequence mass  for a star to become a RSG  is $M_{\rm ZAMS}=31,\,40,\,51,\,38\,$M$_{\odot}$ at $Z=0.001,\,0.004,\,0.008,\,0.02$, respectively. The maximum mass reached by RSGs is $M_{\sc RSG}^{\rm max}=31,\,31,\,41,\,27,\,$M$_{\odot}$, at the same metallicities. 
Finally, Table~\ref{tab:stars} shows that single stellar evolution is always more efficient in producing RSG stars, compared to binary evolution. This directly results from stellar stripping in binary systems, where the outer layers of the binary components are likely peeled off by mass transfer episodes, thereby preventing the formation of cool supergiants.

\subsection{Binary stellar evolution}\label{sec:bin}

Table~\ref{tab:stars}  shows that 
binary evolution both provides an efficient pathway to produce WRs even at low metallicity, and leads to a higher fraction of WRs  across all considered metallicities. The right-hand part of Figure~\ref{fig:HR} shows an example of WR evolution of post-accretion stars at metallicity $Z=0.001$, 
according to the CHE10preMS and NoCHEpreMS models. 

In the evolutionary model without CHE, the main formation channel for WRs is envelope stripping by Roche lobe overflow, with over $65\%$ of WRs originating from primary stars (Table~\ref{tab:stars}). Primary stars evolve more rapidly than secondaries, expanding beyond their Roche lobe and initiating to donate mass to the companion. By the end of this process, the envelope of the primary is partially or completely lost and the star likely evolves into a WR. This is the fate of most primary stars in our interacting binaries, and constitutes the primary process to form WRs in our NoCHEpreMS model. 

Table~\ref{tab:stars} shows that primary stars are also the most likely component to evolve into RSGs. This is the natural outcome for non-interacting binaries in which the primary star has a main sequence mass below $\approx30\,$M$_{\odot}$ while the secondary likely avoids evolving into RSG due to its lower mass. Conversely, if the binary components have nearly equal masses, a non-interacting system may also experience an evolutionary stage with both components in the RSG or WR phase. These systems are rare across all metallicities in our models, as indicated in Table~\ref{tab:stars}.

In the NoCHEpreMS model, less than $30\%$ of all secondary stars evolve into WRs. This occurs only when the secondary star donates mass through Roche lobe overflow and survives enough time in this phase to get completely peeled off (as in the $100, 60, 50, 20\,$M$_{\odot}$ tracks in Figure~\ref{fig:HR}). When a secondary star accretes mass from the primary, 
its luminosity grows. After this phase, the secondary star spends the rest of the evolution with its hydrogen envelope intact. This is a direct consequence of wind weakness at low metallicity. As a result, the secondary star starts to expand and evolves toward lower temperatures. 

The main difference between models without CHE and models with CHE lies in the evolution of the secondary star. 
In the models with CHE, secondary stars are the most common WR progenitors. In these binaries, the primary star fills its Roche radius and initiates the mass transfer, at the end of which it becomes a peeled star likely appearing as WR, as in the NoCHEpreMS model. If the mass transfer on the secondary fulfills the conditions discussed in Section~\ref{sec:CHE}, the secondary star becomes chemically homogeneous 
and evolves into a pure He star at the end of the main sequence. In contrast, the evolution of the primary star remains unaffected by the CHE of the secondary. This happens because primary stars have already undergone a significant portion of their evolution when the secondary enters the CHE stage. By the time the secondary becomes a WR, the primary has already ended its life as a supernova. Because of the CHE mechanism, secondary stars produce more than $72\%$ and $56\%$ of all the WRs in a binary population with $Z=0.001$ and $Z=0.004$, respectively (Table~\ref{tab:stars}). This production of WRs from accretors almost triplicates the number of binaries containing a WR compared to the model without CHE at $Z=0.001$. 
In summary, CHE enhances the production efficiency of WRs in binary systems by increasing the likelihood of secondary stars evolving into WRs.

Finally, stellar mergers significantly contribute to the production of both WR and RSG stars, as reported in Table~\ref{tab:stars}. This becomes particularly evident in the model without CHE, as stellar radii increase with $Z$, enhancing their probability of interacting with their companion. In contrast, in the CHE model, the fraction of WR stars generated by mergers is lower compared to the model without CHE, reflecting a decrease in the overall occurrence of mergers. When a binary undergoes CHE, both members typically remain compact, reducing their interaction probability and the occurrence of a merger. Our binary populations produce $6\%$ less mergers at all metallicities in the CHE10preMS model than in the NoCHEpreMS model.  Additionally, in the CHE model, fewer secondary stars evolve into RSGs, thereby increasing the overall proportion of RSGs produced through mergers.


\begin{table*}[]
\begin{center}
\caption{Percentage of WRs and RSGs produced in binary and single stellar populations evolved with the fiducial models  (CHE10preMS and NoCHEpreMS) at different metallicity.}
\begin{tabular}{@{}l|ccc|ccc|cc|cc|cc|@{}}
\cmidrule(l){2-13}
\multicolumn{1}{l|}{} & \multicolumn{3}{c|}{Z=0.001} & \multicolumn{3}{c|}{Z=0.004} & \multicolumn{2}{c|}{Z=0.008} & \multicolumn{2}{c|}{Z=0.02} & \multicolumn{2}{c|}{Z=0.04} \\
\cmidrule(l){2-13}
\multicolumn{1}{l|}{}  & \multicolumn{1}{c}{CHE} & \multicolumn{1}{c}{NoCHE} & \multicolumn{1}{c|}{Sing} & \multicolumn{1}{c}{CHE} & \multicolumn{1}{c}{NoCHE} & \multicolumn{1}{c|}{Sing}  & \multicolumn{1}{c}{NoCHE} & \multicolumn{1}{c|}{Sing} & \multicolumn{1}{c}{NoCHE} & \multicolumn{1}{c|}{Sing} & \multicolumn{1}{c}{NoCHE} & \multicolumn{1}{c|}{Sing} \\ 
\toprule
\multicolumn{1}{|l|}{ P$_{\rm WR}$ }       &9.5  &3.6  &0     &12.2  &7.0   &1.9  &8.7   &4.6   &10.7  &8.6  &12.6  &12.2  \\
\multicolumn{1}{|l|}{ P$_{\rm WR bin}$ }   &15.9 &5.4  &      &18.4  &10.6  &     &13.6  &      &17.1  &     &19.8  &  \\
\multicolumn{1}{|l|}{ P$_{\rm WR prim}$ }  &25.5 &65.4 &      &34.6  &57.4  &     &55.0 &       &53.5 &      &47.0  &  \\ 
\multicolumn{1}{|l|}{ P$_{\rm WR sec}$ }   &72.9 &29.9 &      &56.6  &26.6  &     &23.0  &      &22.5  &     &25.0  &  \\ 
\multicolumn{1}{|l|}{ P$_{\rm WR merg}$ }  &1.6  &4.7 &       &8.8  &16.0   &     &22.0  &      &24.0  &     &28.0  &  \\ 
\multicolumn{1}{|l|}{ P$_{\rm WR-WR}$ }    &0.3  &0.2  &      &0.7   &0.5   &     &1.2   &      &1.8   &     &2.8   &  \\ 
\midrule
\multicolumn{1}{|l|}{ P$_{\rm RSG}$ }      &13.5 &14.7 &27.5  &14.4  &15.5  &29.2 &16.7  &29.6  &17.3  &15.6 &19.3  &39.2  \\
\multicolumn{1}{|l|}{ P$_{\rm RSG bin}$ }  &23.7 &25.6 &      &25.4  &26.5  &     &28.7  &      &29.8  &     &33.6  &  \\
\multicolumn{1}{|l|}{ P$_{\rm RSG prim}$ } &48.4 &44.5 &      &45.1  &42.1  &     &43.0  &      &43.5  &     &39.5  &  \\ 
\multicolumn{1}{|l|}{ P$_{\rm RSG sec}$ }  &20.4 &26.2 &      &18.3  &22.9  &     &21.6  &      &22.7  &     &26.6  &  \\
\multicolumn{1}{|l|}{ P$_{\rm RSG merg}$ } &31.2 &29.3 &      &36.6  &35.0  &     &35.4  &      &33.0  &     &33.9  &  \\ 
\multicolumn{1}{|l|}{ P$_{\rm RSG-RSG}$ }  &0.03 &0.03 &      &0.2   &0.2   &     &0.3   &      &0.4   &     &0.3   &  \\ 
\bottomrule
\end{tabular}
\label{tab:stars}
\end{center}
\footnotesize{From top to bottom line: percentage of stars in the sample that experience a WR phase (P$_{\rm WR}$), binaries with at least one WR (P$_{\rm WRbin}$),  primary stars that become WRs (P$_{\rm WR prim}$),  secondary stars that become WRs  (P$_{\rm WR sec}$),  stellar mergers that evolve into WRs (P$_{\rm WR merg}$),  binary systems experiencing a WR--WR phase (P$_{\rm WR-WR}$),  stars that experience a RSG phase (P$_{\rm RSG}$), binaries with at least one RSG (P$_{\rm RSGbin}$), primary stars that evolve into RSGs (P$_{\rm RSG prim}$), secondary stars that evolve into RSGs  (P$_{\rm RSG sec}$), stellar mergers that become RSGs  (P$_{\rm RSG merg}$),  binary systems experiencing a RSG-RSG phase (P$_{\rm RSG-RSG}$). }
\end{table*}


\subsection{RSG to WR ratio}

\begin{figure*}
   \centering
   \includegraphics[width=0.95\textwidth]{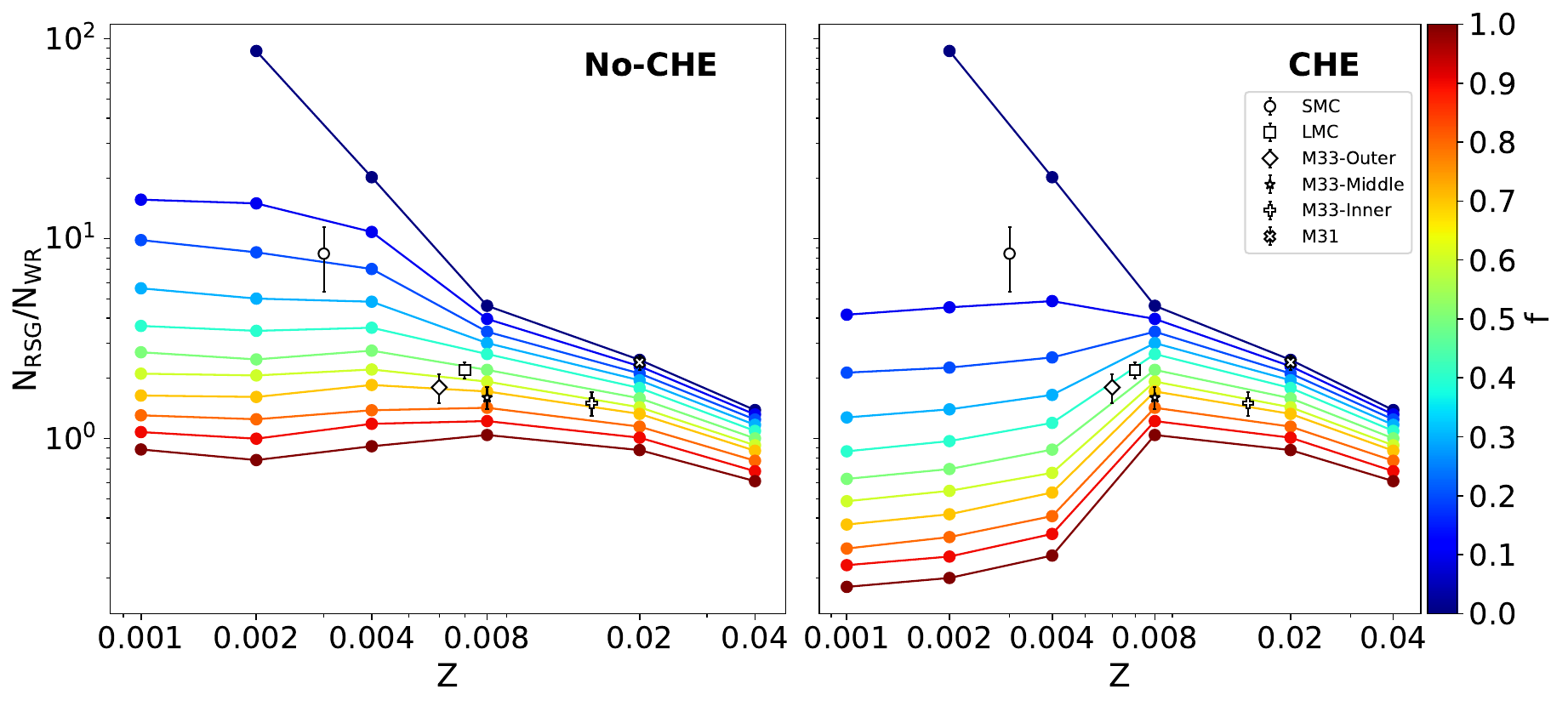}
      \caption{Rate of RSG over WR stars as a function of  metallicity and \ binary fraction. The left-hand (right-hand) side panel reports the results for the simulations evolved without (with) CHE. The markers show the ratio observed for different galaxies reported in \cite{Massey2021a}.}
         \label{fig:rate_WR}
   \end{figure*}

Figure~\ref{fig:rate_WR} shows the effect of metallicity,  binary fraction, and  CHE on the RSG-to-WR ratio in a  stellar population. The ratio is defined as the number of RSGs over the number of WRs integrated assuming a constant star formation rate\footnote{\cite{Massey2021a} have shown in their study that the RSG to WR ratio is only marginally sensitive to different star formation rate models.}. We tested 11 binary fractions $f=2\,N_{\rm binaries}/N_{\rm stars}$ ranging from $f=0$, i.e. a population composed solely of single stars, up to $f=1$, i.e. a population of stars in binaries.

The Figure shows that when the population is only composed of single stars, the ratio $N_\mathrm{RSG}/N_\mathrm{WR}$ monotonically decreases with increasing metallicity. This happens because the minimum mass to form a WR decreases with $Z$, while the lifetime spent by a star in the WR phase increases with increasing $Z$. 
Moreover, the number of RSGs 
drops with increasing $Z$, because at high metallicity the most massive stars get stripped by their winds and evolve into WRs.

The same Figure shows that the higher the binary fraction is, the lower the overall RSG-to-WR ratio. Binary evolution produces a larger number of stripped stars by mass transfer episodes, meaning more WRs and fewer RSGs. Even a modest binary fraction can significantly reduce the ratio at low metallicity. This stems from the inefficiency of producing WRs through single stellar evolution when $Z$ is low, because stellar winds are not strong enough to strip the stellar envelope. 
The binary fraction not only reduces the ratio but also alters its pattern with metallicity. While at low values of $f$, the ratio still exhibits a monotonically decreasing trend with $Z$, at $f=1$ the ratio remains almost constant with increasing metallicity up to $Z=0.008$, beyond which it begins to decline.

If we consider the models where CHE is active, the RSG-to-WR ratio drops when the metallicity is $\leq0.004$ in all the populations with $f>0$. As we discussed in section~\ref{sec:bin}, CHE is extremely efficient in forming WRs. CHE suppresses the evolution of a significant fraction of secondary stars into RSGs, instead causing them to evolve into WRs, as also shown in Table~\ref{tab:stars}. 
This effect becomes more pronounced in stellar populations with both a large binary fraction and low metallicity.

Finally, Figure~\ref{fig:rate_WR} displays the RSG-to-WR ratios observed for some local galaxies, as reported by \cite{Massey2021a}. Direct comparisons with our synthetic data should be approached with caution, as the observed ratios heavily depend on the completeness of the WR and RSG catalogs, while our ratios are influenced by the definitions of WRs and RSGs. Nevertheless, the plot provides an estimate of the binary fraction in these systems. For instance, our results suggest a small binary fraction ($0<f<0.3$) for the Small Magellanic Cloud (SMC), and a higher binary fraction ($0.3<f<0.8$) for the Large Magellanic Cloud (LMC), depending on the model with and without CHE. 

Overall, the dependence of our RSG-to-WR ratio on metallicity for the NoCHEpreMS model is remarkably similar to those reported by \cite{Eldridge2008} and \cite{Massey2021a}. Our definition of WR is more conservative if compared to the one used in such works, as we classify a star as WR only if it possesses an optically thick wind. This likely compensates for the tendency of the {\sc parsec} tracks \citep{Costa2019} to produce a larger number of bluer stars with respect to {\sc bpass} \citep{Eldridge2017}.

   \begin{figure*}
   \centering
   \includegraphics[width=0.95\textwidth]{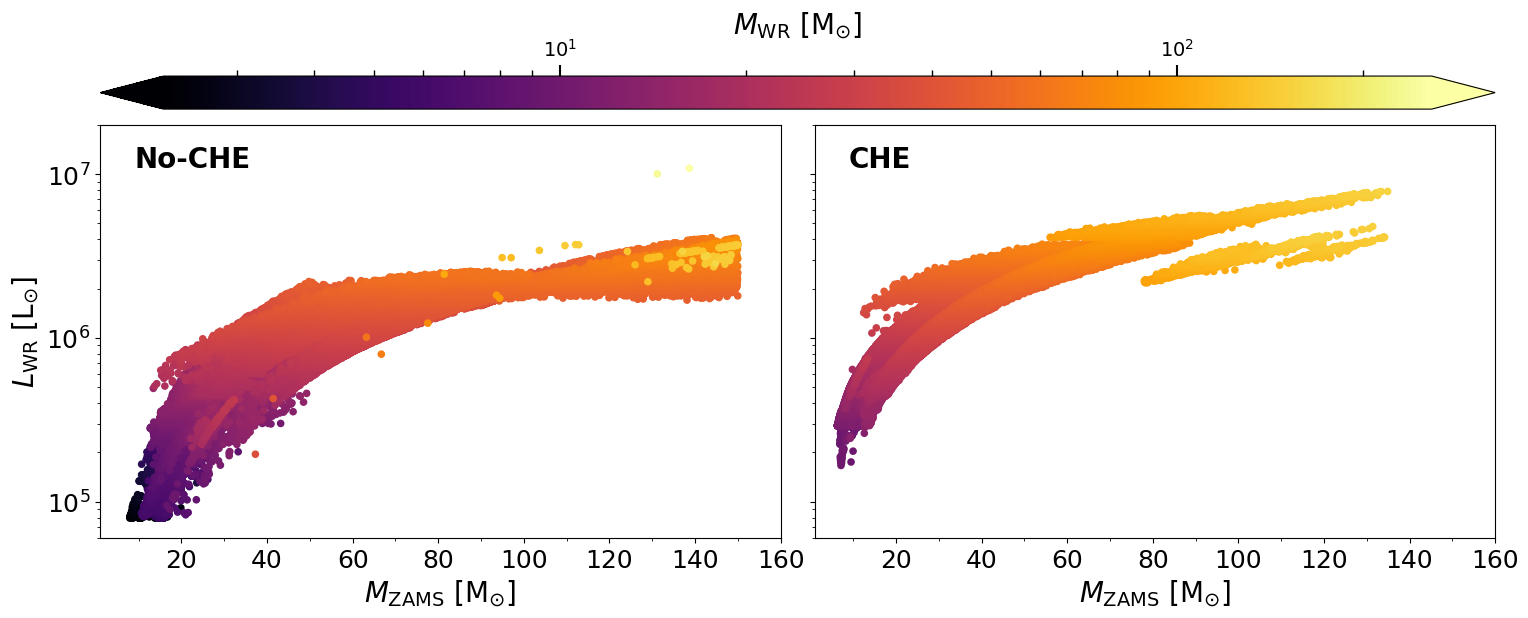}
      \caption{WR luminosity ($L_\mathrm{WR}$)  versus the zero-age-main-sequence mass of their progenitors ($M_\mathrm{ZAMS}$).  The colour scale indicates the WR mass  ($M_\mathrm{WR}$). $L_\mathrm{WR}$ and $M_\mathrm{WR}$ are calculated at the time the star becomes a WR. The left-hand panel shows  all the WRs produced in the NoCHEpreMS model, while the right panel shows only the WRs produced by CHE in the CHE10preMS model. The lower luminosity limit in the NoCHEpreMS models arises from the luminosity threshold we set to classify a star as a WR (Sec.~\ref{sec:def}). The lower luminosity limit in the CHE10preMS model derives from the minimum mass requirement for a star to undergo CHE (Sec.~\ref{sec:CHE}). The plot displays the results for our models at $Z=0.001$. We find similar results at $Z=0.004$.}
         \label{fig:Luminosity}
   \end{figure*}


\subsection{WR luminosity}\label{sec:Lwr}

Figure~\ref{fig:Luminosity} compares the WR luminosity and mass in models with  and  without CHE at $Z=0.001$. 
These two families of WRs cover two distinct regions. At the same zero-age-main-sequence mass, WRs born through CHE are more massive and more luminous than their counterparts formed through envelope-stripping due to binary evolution processes. 

These differences arise from the nature of their formation path. In the standard formation scenario, WRs form in binary systems when one of the two stars fills its Roche lobe, starts to donate mass to the companion, and gets stripped of its outer envelope. What remains is the naked core of the original star, which has partially or completely lost a large fraction of its mass, including its envelope. In contrast, WRs formed through CHE retain their oversize accreted envelopes, as the outer hydrogen layers become fully mixed with the stellar interior. Consequently, WRs resulting from this mixing process exhibit larger mass and luminosity compared to their stripped, pure helium counterparts.  

The maximum luminosity of a WR formed by envelope stripping in our NoCHEzams model is $\approx3.7\times10^{6}\,$L$_{\odot}$, corresponding to stars with $M_{\rm ZAMS}>140\,$M$_{\odot}$. Above this luminosity limit, WRs can only be formed in our models either by CHE or by stellar mergers. The upper-right corner of Figure~\ref{fig:Luminosity} is populated by a few massive WRs, exceeding $200\,$M$_{\odot}$, which result from the merger of two stars in a binary system with an initial zero-age-main-sequence mass exceeding $100\,$M$_{\odot}$. 


\begin{table*}[]
\begin{center}
\caption{Percentages of BBH, BHNS, and BNS systems, mergers, and CHE binaries as a function of the model and the metallicity. For additional results including all the models presented in Table~\ref{tab1:runs}, we refer the reader to Table~\ref{tab:cob2} in Appendix~\ref{sec:app1}.}
\begin{tabular}{@{}lll|lll|lll|lll|@{}}
\cmidrule(l){4-12}
\multicolumn{3}{l|}{}                                              & \multicolumn{3}{c|}{BBH} & \multicolumn{3}{c|}{BHNS} & \multicolumn{3}{c|}{BNS} \\ \midrule
\multicolumn{1}{|l|}{Z}                  & \multicolumn{1}{l|}{name} & P$_{\rm CHE}$ &   P$_{\rm cob}$    &   P$_{\rm merg}$    &   P$_{\rm merg}^{\rm CHE}$    &   P$_{\rm cob}$    &   P$_{\rm merg}$    &  P$_{\rm merg}^{\rm CHE}$   &   P$_{\rm cob}$  &  P$_{\rm merg}$   &  P$_{\rm merg}^{\rm CHE}$   \\ \midrule
\multicolumn{1}{|l|}{\multirow{2}{*}{0.001}}&  \multicolumn{1}{l|}{NoCHEpreMS}   & 0     &   3.77    &   0.54    &   0        &   0.84    &   0.24    &   0        &   0.24    &   0.2     &   0    \\
\multicolumn{1}{|l|}{}&  \multicolumn{1}{l|}{CHE10preMS}   & 14.69 &   5.24    &   0.26    &   21.01    &   5.56    &   0.12    &   10.25    &   0.09    &   0.07    &   0    \\
\midrule
\multicolumn{1}{|l|}{\multirow{2}{*}{0.004}}&  \multicolumn{1}{l|}{NoCHEpreMS}    & 0    &   3.81    &   0.82    &   0    &   0.89    &   0.33    &   0    &   0.4     &   0.21    &  0   \\
\multicolumn{1}{|l|}{}&  \multicolumn{1}{l|}{CHE10preMS}  & 14.31  &   5.48    &   0.07    &  15.34 &   4.39    &   0.04    &  17.99 &   0.12    &   0.05    &  0   \\
\bottomrule
\end{tabular}
\label{tab:cob}
\end{center}
\footnotesize{From left to right column: metallicity ($Z$), name of the model, percentage of binaries that experience CHE in the simulation set (P$_{\rm CHE}$), percentage of compact object binaries produced in the simulation set (P$_{\rm cob}$), percentage of compact object binary mergers produced in the simulation set (P$_{\rm merg}$), and percentage of compact object binary mergers that evolve through CHE among all mergers of the simulation set (P$_{\rm merg}^{\rm CHE}$). The latter three columns are repeated  
for 
BBHs, BHNSs, and BNSs respectively. We call compact  binary mergers those compact object binaries that have a delay time shorter than the Hubble time (13.8 Gyr), i.e. that possibly merge within the lifetime of the Universe.}
\end{table*}


\section{Results: Compact object binaries}

In this Section, we focus on the formation of compact object binaries and compact binary mergers at low metallicity. 
Specifically, we call compact  binary mergers those compact object binaries that have a delay time shorter than the Hubble time (13.8 Gyr), i.e. that possibly merge within the lifetime of the Universe. Conversely, our definition of compact object binaries also includes BBHs, BHNSs, and BNSs with a delay time longer than the Hubble time.

Table~\ref{tab:cob} reports the fraction of binaries with one star undergoing CHE as a function of the simulation model and  metallicity. When CHE is considered, approximately one in every 7 of our secondary stars evolves into a chemically homogeneous star, both at $Z=0.001$ and $Z=0.004$. Since these stars are all progenitors of compact objects, this significantly influences the formation of compact binaries and compact binary mergers. Table~\ref{tab:cob} underscores four primary implications of the CHE scenario on the formation of compact object binaries and their mergers:

\begin{itemize}
    \item CHE enhances the formation of black holes originating from the secondary star, significantly increasing the fraction of BBH and BHNS systems. This is particularly evident for BHNSs: at $Z=0.001$ we form approximately seven times more BHNSs than in the model without CHE. The number of BHNSs exceeds even that of BBHs.\\
    
    \item The CHE10preMS model  reduces the formation of BNSs. This occurs because, in this model, metal-poor stars with ZAMS masses as low as 10 M$_\odot$ (i.e., neutron star progenitors) can become chemically homogeneous after mass accretion and develop larger stellar cores. Consequently, many stars that would have formed neutron stars instead  conclude their life as black holes.\\

    \item The most striking effect shown in the Table is that CHE  suppresses the number of all compact binary mergers produced by binary evolution. The most optimistic scenario occurs with the CHE10preMS model at a metallicity of $Z=0.001$, where the fraction of BBH and BHNS mergers is just $1/2$ of the fraction of mergers produced in the NoCHEpreMS model. At  $Z=0.004$, BBH and BHNS mergers are 
    one order of magnitude less frequent in the simulation CHE10preMS than in the model without CHE. This drop in the formation of merging systems affects  BBHs, BHNS, and BNSs at all metallicities where CHE is active. 
    This happens because of the compact radii of CHE stars. When the secondary star becomes almost homogeneous, its radius remains frozen along its evolution, and the two stars in the binary hardly interact anymore. By the time the secondary starts burning helium and evolves into a WR, the primary is either a pure helium star stripped of its envelope, or has already evolved into a compact object. The binary is then composed of two compact stars that cannot trigger any mechanism to efficiently reduce their orbital separation, such as further mass transfer episodes or a common envelope phase. The system eventually evolves into a binary compact object with an orbital separation that  is too large to lead to efficient gravitational wave emission and orbital decay. Thus, the newborn compact object binary cannot merge within the lifetime of the Universe.\\

    \item Finally, in the models where CHE is active, the fraction of BBH and BHNS mergers arising from a system with one CHE star is always relatively low. It ranges from approximately $10\%$ of the mergers for BHNSs produced at $Z=0.001$ to nearly $21\%$ of all BBH mergers at $Z=0.001$. No BNS mergers form through the CHE channel in our runs. BNS mergers typically require that their progenitors undergo at least one common-envelope phase \citep[e.g.][]{Iorio2023}, a process that happens rarely when the secondary star evolves homogeneously. 
    
\end{itemize}

   \begin{figure*}
   \centering
   \includegraphics[width=0.95\textwidth]{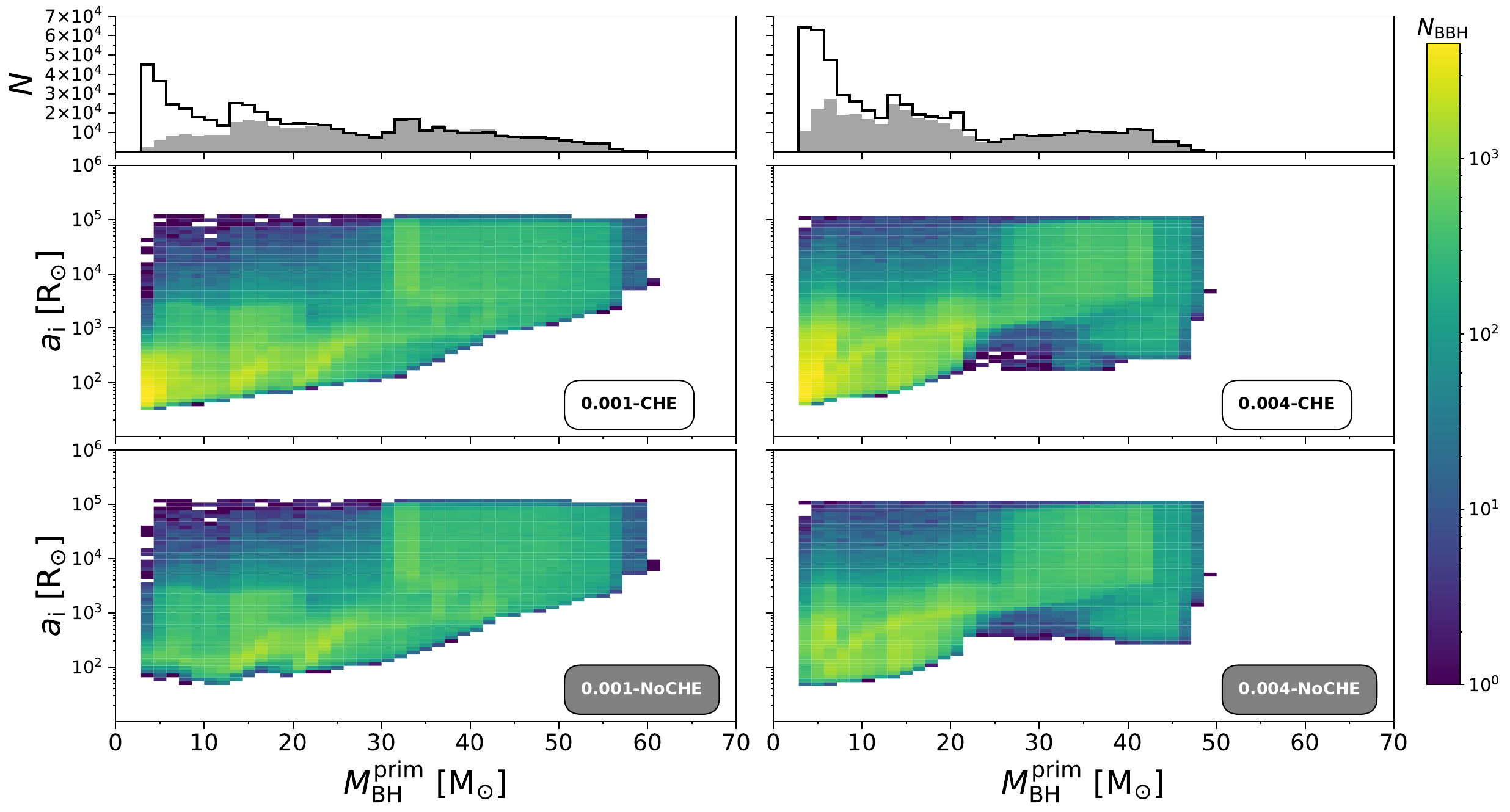}
   \includegraphics[width=0.95\textwidth]{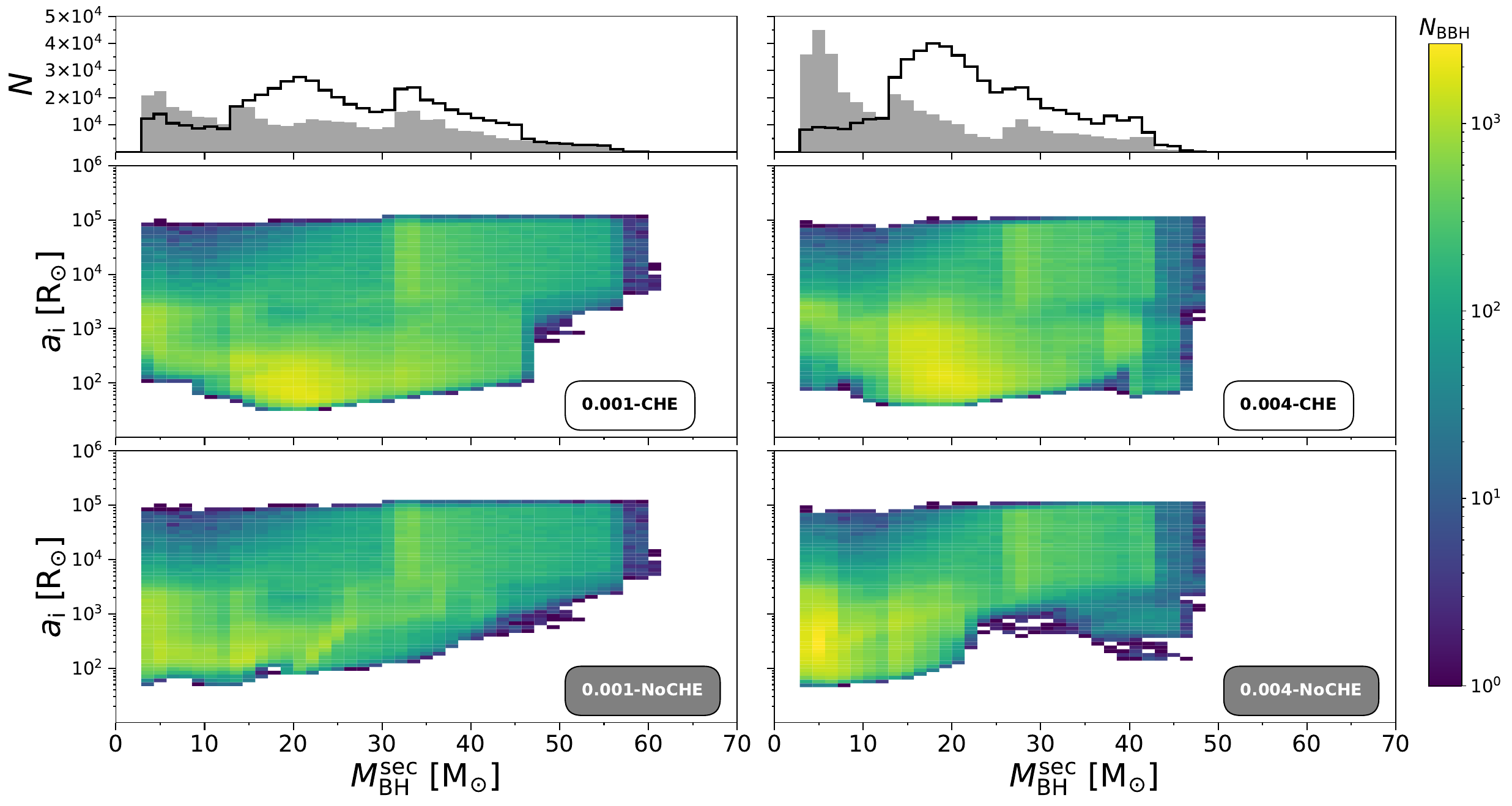}
      \caption{Masses of the black holes in BBH systems produced by the primary stars (upper panels), and secondary stars (lower panels) as a function of the initial semi-major axis of the progenitor binaries. The 2d-histograms show the number of BBHs per bin produced in the CHE10preMS models and the NoCHEpreMS models at metallicities $Z=0.001$ and $Z=0.004$. The marginal histograms show the distributions of the black hole masses for the CHE10preMS (black histograms) and the NoCHEpreMS (grey-filled histograms) models at $Z=0.001$ and $Z=0.004$.}
         \label{fig:BBH}
   \end{figure*}

   \begin{figure*}
   \centering
   \includegraphics[width=0.95\textwidth]{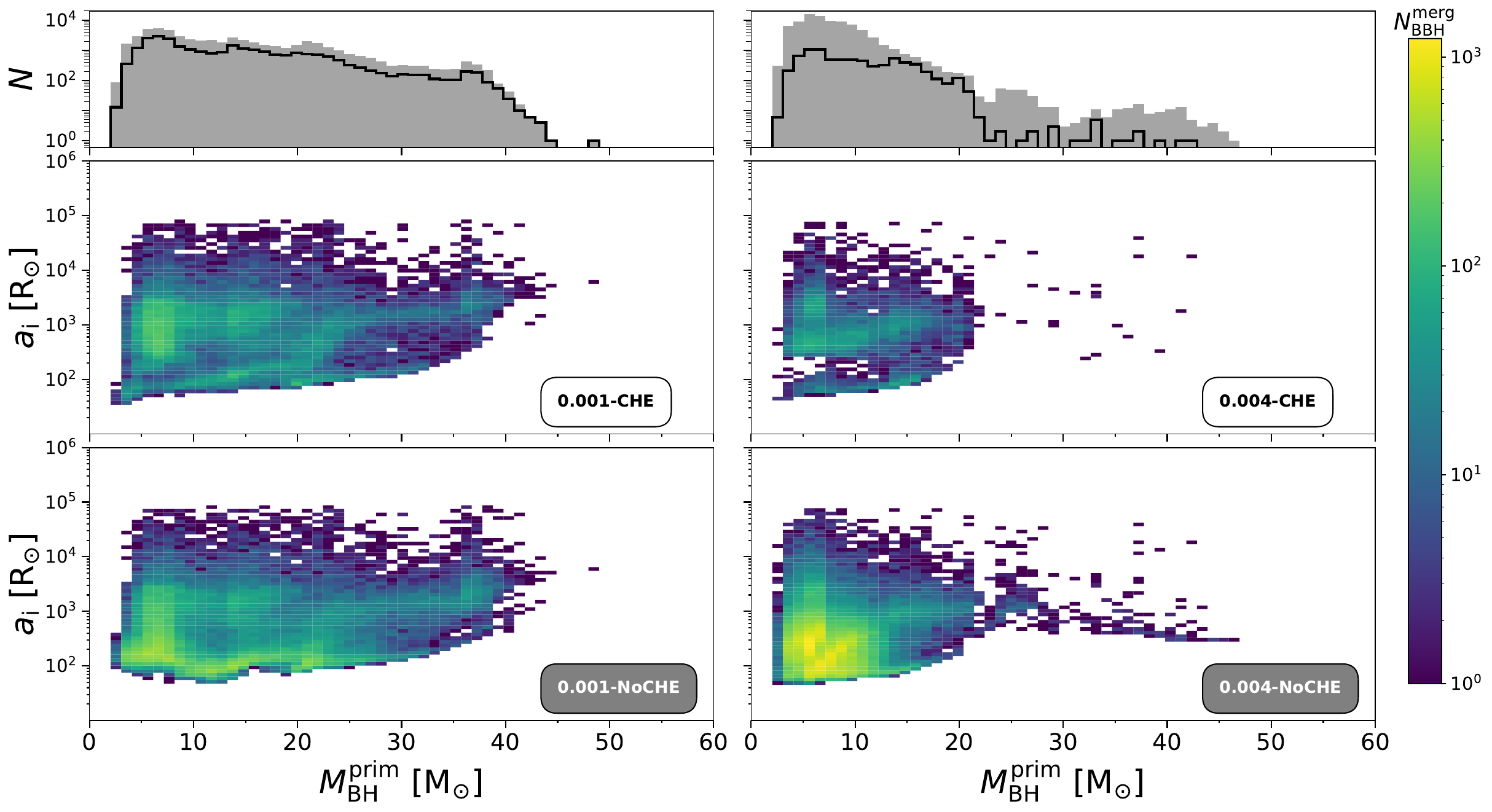}
   \includegraphics[width=0.95\textwidth]{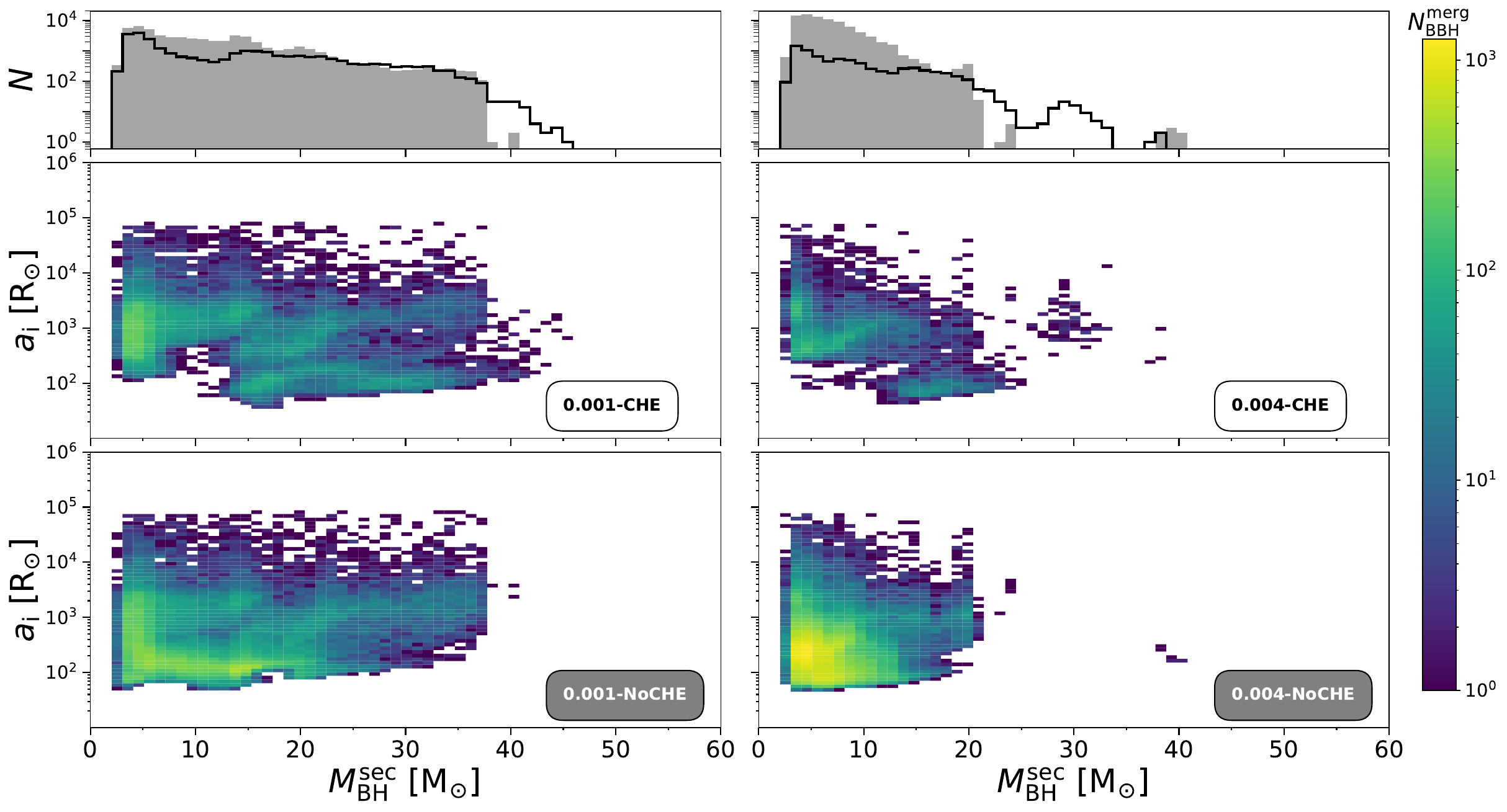}
      \caption{Same as Fig.~\ref{fig:BBH} but for BBH mergers only.}
         \label{fig:BBH_mergers}
   \end{figure*}


In the upcoming sections, we discuss the properties of  BBHs and BHNSs produced in our fiducial models CHE10preMS and NoCHEpreMS models. We present a further comparison with other models in Appendix~\ref{sec:app1}.

\subsection{Binary black holes}\label{sec:bbh}

Figure~\ref{fig:BBH} presents the masses of black holes in BBH systems, alongside the initial semi-major axis of their progenitor binary, for both the CHE10preMS and NoCHEpreMS models at metallicities of $Z=0.001$ and $Z=0.004$. 
All BBH populations have an initial semi-major axis larger than a few ten solar radii, below which binary evolution is likely to lead to a merger. The black hole masses range from a minimum of $3\,$M$_{\odot}$ to around $60\,$M$_{\odot}$ at $Z=0.001$ due to pulsational and pair-instability supernovae, and up to $50\,$M$_{\odot}$ at $Z=0.004$, with stellar winds limiting the formation of more massive black holes.

This Figure reveals that the models with CHE result in a significantly higher fraction of BBHs than those without CHE (Table~\ref{tab:cob}). As discussed in Section~\ref{sec:Lwr}, WR stars formed through CHE are more massive at birth than their non-CHE counterparts. Due to limited wind mass loss at low metallicity combined with the efficient rotational mixing process, a significant portion of their mass is retained until the pre-supernova phase, forming more massive compact objects. In the case of BBH systems, up to $31\%$ of those produced exclusively through CHE would have evolved into BHNS binaries in the model without CHE. All the other BBHs produced exclusively in the CHE model would have resulted in either a stellar merger or the destruction of one star by unstable mass transfer when evolving without CHE. 

In the NoCHEpreMS models, the most and least massive black holes in BBH systems are typically produced by the evolution of the primary and secondary star respectively, as shown by the marginal panels in Figure~\ref{fig:BBH}. 
In contrast, when CHE is active, 
the secondary star becomes the progenitor of the most massive black hole in the BBH system, while the primary star produces the least massive black hole. This results in an overabundance of black holes produced by the secondary star with masses greater than $12\,$M$_{\odot}$, peaking at approximately $20\,$M$_{\odot}$. At the same time, black holes with masses below $\approx12\,$M$_{\odot}$ are less abundant among secondary remnants than in the NoCHEproMS model. This drop occurs either because the progenitor stars were not massive enough to undergo CHE, or, if they did experience CHE, they eventually formed more massive black holes, shifting to the upper portion of the distribution.

In contrast, black holes formed by the primary star exhibit a similar distribution in both models, with and without CHE, except at lower masses where the black hole distribution in the CHE10preMS model presents a large peak below $\approx10\,$M$_{\odot}$. These black holes are part of the BBH systems produced exclusively through CHE and result from binaries where the secondary star would not have evolved into a black hole in the absence of CHE.

Figure~\ref{fig:BBH} also shows that most CHE BBHs originate from progenitor binaries with relatively short initial orbital separations. For CHE to occur, the mass transfer process must be efficient and stable over a sufficient timescale to allow the secondary star to spin up and become chemically homogeneous. This is typically achievable only in binary systems with low to intermediate orbital separations, where the primary star can fill its Roche lobe and initiate mass transfer without the risk of triggering a merger.\\

Figure~\ref{fig:BBH_mergers} is the same as Fig.~\ref{fig:BBH}, but shows only BBH mergers (instead of all BBHs). The figure highlights that the CHE formation channel is significantly less efficient in producing BBH mergers than the standard binary evolution channel, as also shown in Table~\ref{tab:cob}. The CHE model results in fewer BBH mergers, with $\leq 21\%$ formed through CHE (Table~\ref{tab:cob}), primarily influencing the high-mass end of the black hole mass distribution. Black holes produced by CHE favors large masses for the secondary star remnant.


   \begin{figure*}
   \centering
   \includegraphics[width=0.95\textwidth]{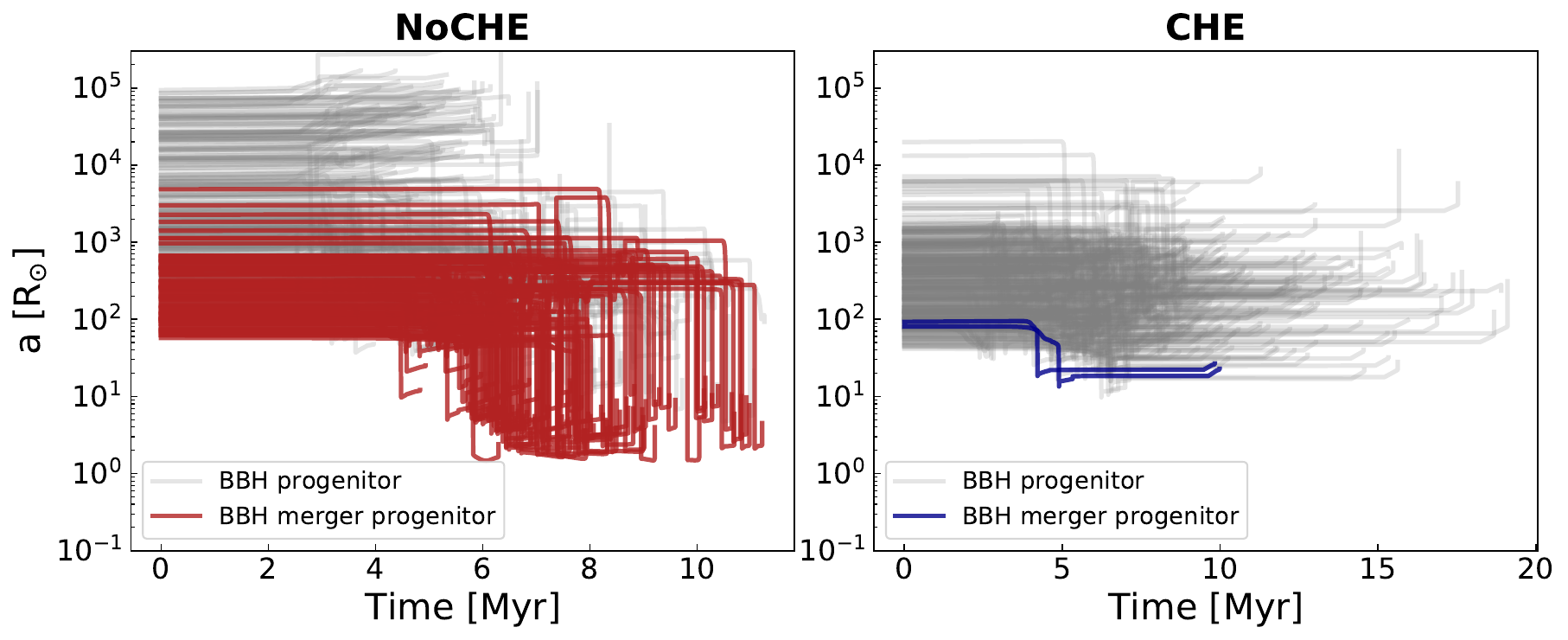}
      \caption{Evolution of the semi-major axis over time for a sample of BBH progenitors in models with (right-hand panel) and without (left-hand panel) CHE. The evolution traces from the birth of the binary to the formation of the BBH (grey lines) or a BBH merger (red and blue lines). The two panels depict models at $Z=0.004$, with similar results found at $Z=0.001$.}
         \label{fig:sma_evol}
   \end{figure*}

   \begin{figure*}
   \centering
   \includegraphics[width=0.49\textwidth]{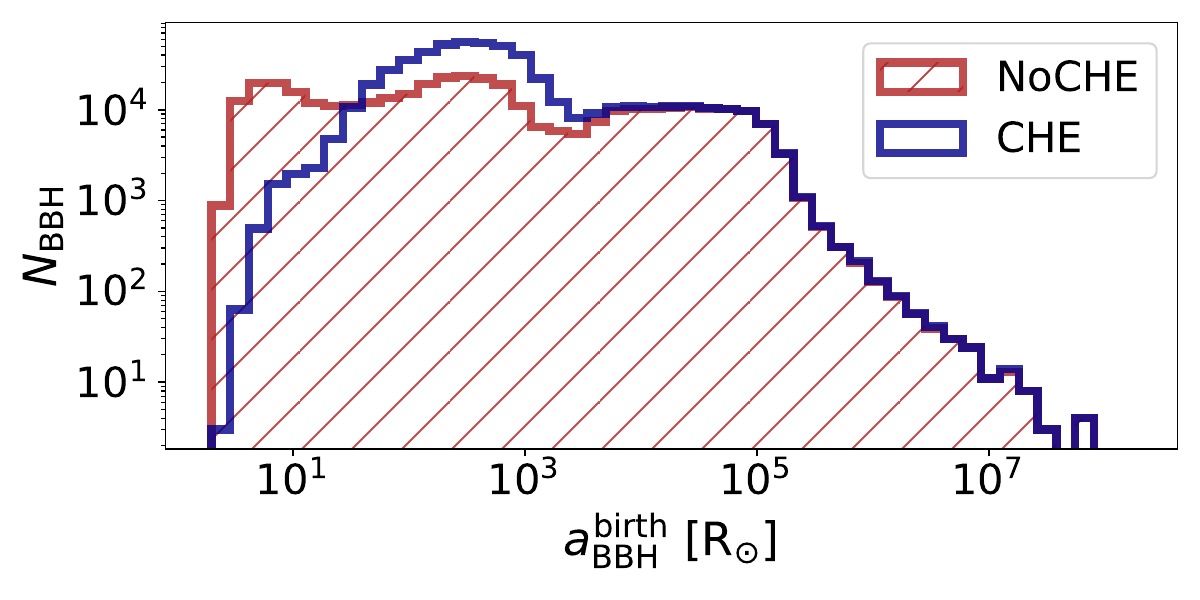}
   \includegraphics[width=0.49\textwidth]{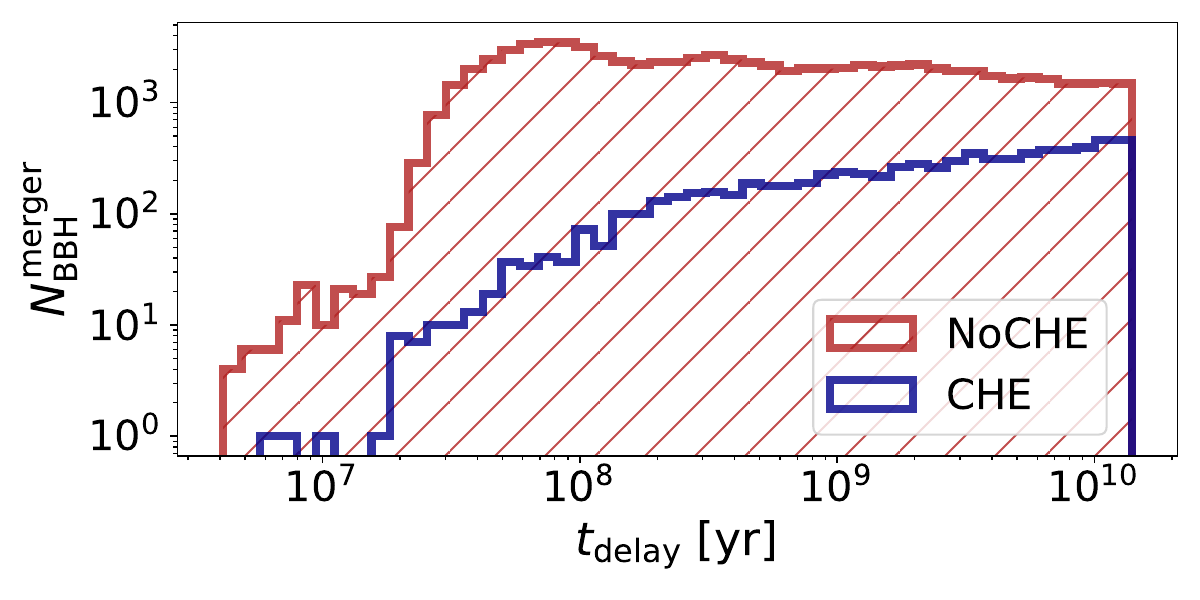}
      \caption{Left panel: Distribution of the semi-major axis at the formation of the second black hole for all BBH systems in models with (blue histogram) and without (red histogram) CHE. Right panel: Delay time distribution of all BBH mergers in models with (blue histogram) and without (red histogram) CHE. Both panels present results for models with $Z=0.004$. }
         \label{fig:a_birth}
   \end{figure*}

   \begin{figure*}
   \centering
   \includegraphics[width=0.95\textwidth]{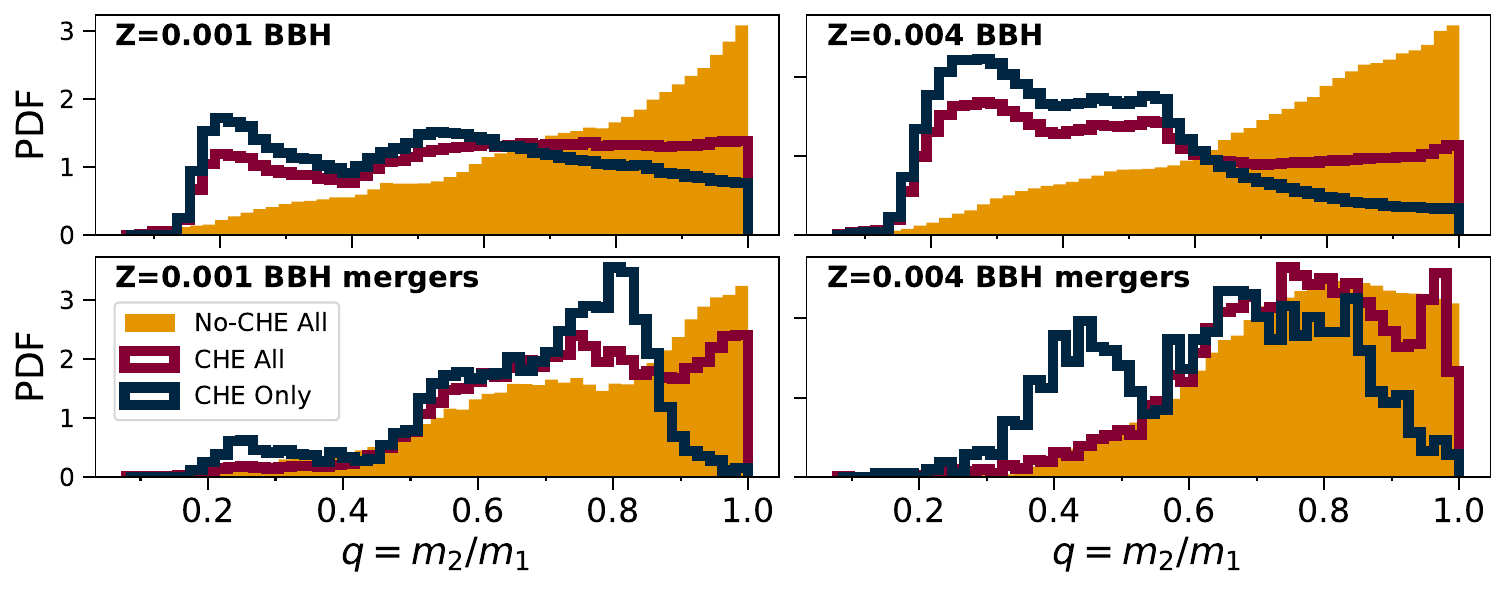}
      \caption{Mass ratio distribution for the BBHs (upper panels) and BBH mergers (lower panels) in the NoCHEpreMS (orange-filled histograms) and CHE10preMS (red histograms) models at $Z=0.001$ (left) and $Z=0.004$ (right). The plot also shows the distribution only of the binaries in the CHE10preMS models that experienced CHE (blue histograms).
              }
         \label{fig:qplot}
   \end{figure*}

The reason for this dearth of BBH mergers in the CHE model becomes evident when examining Figure~\ref{fig:sma_evol}. The two panels show the time evolution of the semi-major axis for a sample of BBH progenitors. In the NoCHEpreMS model, the progenitors of BBH mergers are characterized by a drop in the semi-major axis at an age $\approx6-10\,$Myr, which is due to a common-envelope phase. This evolutionary stage results in a dramatic contraction of the orbit over a short timescale, which appears as a vertical path in the orbital evolution line on the left-hand panel. The common envelope phase is essential for producing BBHs with sufficiently short orbital separations to allow gravitational waves to become efficient.

In contrast, BBH progenitors in the CHE10preMS models do not exhibit such a dramatic drop in their semi-major axes, as shown in the right-hand panel of Figure~\ref{fig:sma_evol}. Chemically homogeneous stars remain compact throughout their evolution, rarely filling their Roche lobe. Consequently, most binary systems do not undergo any subsequent mass transfer phase. In the absence of a common envelope or late mass transfer to significantly reduce the orbital separation, BBHs formed through CHE can only merge under two specific conditions: (i) if the progenitor binary star has a relatively small initial semi-major axis ($a_{\rm i} < 100,\mathrm{R}{\odot}$), or (ii) if the initial semi-major axis is intermediate ($a{\rm i} > 400,\mathrm{R}_{\odot}$) and the natal kick of the secondary star induces the formation of an eccentric BBH system.

The impact of these two distinct evolutionary pathways for BBH progenitors, with and without CHE, are visible in both panels of Figure~\ref{fig:a_birth}. The left-hand panel shows the semi-major axis at the formation time of the second black hole, for all BBHs in the CHE10preMS and NoCHEpreMS models at $Z=0.004$. In the model without CHE, most progenitors of BBH mergers undergo a common envelope phase, leading to orbital separations $a^\mathrm{birth}_\mathrm{BBH}\lesssim{}10\,$R$_{\odot}$. Conversely, BBHs produced through CHE form with a significantly larger semi-major axis (peaking at almost $10^{3}\,$R$_{\odot}$), due to the absence of late mass-transfer episodes. These BBHs are born with semi-major axes too large to efficiently emit gravitational waves, preventing most of them from reaching coalescence within the age of the Universe. At the same time, BBH systems that merge within the age of the Universe through CHE require more time to reach coalescence than those formed through the standard evolutionary scenario. This is illustrated in the right-hand panel of Figure~\ref{fig:a_birth}, which shows the delay time distribution for BBH mergers in the CHE10preMS and NoCHEpreMS models at $Z=0.004$. Delay time is the period between the formation of a stellar binary and the merger of the resulting BBH system. Because of the larger semi-major axis at birth, BBH mergers produced through CHE are characterized by larger delay times than their NoCHE counterpart mergers. This also largely impacts the merger rate density evolution with redshift, as discussed in Section \ref{sec:MRD}.\\

Finally, CHE strongly affects the mass-ratio distribution of BBHs. Figure~\ref{fig:qplot} shows that BBHs formed through CHE tend to exhibit an asymmetric mass ratio within the range of $0.2-0.6$ at both $Z=0.001$ and $Z=0.004$. In the standard evolutionary channel, mass transfer generally tends to equally redistribute the mass of the system between the two stars, such that the final mass ratio of BBH mergers tends to $1$. In contrast, when the secondary stars evolve chemically homogeneous, most of the mass accreted from the primary gets fully mixed within the stellar interior of the secondary star. This mass is neither redistributed within the system through subsequent mass transfer episodes nor lost during common envelope phases. The secondary retains most of the accreted mass from the primary, and it eventually becomes the most massive component of the binary system. In some cases, the secondary star accretes enough mass to become more than twice as massive as the post-mass transfer primary star.


This behavior was anticipated in Figure~\ref{fig:BBH}, which shows that the secondary star is likely to produce the most massive black hole in the system when evolving through CHE. It is important to note that most binary progenitors of these asymmetric BBH systems formed through CHE initially have a mass ratio close to $1$. This is because a near-equal mass ratio is necessary for the system to undergo a stable mass transfer phase, enabling accretion-induced CHE. In contrast, systems with a lower mass ratio are more likely to experience unstable mass transfer, preventing the binary from undergoing CHE.

On the other hand, BBH mergers do not follow the same distributions as the rest of BBHs. This is because only a relatively low fraction of the BBH merger population can reach the merger after experiencing CHE (see Table~\ref{tab:cob}). Even with CHE, asymmetric BBH mergers remain relatively rare. The fraction of BBH mergers with $q<0.4$ is typically under $4\%$ of the total BBH merger population. 

\subsection{Black Hole - Neutron Star binaries}

\begin{figure*}
   \centering
   \includegraphics[width=0.95\textwidth]{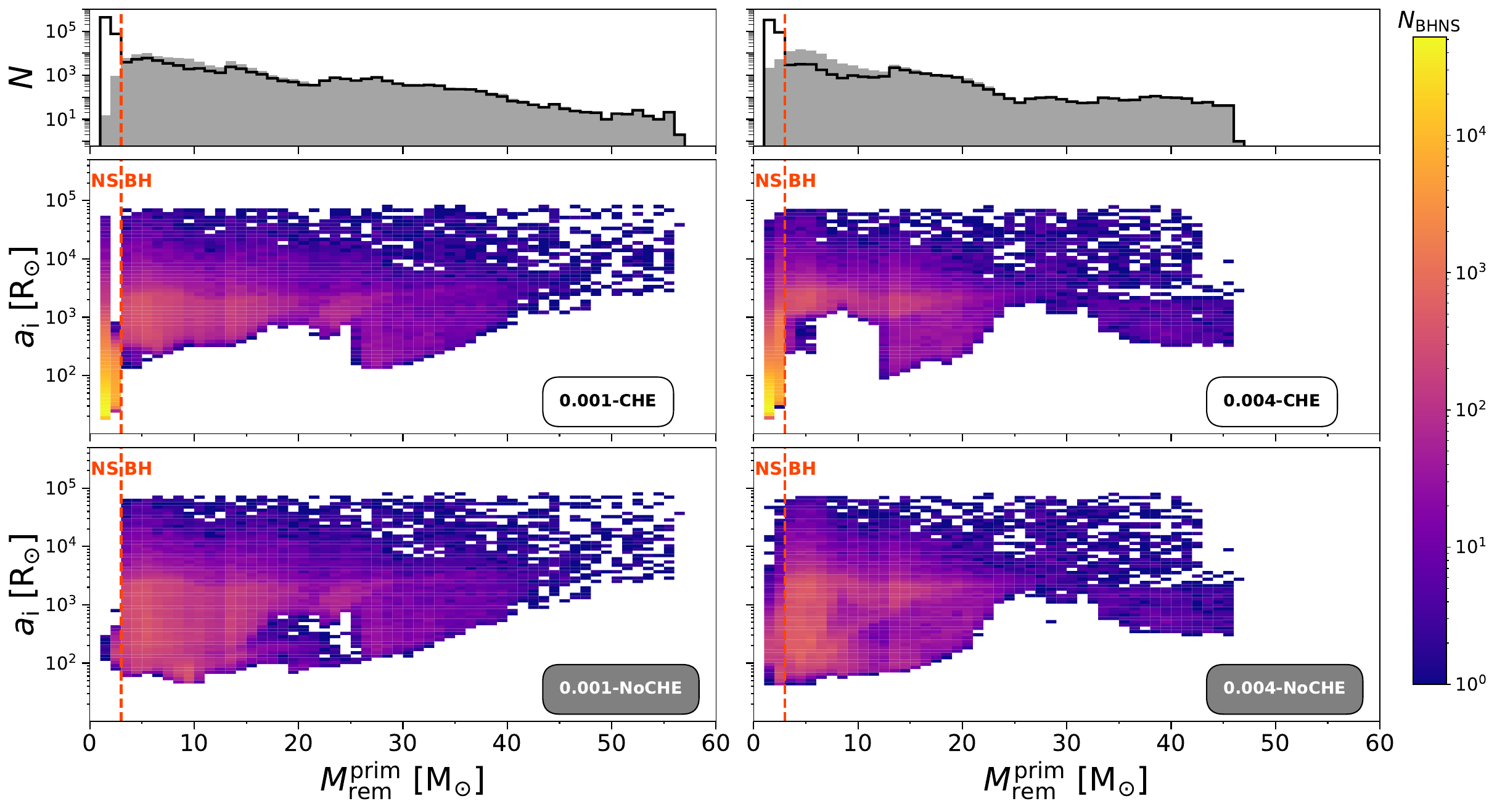}
   \includegraphics[width=0.95\textwidth]{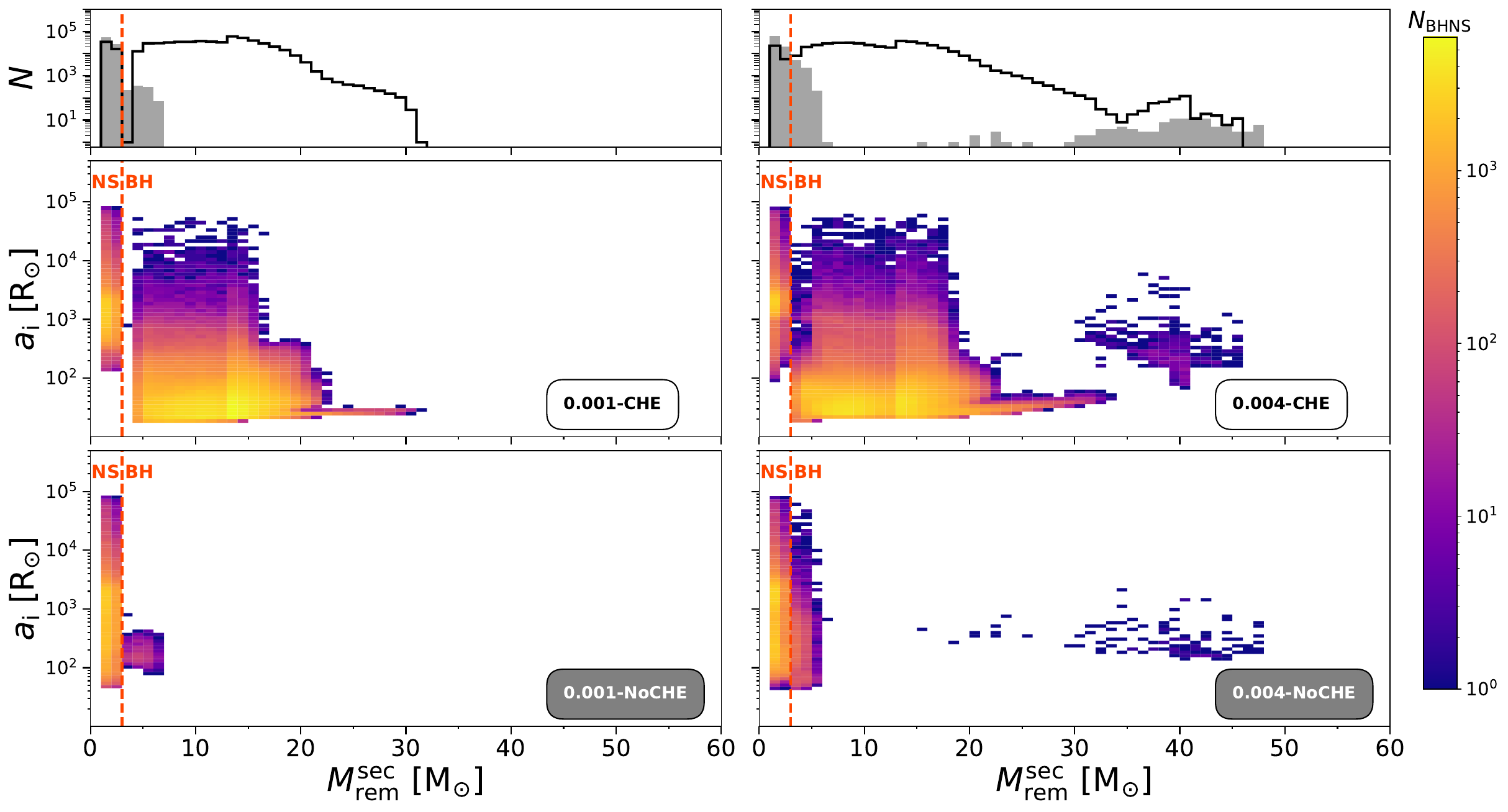}
      \caption{Mass of the black holes and neutron stars in BHNSs as a function of the initial semi-major axis of the progenitors. The upper (lower) panels show the mass of the compact object produced by the primary (secondary) star. The orange dashed line represents the threshold at $3\,$M$_{\odot}$: compact objects below this limit are classified as neutron stars, while those above it are black holes. The 2d-histograms show the number of BHNSs per bin produced in the CHE10preMS  and  NoCHEpreMS models at metallicities $Z=0.001$ and $Z=0.004$. The marginal histograms show the distributions of the compact remnant masses for the CHE10preMS (black histograms) and  NoCHEpreMS (grey-filled histograms) models at $Z=0.001$ and $Z=0.004$.}
         \label{fig:BHNS}
   \end{figure*}

Figure~\ref{fig:BHNS} displays the masses of the compact object components in all the  BHNS systems as a function of the initial orbital separation of their binary progenitor in the CHE10preMS and NoCHEpreMS models, at both $Z=0.001$ and $Z=0.004$. The Figures highlight the significant growth in the number of NSBH formed through CHE, as also reported in Table~\ref{tab:cob}.

BHNSs formed in the NoCHEpreMS model follow the traditional formation pathway in which the primary, stripped of its outer envelope, collapses into a black hole, while the secondary, after possibly undergoing a further mass transfer phase, evolves into a neutron star. This formation channel is followed by almost $99\%$ of the BHNSs at $Z=0.001$, and by $91\%$ of the BHNSs at $Z=0.004$. 

In contrast, in the CHE10preMS model, the secondary star is generally the progenitor of the black hole while the primary star evolves into a neutron star. This happens because the secondary, undergoing CHE evolution, retains more mass and eventually collapses into a black hole, while the primary has already evolved into a neutron star. This is the evolution history of more than $91\%$ and $93\%$ of the BHNS systems at $Z=0.001$ and $Z=0.004$. In these systems, if the secondary star had evolved without CHE, it would have produced a neutron star instead of a black hole. Without CHE, up to $5\%$ of these systems would have evolved into a BNS instead of a BHNS, while the others would have resulted in a stellar merger or the destruction of one star by unstable mass transfer.

\begin{figure*}
   \centering
   \includegraphics[width=0.95\textwidth]{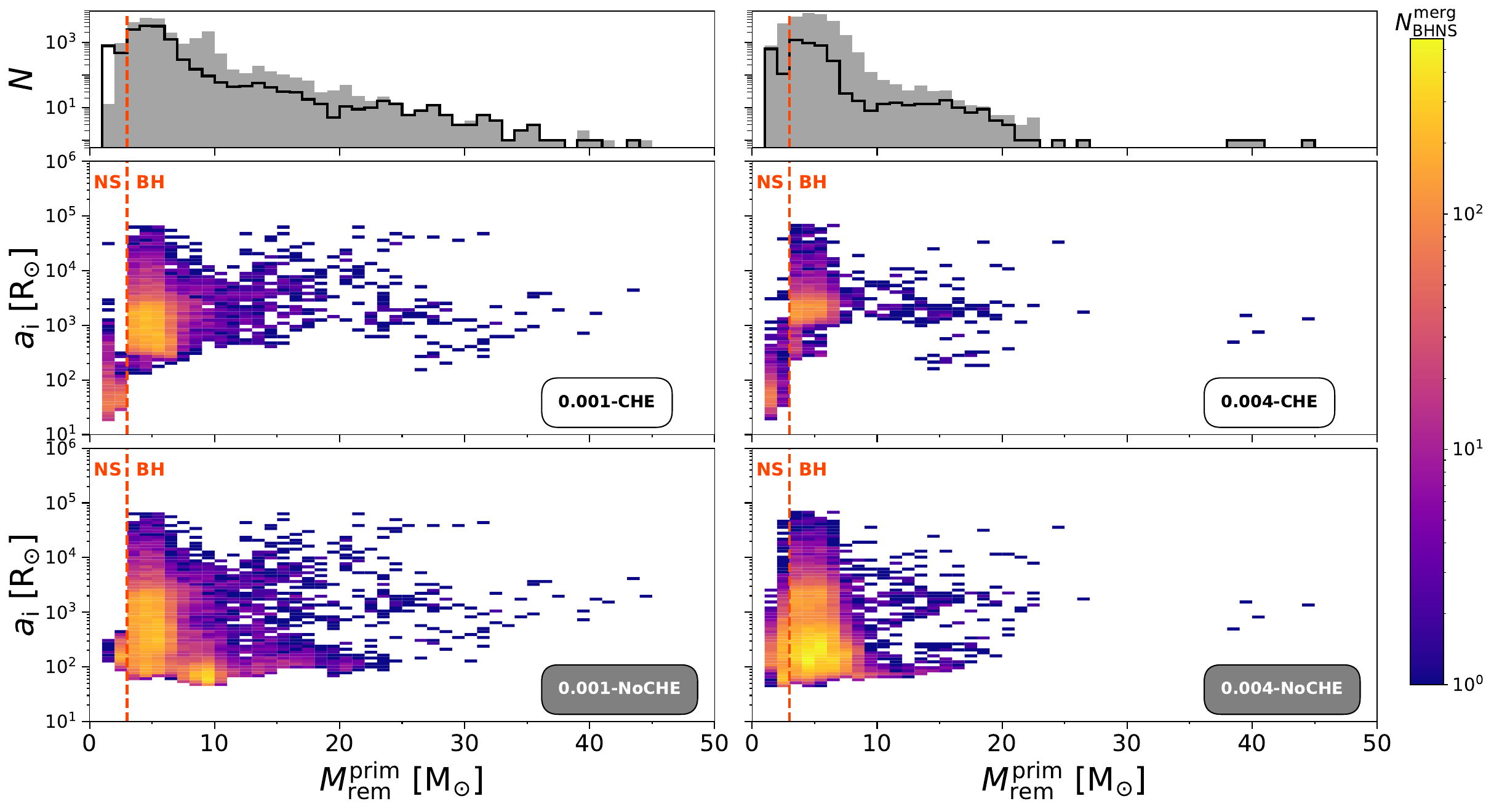}
   \includegraphics[width=0.95\textwidth]{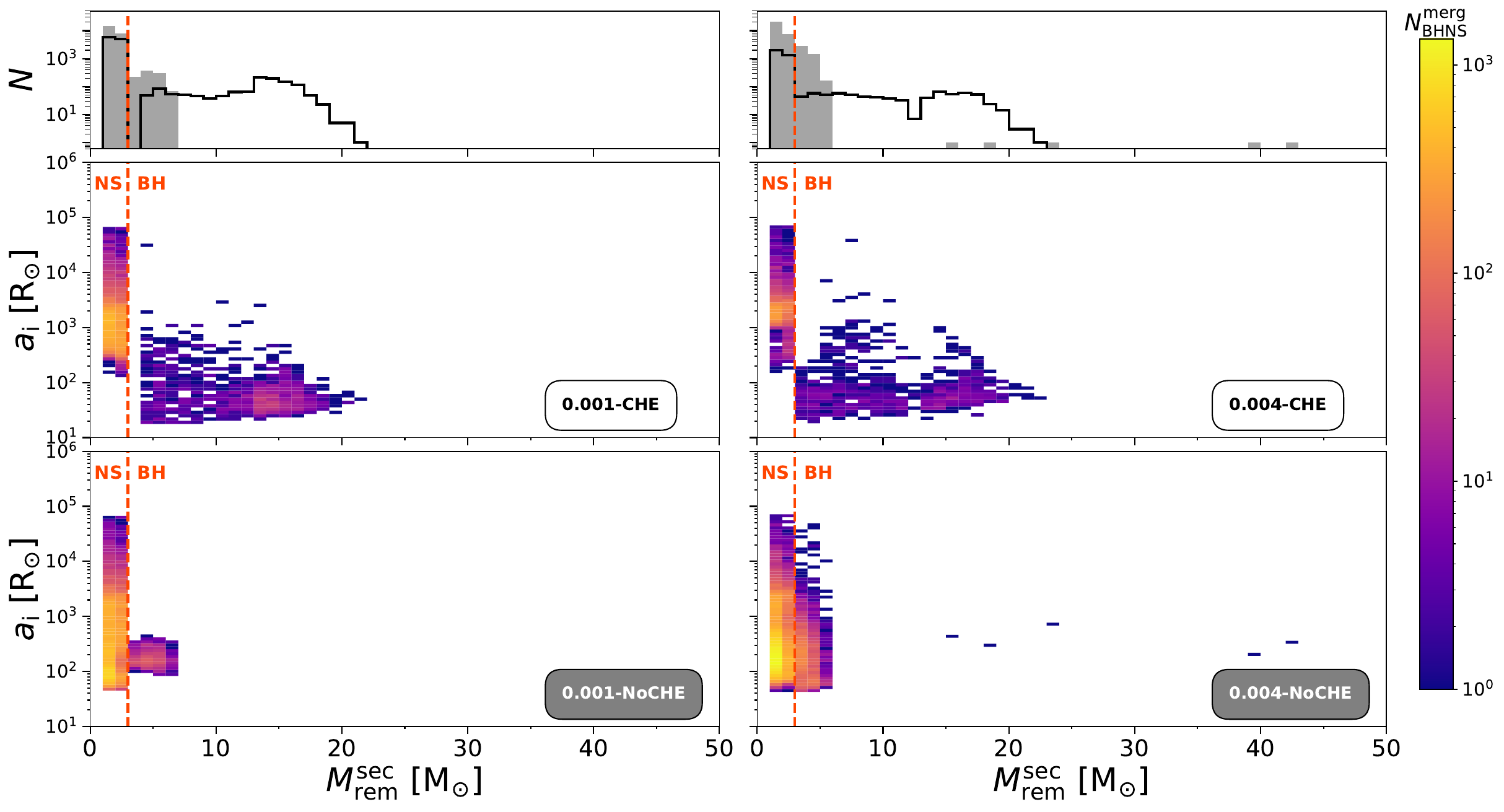}
      \caption{Same as Fig.~\ref{fig:BHNS} but for BHNS mergers only.
              }
         \label{fig:BHNS_mergers}
   \end{figure*}

BHNSs formed through CHE face the same fate as CHE-BBH systems: after the initial mass transfer episode, the two stars in the binary stop interacting, and the orbital period remains frozen at large values. This  suppresses the formation of BHNS mergers, as shown in Figure~\ref{fig:BHNS_mergers} and  Table~\ref{tab:cob}. Nevertheless, CHE is the primary formation channel for BHNSs with black hole mass in the range $10-20\,$M$_{\odot}$.

\section{Discussion}

\subsection{Accretion vs tidally-induced CHE}\label{sec:discussion}

Here, we have focused on accretion-induced spin-up during Roche lobe overflow, which is generally considered the primary mechanism for triggering CHE in massive stars \citep[e.g.][]{Ghodla2023}. This assumption is supported by the findings of \cite{Sana2012}, which indicate that up to $40\%$ of massive stars in binary systems may experience spin-up due to accretion from their companions. In principle, CHE can also be achieved in both stars of a tight binary system through tidal interactions \citep[][]{DeMink2009}. In this scenario, the binary reaches tidally locked equilibrium, where the orbital separation between the two stars is reduced by tidal forces to the point where the orbital period is equal to the rotational period of the two stars. As a result, the two stars undergo rapid spin-up, experience rotational mixing, and possibly transition into the CHE state. The main difference compared to accretion-induced CHE is that the tidal spin-up mechanism leads to the formation of more massive BBH systems with nearly equal-mass components \citep[][]{Mandel2016,Marchant2016}. Additionally, the tidal spin-up mechanism does not hinder the formation of BBH mergers, as it exclusively occurs in binaries with already short orbital separations.

\cite{Riley2021} explored the formation of massive BBH mergers through tidal spin-up CHE. They find that only $\sim0.2\%$ of their binary population has at least one component that evolves through CHE. This is because only extremely tight binaries where the two stars are tidally locked can achieve CHE. These conditions are generally met when the period of the binary is smaller than $2$ days and both stars are more massive than $\approx20\,$M$_{\odot}$ \citep{Song2016}. In our binary population, only $3\%$ of the binaries have an initial orbital period below $2$ days. Additionally, in our CHE10preMS model at $Z=0.001$, only $\approx19\%$ of all the binaries that evolve through CHE have the orbital period below $2$ days when the secondary becomes a chemically homogeneous star. This fraction is reduced to $13\%$ at $Z=0.004$. In most of these systems, at least one of the two stars in the binary is typically in the main sequence and possesses a radiative envelope, making tidal forces relatively ineffective. Hence, tidally induced CHE would have affected only a negligible fraction of our binaries. We will include the effect of CHE induced by tidal forces in a forthcoming paper.

\subsection{Impact of CHE on merger efficiency and merger rate density}\label{sec:MRD}

\begin{figure}
    \includegraphics[width=0.49\textwidth]{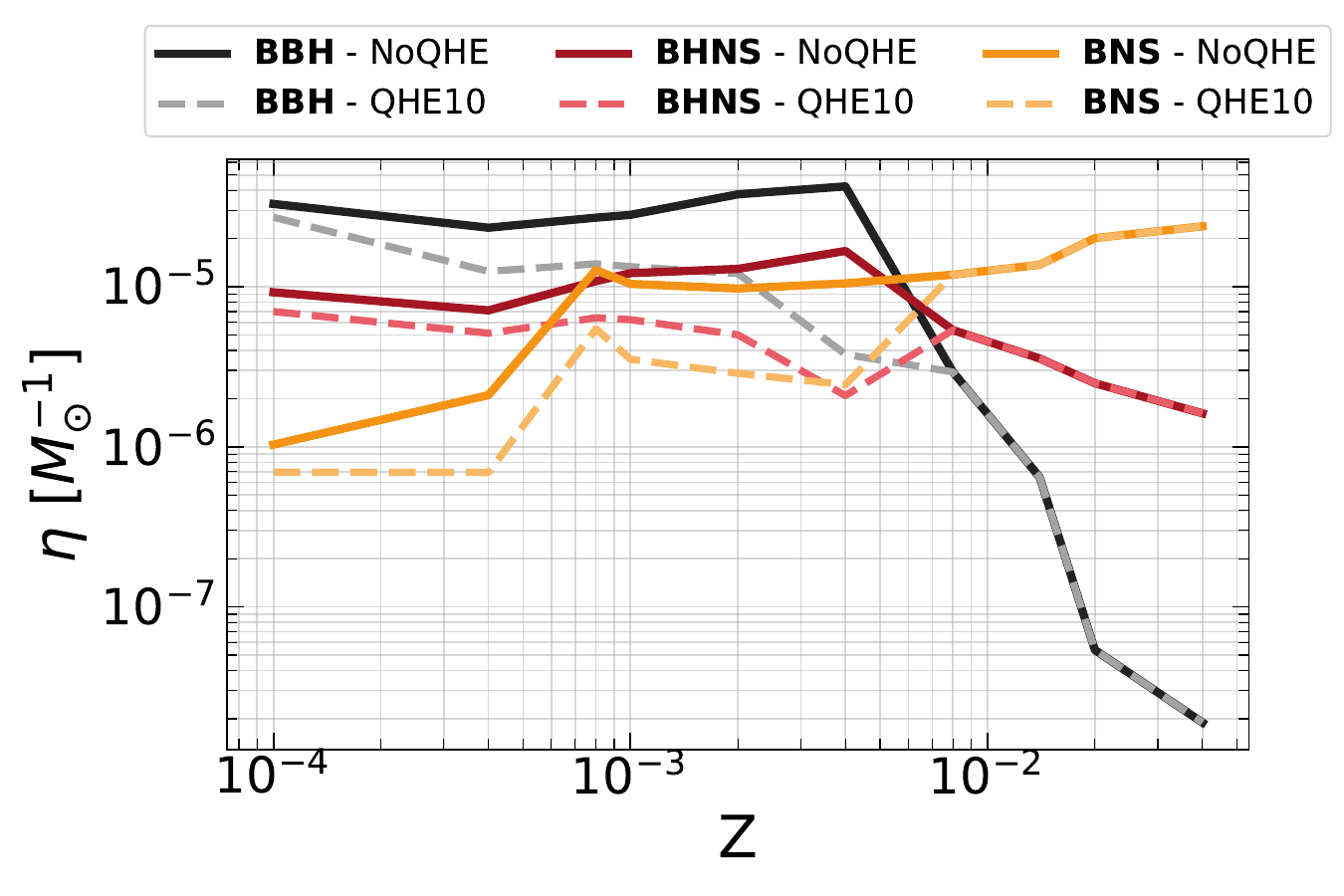}
    \caption{Merger efficiency ($\eta$) of BBHs (black), BHNSs (red), and BNSs (yellow) mergers as a function of metallicity ($Z$). Dashed (solid) lines represent mergers from the NoCHEpreMS (CHE10preMS) model.}
    \label{fig:ME}
\end{figure}


   \begin{figure}
   \centering
   \includegraphics[width=0.47\textwidth]{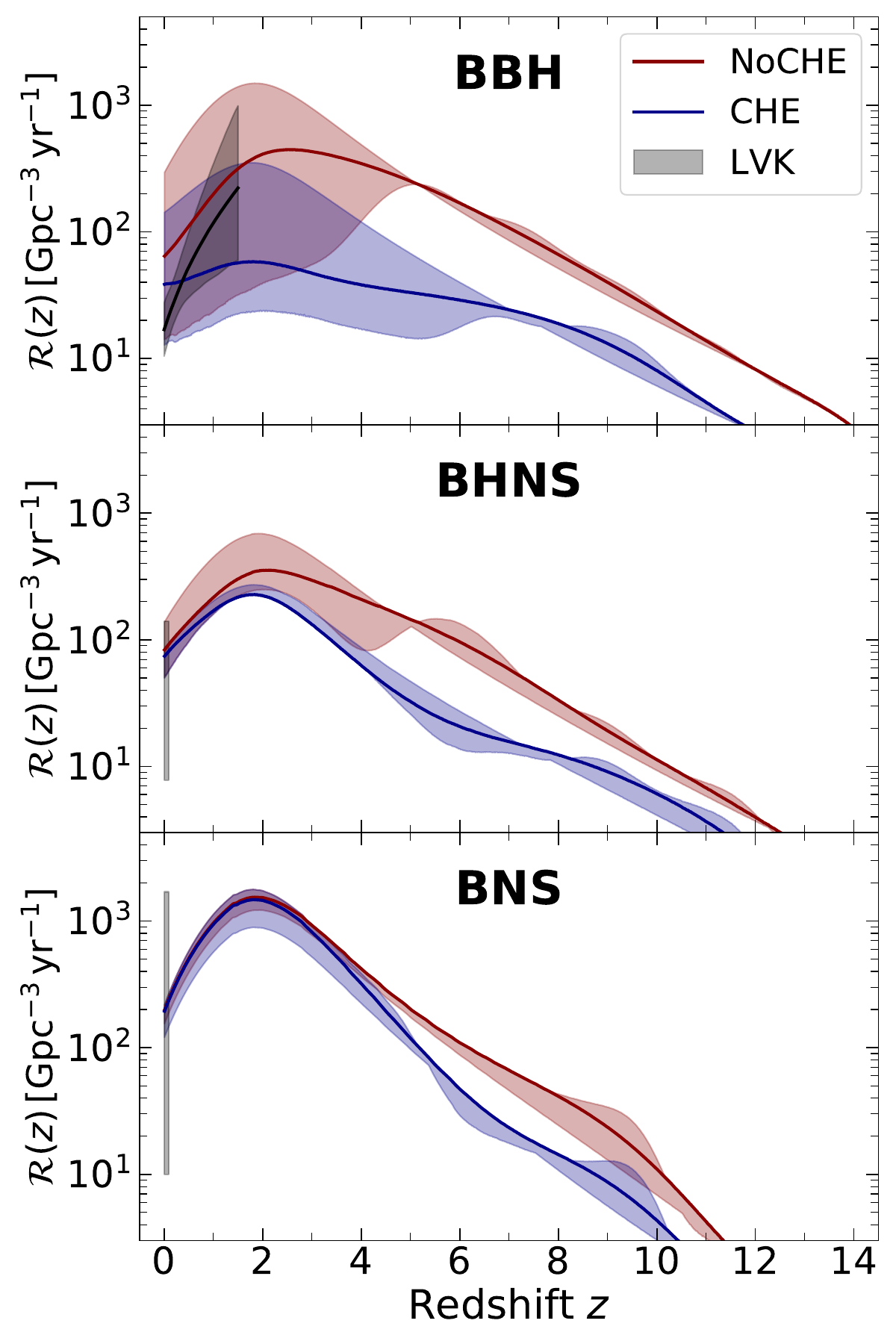}
      \caption{Merger rate density evolution of BBHs (upper panel), BHNSs (central panel), and BNSs (lower panel) as a function of redshift. The red (blue) solid lines represent the merger rate density for mergers produced in the fiducial model NoCHEpreMS (CHE10preMS), assuming a metallicity spread of $\sigma_{Z}=0.3$. The blue and red shaded regions highlight the portions of the plot enclosed by the merger rate density evolution calculated with metallicity spreads of $\sigma_{Z}=0.1$ and $0.7$. The grey shaded region show the $90\%$ credible interval for the merger rate density evolution with redshift (upper panel), and the local merger rate density (central and bottom panel) reported by \cite{Abbott2023}.}
         \label{fig:MRD}
   \end{figure}

How do our findings impact the efficiency of formation channels in producing compact binary mergers? According to our results, we expect that over $14\%$ of a binary population undergoes accretion-induced CHE. This significantly boosts the formation of more massive compact binary systems at low metallicity, while simultaneously reducing the rate of mergers (Table~\ref{tab:cob}). For instance, the production of BBHs increases by a factor of $\approx1.4$ at metallicities $Z=0.001$ and $Z=0.004$, whereas the number of BBH mergers decreases by factors of $\approx2$ and $\approx12$ at the same metallicities, respectively. These results reflect the nature of accretion-induced CHE, which predominantly affects binaries that would have otherwise formed mergers through traditional evolutionary pathways. Instead, CHE leads to the creation of wider, more massive, and more numerous compact binaries that are less likely to merge within the lifetime of the Universe. This reduction in the compact binary merger population also impacts the relative merger efficiency and the merger rate density of both the isolated and dynamical formation channels. 

The merger efficiency $\eta(Z)$ is defined as the total number of compact object binary that merge within a Hubble time $\mathcal{N}_{\rm TOT}(Z)$ divided by the total initial stellar mass $M_{*}(Z)$\footnote{We computed $M_{*}(Z)$ including the correction factors $f_{\rm IMF}=0.252$ and $f_{\rm bin}=0.5$. The first factor accounts for the incomplete sampling of the IMF, while the second accounts that half of the population's mass is stored in binaries. See also \cite{Santoliquido2021} and \cite{Iorio2023}.} of its progenitor population at a given metallicity $Z$. This can be written as
\begin{equation}
    \eta(Z)=\frac{\mathcal{N}_{\rm TOT}(Z)}{M_{*}(Z)}.
\end{equation}
Figure \ref{fig:ME} shows the merger efficiency as a function of metallicity for BBH, BHNS, and BNS mergers in models with and without CHE.  
As  widely discussed in the literature \citep[e.g.,][]{Giacobbo2018,Klencki2018,Iorio2023,VanSon2024}, the merger efficiency of BBHs drops at high $Z$ and increases by orders of magnitude at low $Z$, because of the metallicity-dependence of line-driven stellar winds. Rapid mass loss by winds at high metallicity forces massive stars to become WR and prevents them from filling their Roche lobe: in absence of efficient mass transfer the binary semi-major axis remains too wide to produce merging BBHs \citep[e.g.,][]{Korb2024}. The same happens for BHNS progenitors, even if with less difference between metal-poor and metal-rich binaries. For the stellar tracks we adopt in \textsc{sevn}, we find an opposite trend for BNSs, whose merger efficiency increases at higher metallicity. This trend of the BNS merger efficiency  with metallicity is only present for $\alpha_\textrm{CE}\leq{}1$ and is the consequence of premature mergers of potential BNS progenitors during common envelope \citep{Iorio2023}. 


Comparing the models with and without CHE, we clearly see that CHE reduces merger efficiency, especially at metallicity $Z\approx{0.002-0.004}$. The merger efficiency of BBHs and BHNSs is suppressed by nearly one order of magnitude at $Z=0.004$, whereas for BNSs the quenching is a factor of $\approx 4$. The differences in merger efficiency between the model with and without CHE decrease with metallicity. This happens because, in the CHE model, at very low metallicities only a small fraction of mergers undergo CHE, while the majority evolve through a common-envelope phase (Table~\ref{tab:cob}), which is the same as in the non-CHE model. 


Based on the merger efficiency, we can now estimate the merger rate density $\mathcal{R}(z)$, by assuming a metal-dependent cosmic star formation rate \citep{Madau2017}, as described in Appendix~\ref{app:mrd}. The most important free parameter of the model is the assumed spread $\sigma_\mathrm{Z}$ around the average stellar metallicity.

Figure \ref{fig:MRD} shows the evolution of merger rate density with redshift in the comoving frame for BBH, BHNS, and BNS mergers. Accretion-induced CHE effectively quenches the merger rate density of BBHs along all the cosmic history, while it has a milder effect for BHNSs and BNSs. 

The peak of the merger rate density shifts toward lower redshifts for BBHs and BHNSs in the CHE model compared to the standard binary evolution scenario. This happens because the CHE model  produces BBH and BHNS mergers with longer delay times compared to the model without CHE (Fig.~\ref{fig:a_birth}). As a result, the difference between the two scenarios with and without CHE is minimum at redshift zero. 

Overall, accounting for CHE in the isolated formation channel, we find local merger rate densities of $\mathcal{R}_{\rm BBH} = 39^{+104}_{-26}\,\mathrm{Gpc}^{-3}\,\mathrm{yr}^{-1}$, $\mathcal{R}_{\rm BHNS} = 74^{+12}_{-24}\,\mathrm{Gpc}^{-3}\,\mathrm{yr}^{-1}$, and $\mathcal{R}_{\rm BNS} = 194^{-74}_{+23}\,\mathrm{Gpc}^{-3}\,\mathrm{yr}^{-1}$ if we assume a metallicity spread $\sigma_\mathrm{Z}=0.3$ as fiducial model, with $\sigma_\mathrm{Z}=0.7, 0.1$ for the upper and lower bound, respectively. \cite{Sgalletta2024}  show that most population-synthesis models yield values of the BBH merger rate density that exceed the BBH merger rate density inferred from LIGO--Virgo data at the 90\% credible level \citep{Abbott2023pop}. The excess of predicted over observed merger rate becomes even worse when a more sophisticated model for the cosmic metallicity evolution is adopted. Here, we show that including accretion-induced CHE in binary evolution models is a possible way to relieve this issue. However, the longer delay times of the CHE model also result in a flatter growth of the merger rate density with redshift compared to the one inferred from LIGO-Virgo data \citep{Abbott2023pop}.

The effect of CHE on the dynamical formation channel is beyond the scope of this paper. However, we can speculate on its potential impact. CHE leads to a larger number of BBH and BHNS systems compared to those produced by the standard binary evolution scenario (see Table~\ref{tab:cob}). While most of these binaries are too wide to merge within a Hubble time, if hosted in a dynamically active environment, they could reduce their orbital separation through dynamical hardening and eventually merge. This is particularly relevant in environments with a relatively high binary fraction, such as young star clusters \citep{Mapelli2022}. The overabundance of CHE-BBH and BHNS systems could result in a higher fraction of mergers triggered by dynamical interactions, thereby increasing their merger efficiency in the dynamical channel. We will investigate the effects of CHE on dynamically-assembled compact binary mergers in a forthcoming paper.

\section{Summary}

We have investigated the effects of accretion-induced CHE on both stellar and compact binary populations. 
We specifically focused on the accretion spin-up scenario where CHE is triggered by a Roche-lobe overflow event. 

We find that $\approx14-15\%$ of our binary systems experience CHE at metallicity $Z=0.004$ and $Z=0.001$, respectively. The accretion-induced CHE mechanism reduces the formation of RSGs, instead favoring the production of WRs from secondary stars in binary systems. Our binary population model including CHE produces nearly three times more WRs at $Z=0.001$ and almost two times more WRs at $Z=0.004$ compared to the model without CHE. Overall, the RSG-to-WR ratio of a stellar population decreases with a higher binary fraction, and drops at lower metallicities when CHE is active.

Secondary stars spun-up by accretion are the most common progenitors of WRs 
when we include CHE in our models. 
The WRs produced by CHE are, on average, more massive, more numerous, and more luminous than the other WRs. 
WRs formed via accretion-induced CHE remain more massive than their non-CHE counterparts up to the pre-supernova phase, and eventually produce more massive compact remnants. 

Our simulations show that accretion-induced CHE greatly affects the formation and merger of compact binary systems. CHE significantly enhances the formation of BBH and BHNS systems, with the latter especially growing by a factor of $\approx7$ at $Z=0.001$, becoming even more common than BBHs. At the same time, the number of BBHs is increased by a factor $\approx1.4$ at both $Z=0.001$ and $Z=0.004$. The higher formation efficiency of BHNSs comes at the cost of a substantial decrease in the production of BNS systems, whose number drops by a factor larger than $\approx3$ in the CHE model at $Z=0.004$.

Unlike the "traditional" BHNS formation channel, where the primary star evolves into a black hole and the secondary becomes a neutron star, our CHE models predict the opposite sequence: the primary star evolves into a neutron star, while the secondary forms the black hole. This reversal occurs because the primary star loses mass through mass transfer, while the secondary star accretes mass, becomes quasi-chemically homogeneous, and eventually collapses into a black hole. Similar scenarios where the neutron star forms first have also been reported in other binary evolution studies \citep[e.g.][]{Floor2021,Debatri2021}. However, our results indicate that this outcome becomes significantly more common when CHE is included in binary evolution models.

 We find that most of the BBHs formed through CHE exhibit asymmetric mass ratio ($q\in[0.1,0.6]$). 
In our CHE models, the secondary star gains enough mass to become the new primary in the binary system. As the system does not undergo any further mass transfers, the binary remains asymmetric until the formation of the two compact remnants.

A primary consequence of CHE is that the majority of BBHs, BHNSs, and BNSs formed through this channel have wide orbital separations, preventing them from merging via gravitational-wave emission within a Hubble time. Thus, CHE  suppresses merger efficiencies of BBHs, BHNSs, and BNS mergers. Including CHE in our models at $Z=0.004$ leads to a decrease of the number of BBH, BHNS and BNS mergers by a factor of $\approx11$, $\approx8$, and $\approx4$, respectively.

This decrease of the merger efficiency  happens because, when the secondary undergoes CHE, the system is composed of a primary that has lost its outer envelope and a secondary that evolves with a compact radius. Most of these systems do not undergo additional stable or unstable mass transfer episodes that can shrink the orbit. 
Only binaries with initially tight orbits ($a_i<100\,$R$_{\odot}$), or  binaries with intermediate orbital separations ($a_i>400\,$R$_{\odot}$) that become eccentric via supernova kick successfully evolve into compact binary mergers. We find that $\approx21-15\%$ of the BBH mergers and $\approx10-18\%$ of the BHNS mergers  are produced by the CHE channel at $Z=0.001$ and $Z=0.004$. 

If we interface our population-synthesis models with a metal-dependent cosmic star formation rate history \citep{Madau2017}, we find that the CHE channel yields local merger rate densities of $\mathcal{R}_{\rm BBH} = 39^{+104}_{-26}\,\mathrm{Gpc}^{-3}\,\mathrm{yr}^{-1}$, {$\mathcal{R}_{\rm BHNS} = 74^{+12}_{-24}\,\mathrm{Gpc}^{-3}\,\mathrm{yr}^{-1}$, and $\mathcal{R}_{\rm BNS} = 194^{-74}_{+23}\,\mathrm{Gpc}^{-3}\,\mathrm{yr}^{-1}$ in the isolated formation channel, assuming a spread in metallicity $\sigma_{\rm Z}=0.3$ as fiducial model, and  $\sigma_{\rm Z}=0.7, 0.1$ for the upper and lower bounds, respectively. This result indicates that chemically homogeneous evolution could possibly relieve the tension between the BBH merger rate density inferred from LIGO-Virgo data \citep{Abbott2023pop} and the predictions by population-synthesis models \citep{Sgalletta2024}.

\begin{acknowledgements}
      We thank Sandro Bressan, Jan Eldridge, and Andreas Sander for their useful comments. MM, GC, GI, and MD acknowledge financial support from the European Research Council for the ERC Consolidator grant DEMOBLACK, under contract no. 770017 (PI: M. Mapelli). MD acknowledges financial support from the Cariparo Foundation under grant 55440. MM acknowledges financial support from the German Excellence Strategy via the Heidelberg Cluster of Excellence (EXC 2181 - 390900948) STRUCTURES. EK and MM acknowledge support from the PRIN grant METE under contract No. 2020KB33TP. GI received the support of a fellowship from the la Caixa Foundation (ID 100010434). The fellowship code is LCF/BQ/PI24/12040020. GI also acknowledges financial support under the National Recovery and Resilience  1092  Plan (NRRP), Mission 4, Component 2, Investment 1.4, - Call for tender No. 1093  3138 of 18/12/2021 of Italian Ministry of University and Research funded by the 1094  European Union –NextGenerationEU. GC acknowledges support from the Agence Nationale de la Recherche grant POPSYCLE number ANR-19-CE31-0022. The authors acknowledge support by the state of Baden-W\"urttemberg through bwHPC and the German Research Foundation (DFG) through grants INST 35/1597-1 FUGG and INST 35/1503-1 FUGG.

\end{acknowledgements}


  \bibliographystyle{aa} 
  \bibliography{bibliography.bib} 


\begin{appendix}
\renewcommand{\thetable}{A\arabic{table}}

\begin{table*}[]
\begin{center}
\caption{Percentages of BBHs, BHNSs, and BNSs systems, mergers, and CHE binaries for all the models presented in Table~\ref{tab1:runs} as a function of the metallicity.}
\begin{tabular}{@{}lll|lll|lll|lll|@{}}
\cmidrule(l){4-12}
\multicolumn{3}{l|}{}                                              & \multicolumn{3}{c|}{BBH} & \multicolumn{3}{c|}{BHNS} & \multicolumn{3}{c|}{BNS} \\ \midrule
\multicolumn{1}{|l|}{Z}                  & \multicolumn{1}{l|}{name} & P$_{\rm CHE}$ &   P$_{\rm cob}$    &   P$_{\rm merg}$    &   P$_{\rm merg}^{\rm CHE}$    &   P$_{\rm cob}$    &   P$_{\rm merg}$    &  P$_{\rm merg}^{\rm CHE}$   &   P$_{\rm cob}$  &  P$_{\rm merg}$   &  P$_{\rm merg}^{\rm CHE}$   \\ \midrule
\multicolumn{1}{|l|}{\multirow{6}{*}{0.001}} &  \multicolumn{1}{l|}{NoCHEzams}   & 0 &   3.77    &  0.55     &  0   &   0.84    &  0.24      &   0    &   0.24    &   0.2    &   0    \\
\multicolumn{1}{|l|}{}&  \multicolumn{1}{l|}{NoCHEpreMS}   & 0     &   3.77    &   0.54    &   0        &   0.84    &   0.24    &   0        &   0.24    &   0.2     &   0    \\
\multicolumn{1}{|l|}{}&  \multicolumn{1}{l|}{CHE10zams}    & 14.8  &   5.22    &   0.26    &   20.28    &   5.62    &   0.12    &   10.32    &   0.09    &   0.07    &   0    \\
\multicolumn{1}{|l|}{}&  \multicolumn{1}{l|}{CHE10preMS}   & 14.69 &   5.24    &   0.26    &   21.01    &   5.56    &   0.12    &   10.25    &   0.09    &   0.07    &   0    \\
\multicolumn{1}{|l|}{}&  \multicolumn{1}{l|}{CHE20zams}    & 4.24  &   4.68    &   0.25    &   15.81    &   1.39    &   0.18    &   0.94     &   0.24    &   0.21    &   0    \\
\multicolumn{1}{|l|}{}&  \multicolumn{1}{l|}{CHE20preMS}   & 4.29  &   4.68    &   0.25    &   16.38    &   1.40    &   0.18    &   0.94     &   0.24    &   0.2     &   0    \\ \midrule
\multicolumn{1}{|l|}{\multirow{6}{*}{0.004}} &  \multicolumn{1}{l|}{NoCHEzams}   & 0 &   3.81    &   0.83    &   0    &   0.9    &   0.33    &  0   &   0.4    &   0.2    &  0   \\
\multicolumn{1}{|l|}{}&  \multicolumn{1}{l|}{NoCHEpreMS}    & 0    &   3.81    &   0.82    &   0    &   0.89    &   0.33    &   0    &   0.4     &   0.21    &  0   \\
\multicolumn{1}{|l|}{}&  \multicolumn{1}{l|}{CHE10zams}   & 14.92  &   5.47    &   0.07    &  14.66 &   4.55    &   0.04    &  18.28 &   0.12    &   0.05    &  0   \\
\multicolumn{1}{|l|}{}&  \multicolumn{1}{l|}{CHE10preMS}  & 14.31  &   5.48    &   0.07    &  15.34 &   4.39    &   0.04    &  17.99 &   0.12    &   0.05    &  0   \\
\multicolumn{1}{|l|}{}&  \multicolumn{1}{l|}{CHE20zams}   & 5.84   &   4.88    &   0.07    &  6.96  &   1.79    &   0.15    &  1.75  &   0.38    &   0.19    &  0   \\
\multicolumn{1}{|l|}{}&  \multicolumn{1}{l|}{CHE20preMS}  & 5.9    &   4.89    &   0.07    &  7.29  &   1.81    &   0.14    &  1.69  &   0.38    &   0.2     &  0   \\ \bottomrule
\end{tabular}
\label{tab:cob2}
\end{center}
\footnotesize{From left to right column: metallicity ($Z$), name of the model, percentage of binaries that experience CHE in the simulation set (P$_{\rm CHE}$), percentage of compact object binaries produced in the simulation set (P$_{\rm cob}$), percentage of compact object binary mergers produced in the simulation set (P$_{\rm merg}$), and percentage of compact object binary mergers that evolve through CHE among all mergers of the simulation set (P$_{\rm merg}^{\rm CHE}$). The latter three columns are repeated respectively for BBHs, BHNSs, and BNSs.}
\end{table*}


\section{Supplementary Model Analysis}\label{sec:app1}

In this Appendix, we discuss the results of the remaining models presented in Table~\ref{tab1:runs}. First, we tested the potential effects of using a different initial time for the evolution of our binaries. Additionally, we explored the impact of a different CHE prescription by changing the minimum mass required for a star to trigger CHE (see section~\ref{sec:CHE}). The main differences with our fiducial models are discussed in the following sections and summarized in Table~\ref{tab:cob2}.

\subsection{preMS vs ZAMS}

Our fiducial models follow the evolution of binary and single stars starting from their pre-main sequence (preMS) phase, before core hydrogen ignition. This choice allowed us to account for potential binary interactions that might occur during this early evolutionary stage. Many works in the literature typically initiate the evolution directly from the zero-age-main-sequence (ZAMS), both for computational and historical reasons. To compare our results with present and future works, we repeated our simulations starting the evolution of our stars from the ZAMS. We find no significant differences between the preMS and the ZAMS models, both on the stellar population side and the compact object binaries and their mergers. For instance, Table~\ref{tab:cob2} shows that the differences in binary systems undergoing CHE are marginal, as is the fraction of compact object binaries produced in the simulation.
We find that the preMS models consistently produce $1-3\%$ more stellar mergers in the simulations compared to the ZAMS models when $Z=0.001,\,0.004$. Although this difference in merger fraction is marginally small, many of these colliding stars merge at different evolutionary phases, as illustrated in Figure~\ref{fig:20mergers}. In the ZAMS models, most mergers occur in the binary's advanced stages, either when the secondary star expands after the first supernova, or when the primary evolves away from the ZAMS crossing the Herzsprung gap. This is also true for the preMS model. However, unlike the ZAMS models, in the preMS model, many stellar mergers that previously involved evolved stars are now triggered when both binary components are in the preMS phase. During this evolutionary stage, stars have inflated radii as they contract to reach the ZAMS. Since their radii are comparable to those they develop during the shell hydrogen-burning phase, preMS stars do not need to evolve toward the giant branch to merge with their companion. Consequently, almost all the binaries that merge in the preMS phase are the same systems that merge at later evolutionary stages in the ZAMS model. These systems do not contribute to the production of compact object binaries and only provide a negligible fraction of the WR population (see Table~\ref{tab:stars}). Therefore, our results are only marginally affected by the choice of the initial time for the binary evolution.

\subsection{CHE10 vs CHE20}

In this work, we implemented in {\sc sevn} the accretion-induced CHE prescription presented in \cite{Eldridge2011}. With this prescription, a star evolved in our fiducial models undergoes CHE if its metallicity is $Z\leq0.004$, it has accreted at least $5\%$ of its initial mass, and if it has a minimum mass $M_{\rm min}=10\,$M$_{\odot}$ (see section~\ref{sec:CHE}). Although it is commonly assumed that rotational mixing is more efficient in massive stars \citep[e.g.,][]{Maeder2001,Yoon2006}, the exact mass limit at which a star can spin-up and enter CHE without reaching its critical rotation remains unknown (see \citealt{Ghodla2023} for a discussion). For this reason, we repeated our simulations using the CHE prescription from \cite{Eldridge2017}, which updates the earlier \cite{Eldridge2011} prescription by increasing the minimum mass threshold to $M_{\rm min}=20\,$M$_{\odot}$.\\

Figure~\ref{fig:20rate} presents a comparison of the RSG-to-WR ratio between our fiducial model and the CHE20preMS model. The CHE20preMS model also exhibits a decreasing ratio with metallicity when $Z \leq 0.004$ and at a non-zero binary fraction, but the ratio is shifted to higher values compared to that of the CHE10preMS model. This is because stars more massive than $20\,$M$_{\odot}$ are less common than those in the $10-20\,$M$_{\odot}$ range, resulting in fewer stars undergoing the CHE stage and producing WRs, which in turn might prevent the formation of RSGs. Furthermore, most stars above $20\,$M$_{\odot}$ would likely evolve into WRs even without experiencing CHE. Nonetheless, we still observe that the CHE20preMS model produces more WRs and fewer RSGs compared to the NoCHEpreMS model, although this effect is less pronounced than in the CHE10preMS model. We find that in the CHE20preMS model $4.8\%$ and $8.2\%$ of all the stars in the binary population evolve into WRs at $Z=0.001$ and $Z=0.004$, respectively. The key difference from the CHE10preMS model is that secondary stars now account for only $46.5\%$ at $Z=0.001$ and $37.5\%$ at $Z=0.004$ of the WRs in the sample, no longer serving as the primary progenitors of WRs in binaries. This results directly from the reduced number of binary systems experiencing CHE: in the CHE20preMS model, only one in every $\approx23$ and $\approx17$ binaries contain a CHE star when $Z=0.001$ and $Z=0.004$, respectively, as reported in Table~\ref{tab:cob2}.\\

Figure~\ref{fig:20Luminosity} shows that even with a relaxed criterion on the minimal mass for a star to enter the CHE phase, we consistently observe that CHE-produced WRs are, on average, characterized by higher luminosities and masses at birth compared to their non-CHE counterparts. The only difference with the WRs produced by the CHE10preMS model is that now the less massive star to produce a WR through CHE has approximately $10\,$M$_{\odot}$ at ZAMS, while the less massive WR produced is $16\,$M$_{\odot}$. After entering the CHE stage, the newborn WRs experience wind mass loss which decreases their total mass below $20\,$M$_{\odot}$. This sets a lower limit in the luminosity of these CHE-WRs that corresponds to $\approx4.5\times10^{5}\,$L$_{\odot}$, as can be seen in the right panel of Figure~\ref{fig:20Luminosity}.\\

Similar to the CHE10preMS model, the CHE20preMS model's overproduction of more massive WRs results in the formation of more massive remnants. This also affects the formation of compact object binaries and their mergers, as can be seen in Table~\ref{tab:cob2}. Although the CHE20preMS model is more effective than the NoCHEpreMS model at producing compact object binaries, it is still less efficient than the CHE10preMS model in forming BBHs and BHNSs. Specifically, the increase in BHNS systems is less pronounced than the impressive growth observed in the CHE10preMS model, resulting in BHNS systems being less numerous than BBHs. At the same time, the fraction of BNS systems in the CHE20preMS model remains nearly equal to that in the NoCHEpreMS model. This is because most neutron star progenitors are less massive than $20\,$M$_{\odot}$ and evolve into neutron stars without being influenced by CHE. On the contrary, most of the stars experiencing CHE in the CHE20preMS model are black hole progenitors. We find that CHE reduces the formation of BBH mergers at nearly the same rate as in the CHE10preMS model, while it only mildly affects the fraction of BHNS mergers.\\

This growth in BBH systems is also evident in Figure~\ref{fig:20BBH}, which reports the mass of the black holes belonging to BBHs and produced by the primary and secondary stars in the progenitor binary as a function of its initial semi-major axis. The plots specifically highlight the differences between the NoCHEpreMS and the CHE20preMS models at $Z=0.001$ and $Z=0.004$. Similar to the CHE10preMS model, we observe an overproduction of BBHs, with the most massive black holes produced by secondary stars concentrated in the area of the diagram where progenitor binaries begin their evolution with relatively low orbital separation ($a_{\rm i}<100\,$R$_{\odot}$). The black hole mass spectra are very similar to those in the CHE10preMS model, as most progenitor binaries of these compact objects undergo CHE in both models. Similar considerations are also valid for the BBH merger populations presented in Figure~\ref{fig:20BBHmergers}, which appears to be almost identical to the BBH merger population produced by the CHE10preMS model in Figure~\ref{fig:BBH_mergers}. The main differences between these two merger populations are more evident in Figure~\ref{fig:20qplot}. The figure shows that while BBH mergers produced via CHE in the CHE20preMS model still prefer non-equal black hole components in the range $q=0.6-0.8$. However, the distributions lose the low mass ratio tail present in the distribution of CHE10preMS mergers.\\


Figure~\ref{fig:20BHNS} presents the BHNS population, illustrating the masses of the compact objects formed by the primary and secondary stars as a function of the initial orbital separation of the progenitor binary. The plots display BHNS produced in the CHE20preMS and the NoCHEpreMS model at $Z=0.001$ and $Z=0.004$. As in the CHE10preMS model, when CHE is triggered, the secondary star always becomes the progenitor of the black hole. The main difference here is that secondary stars that experience CHE can produce black holes only more massive than $\approx16\,$M$_{\odot}$. This occurs because, in the CHE20preMS model, the minimum mass for a star to experience CHE is higher than in the CHE10preMS model, resulting in more massive remnants from stars undergoing CHE. The gap observed between lower-mass black holes (i.e. $<7\,$M$_{\odot}$) and these more massive black holes arises because stars within this mass range do not experience CHE and thus produce less massive remnants, likely neutron stars. This is also valid for the case of BHNS mergers, as shown in Figure~\ref{fig:20BHNSmergers}. Mergers in which the black hole is produced by the secondary star can only occur through CHE at $Z=0.001$, whereas they make up the majority of cases at $Z=0.004$.\\

Figure~\ref{fig:20ME} presents the merger efficiency as a function of metallicity, while Figure~\ref{fig:20MRD} shows the evolution of the merger rate density with redshift for BBH, BHNS, and BNS mergers in the CHE20preMS model. Both Figures reflect the results also visible in Table~\ref{tab:cob2}. The BBH merger efficiency and merger rate density yield almost identical results to the fiducial CHE10preMS model. In contrast, both the BNS merger efficiency and merger rate density exhibit completely different behavior compared to the CHE10preMS model. When \( M_{\rm min} = 20\,\mathrm{M}_\odot \) is assumed, most neutron star progenitors evolve unaffected, avoiding the CHE scenario. As a result, the merger efficiency and merger rate density curves for the CHE20preMS model nearly overlap with those of the NoCHEpreMS model. This also affects BHNS mergers, where the reduction in merger efficiency and merger rate density is milder compared to the CHE10preMS case.


\begin{figure*}
\centering
\includegraphics[width=0.95\textwidth]{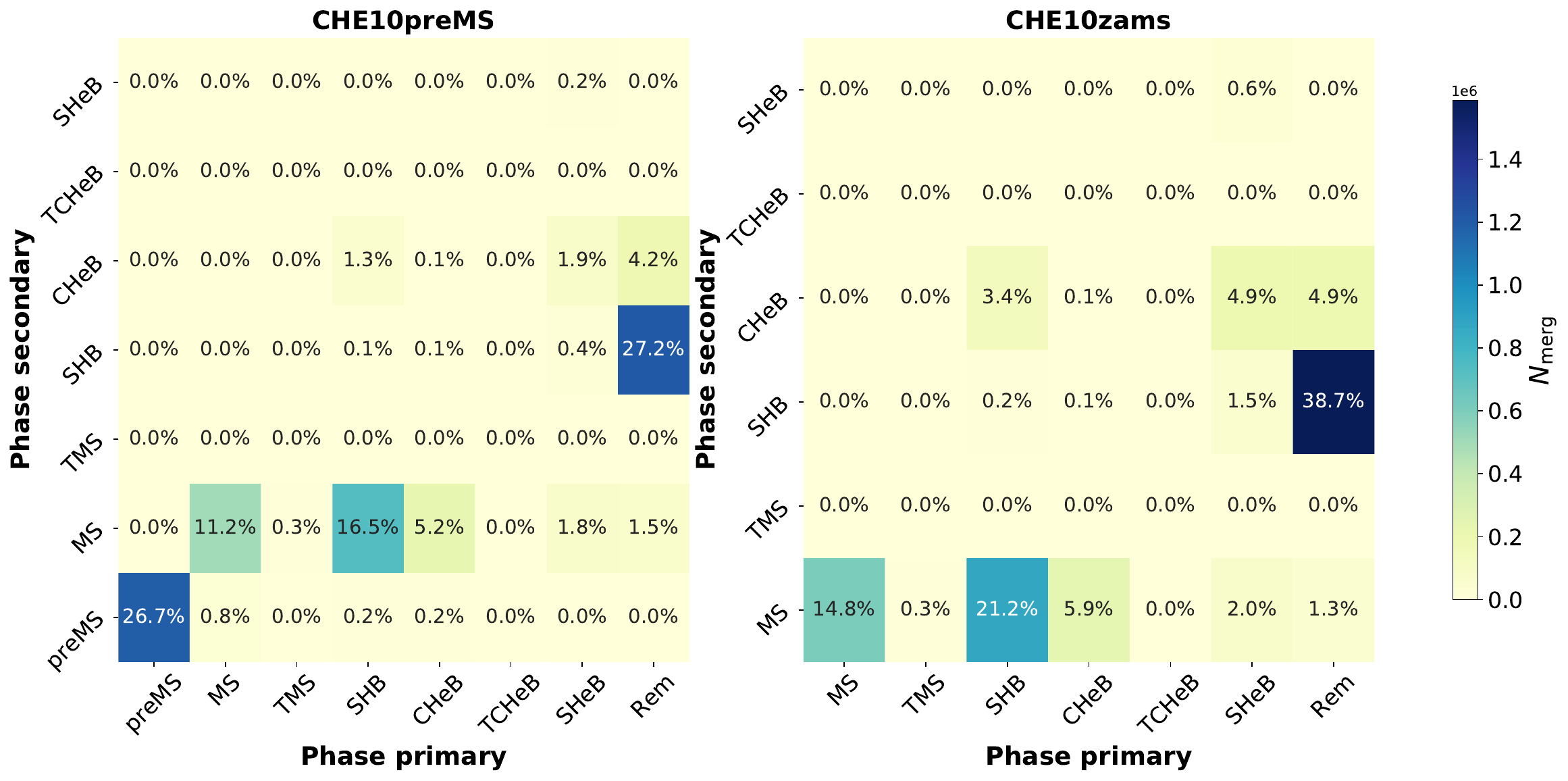}
    \caption{Percentage of stellar mergers with the primary and secondary stars in a given stellar phase among all the mergers for the CHE10preMS (fiducial model, left) and CHE10zams (right) models at $Z=0.004$. The stellar phases are divided as in \cite{Iorio2023}: pre-main sequence (preMS), main sequence (MS), terminal main sequence (TMS), shell hydrogen burning (SHB), Core helium burning (CHeB), terminal core helium burning (TCHeB), shell helium burning (SHeB), and remnant phase (Rem). The colourbar shows the number of stellar mergers in each bin. We find similar results in the models with $Z=0.001$.}
        \label{fig:20mergers}
\end{figure*}


\begin{figure*}
\centering
\includegraphics[width=0.95\textwidth]{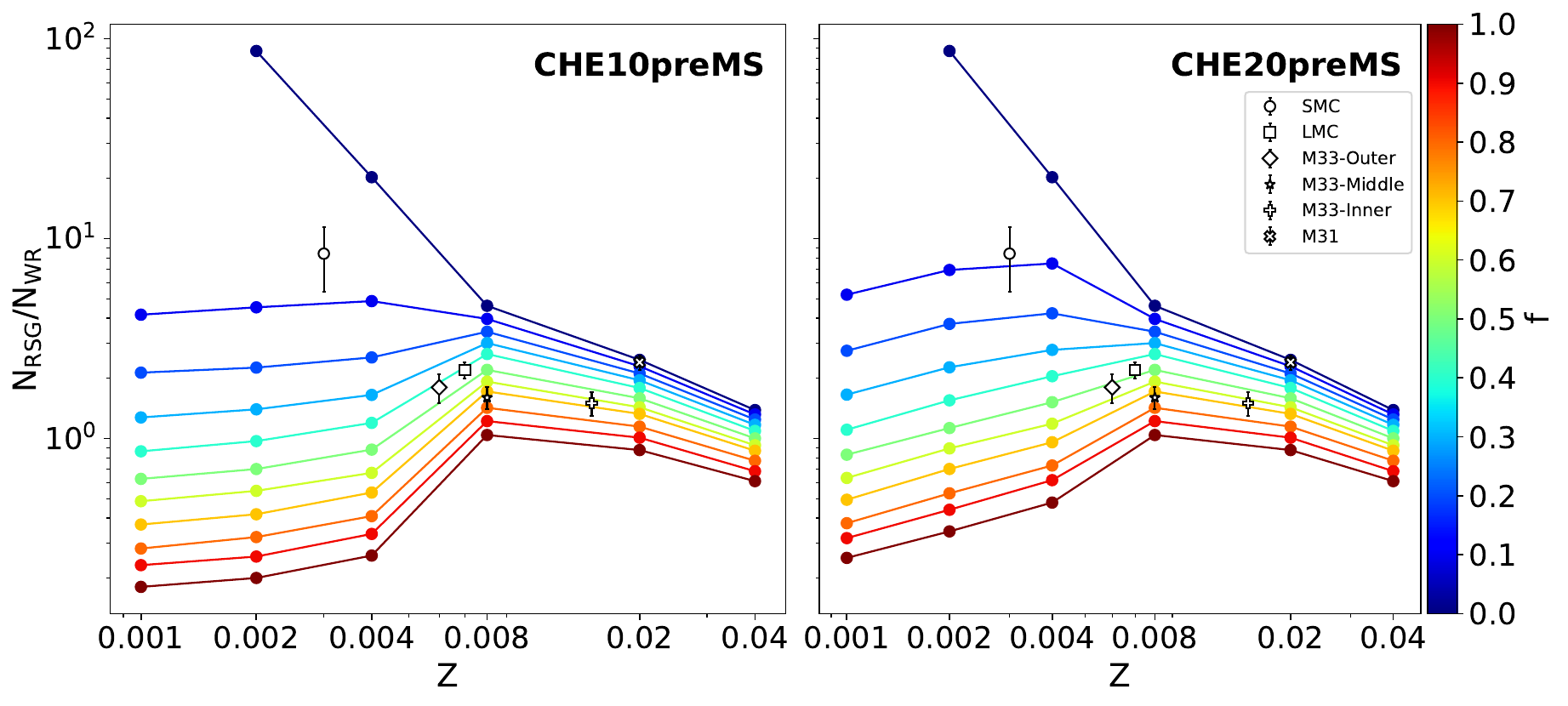}
    \caption{Rate of RSG over WR stars as a function of the metallicity and the binary fraction (see Fig.~\ref{fig:rate_WR}). The left-hand plot shows our fiducial model CHE10preMS, while the right-hand plot shows model CHE20preMS. }
        \label{fig:20rate}
\end{figure*}


\begin{figure*}
\centering
\includegraphics[width=0.95\textwidth]{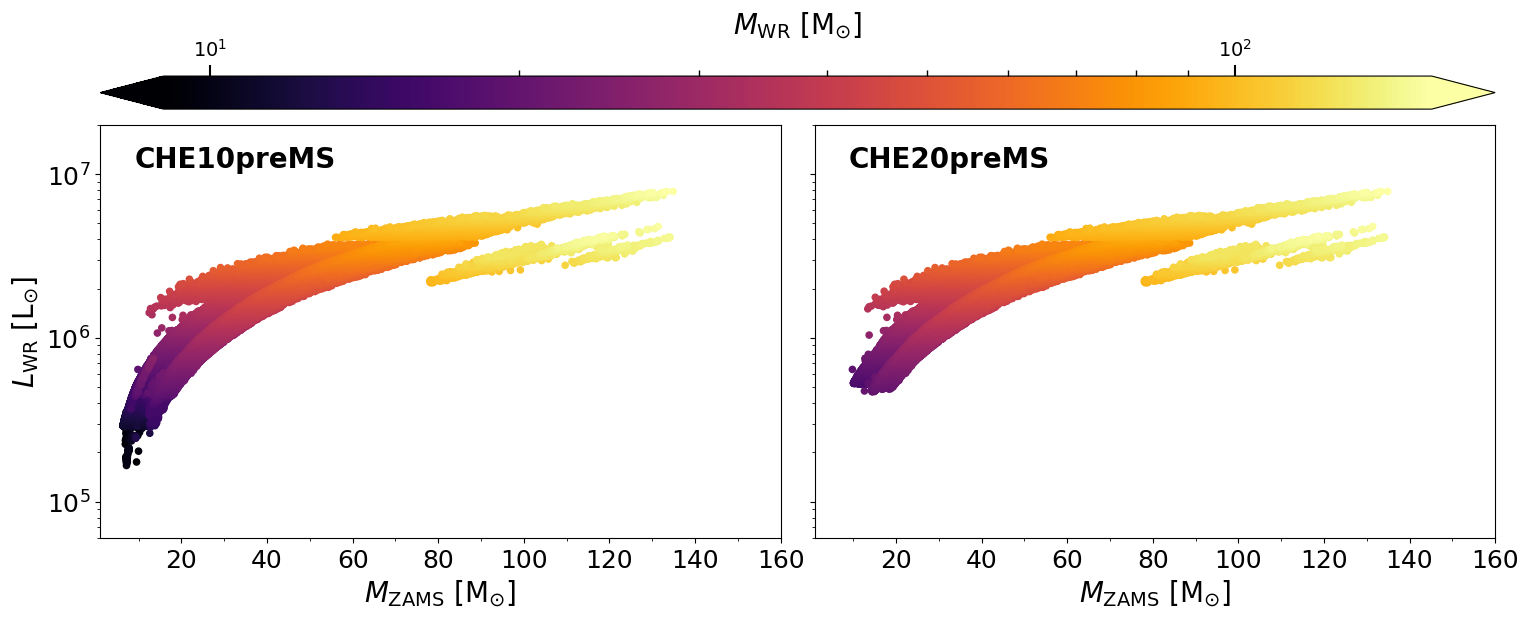}
    \caption{Same as Figure~\ref{fig:Luminosity} but for CHE10preMS (fiducial model, left) and CHE20preMS (right).}
        \label{fig:20Luminosity}
\end{figure*}


\begin{figure*}
\centering
\includegraphics[width=0.95\textwidth]{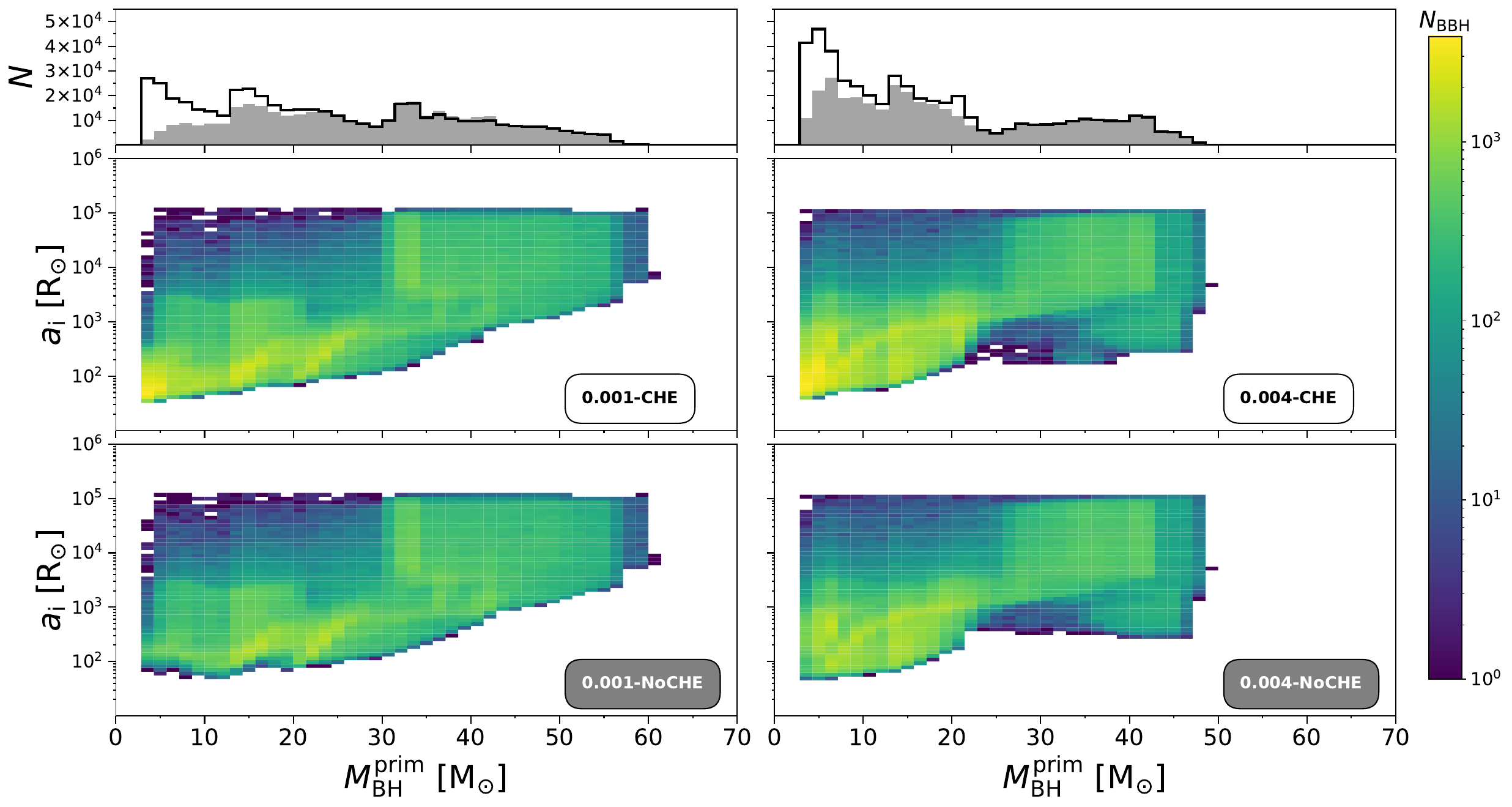}
\includegraphics[width=0.95\textwidth]{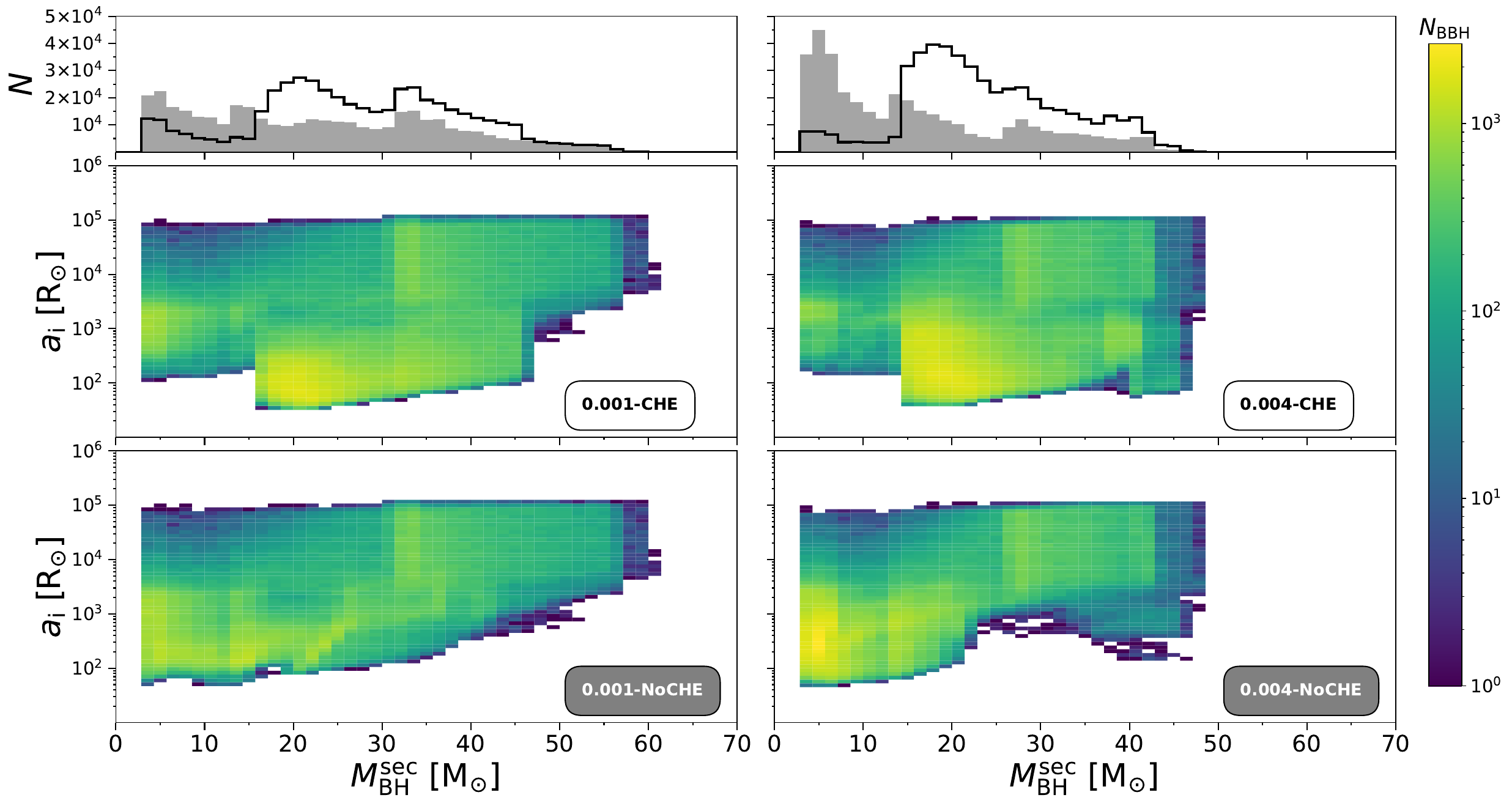}
    \caption{Same as Figure~\ref{fig:BBH} but for the CHE20preMS model.}
        \label{fig:20BBH}
\end{figure*}


\begin{figure*}
\centering
\includegraphics[width=0.95\textwidth]{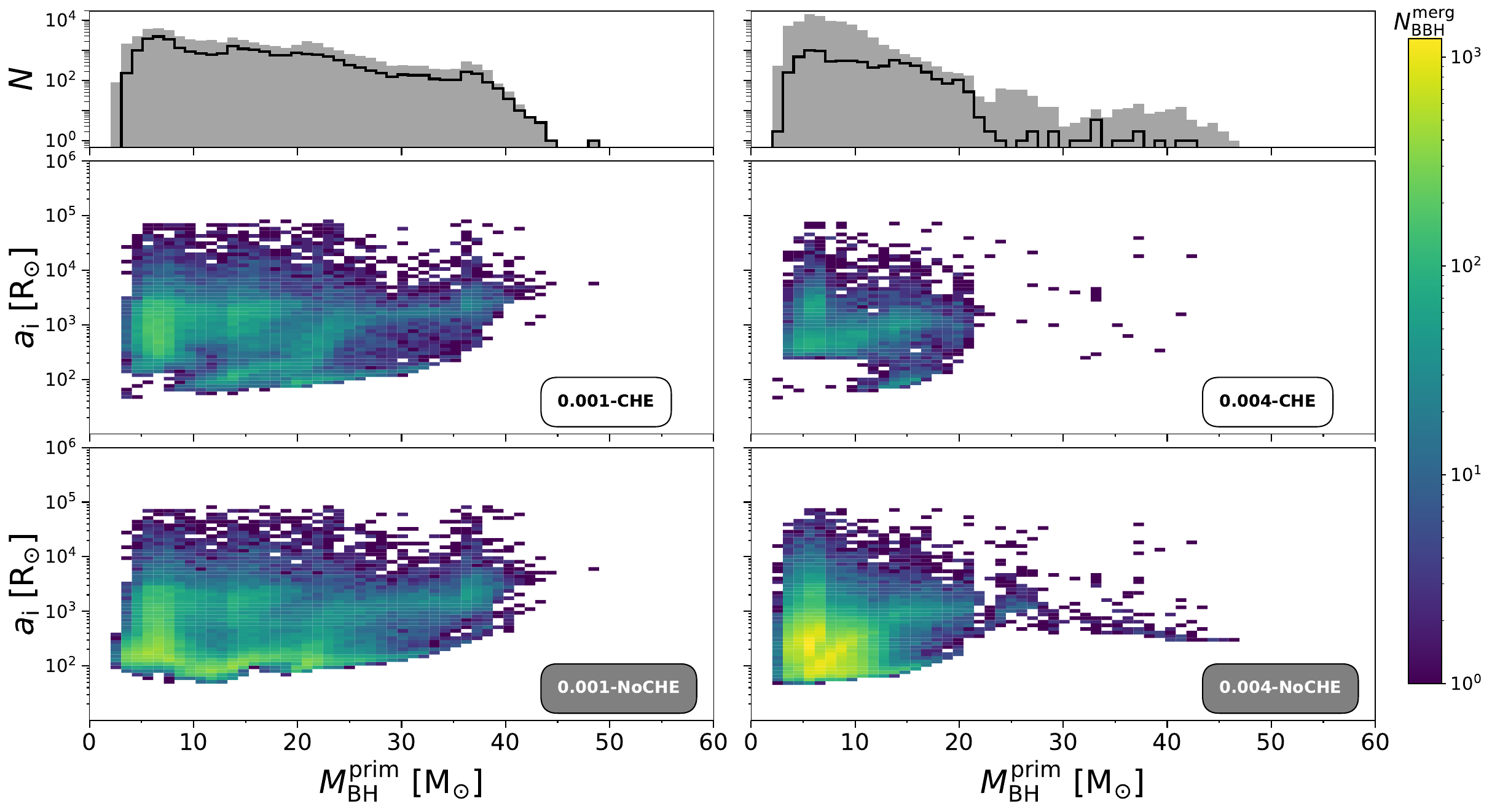}
\includegraphics[width=0.95\textwidth]{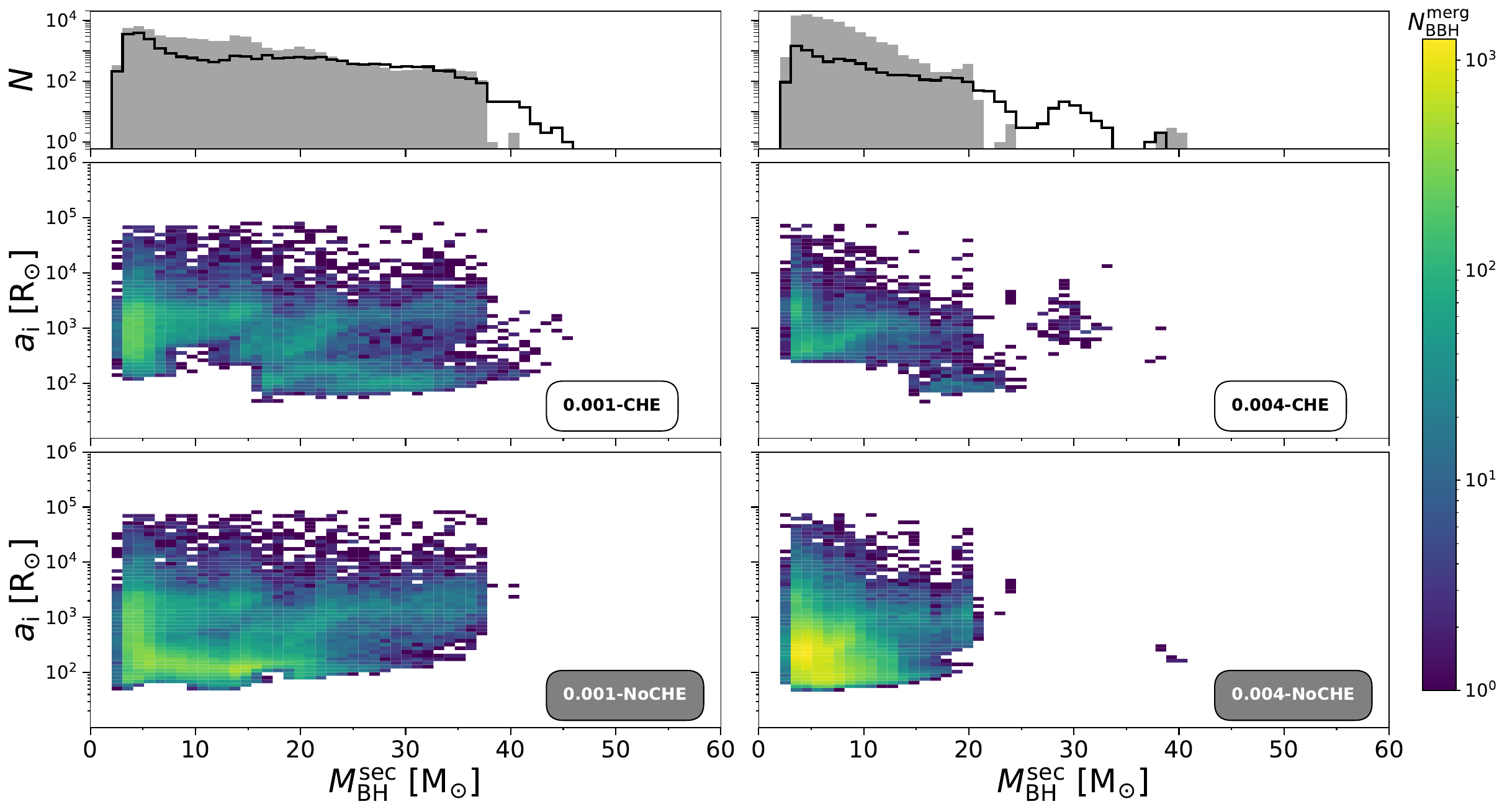}
    \caption{Same as Figure~\ref{fig:BBH_mergers} but for the CHE20preMS model.}
        \label{fig:20BBHmergers}
\end{figure*}


\begin{figure*}
\centering
\includegraphics[width=0.95\textwidth]{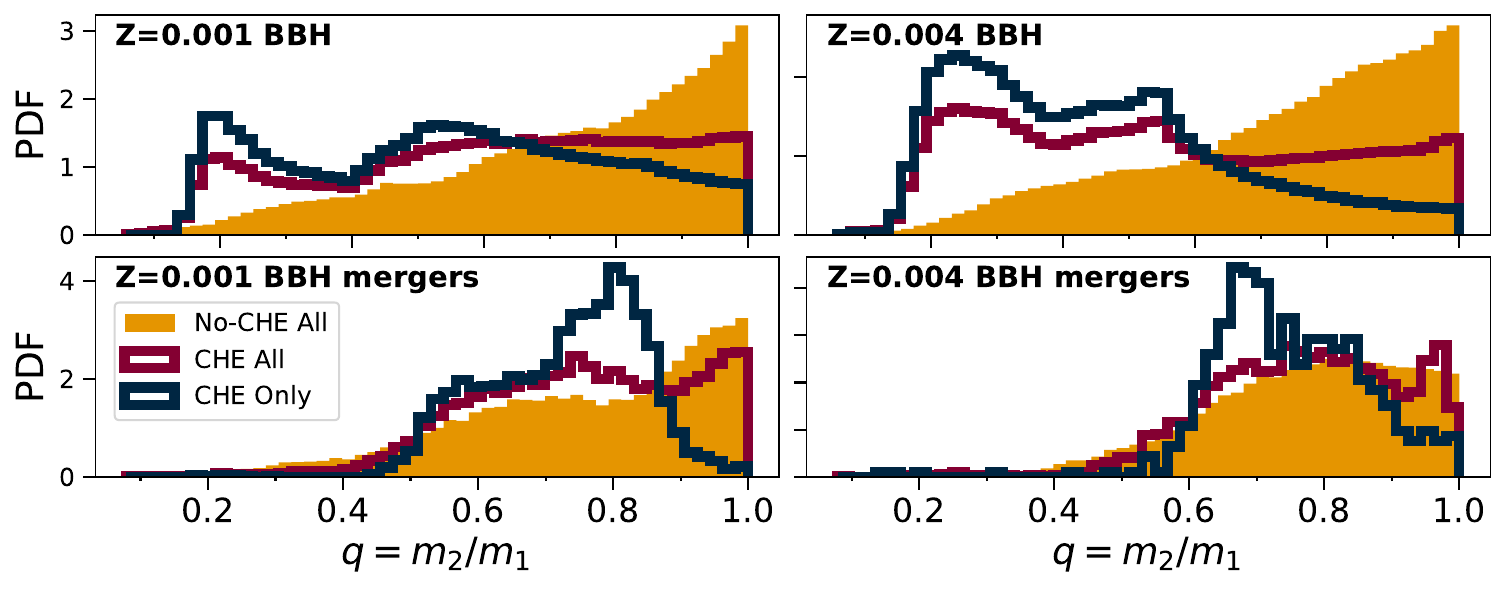}
    \caption{Same as Figure~\ref{fig:qplot} but for the CHE20preMS model.}
        \label{fig:20qplot}
\end{figure*}


\begin{figure*}
\centering
\includegraphics[width=0.95\textwidth]{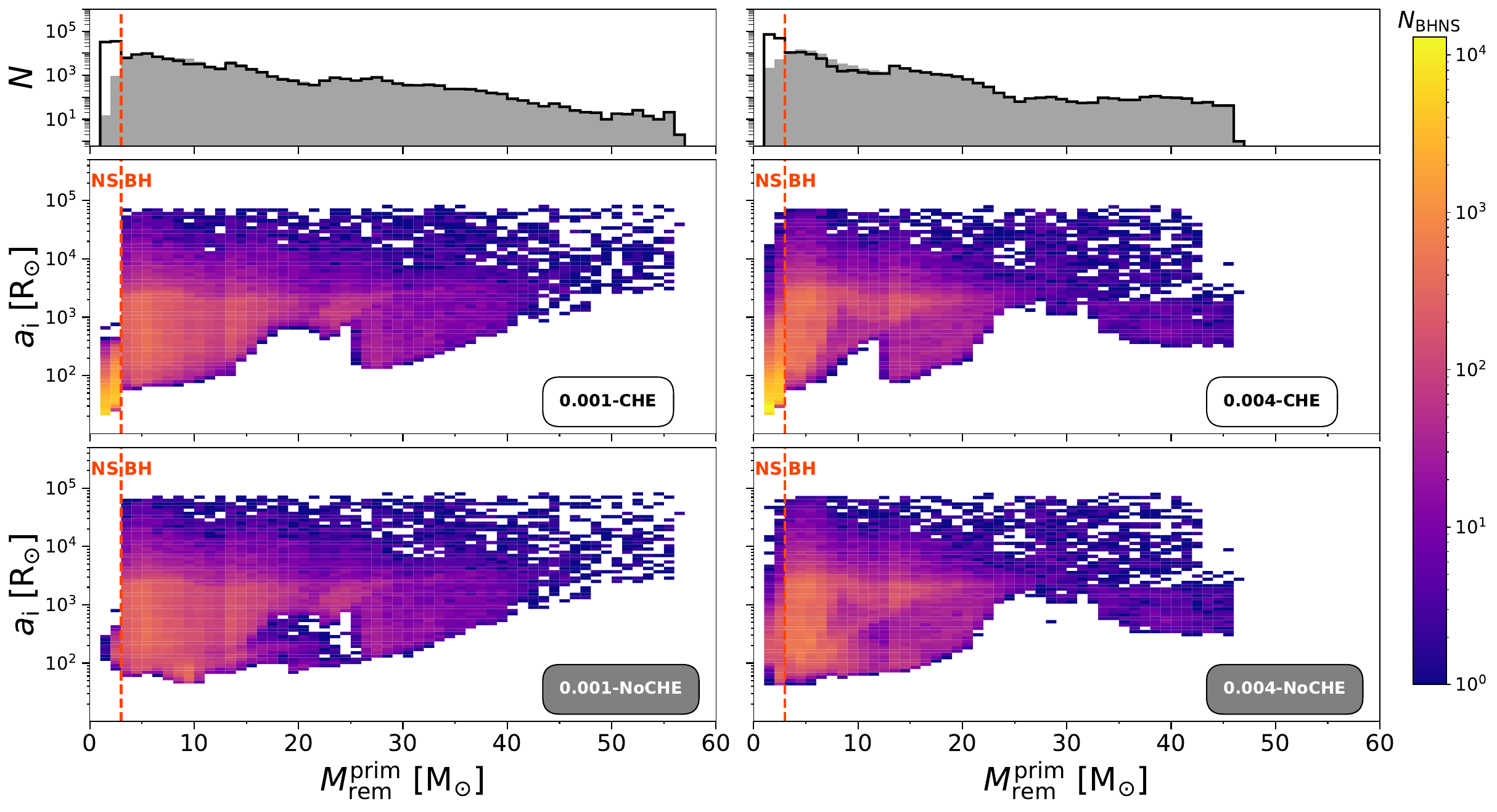}
\includegraphics[width=0.95\textwidth]{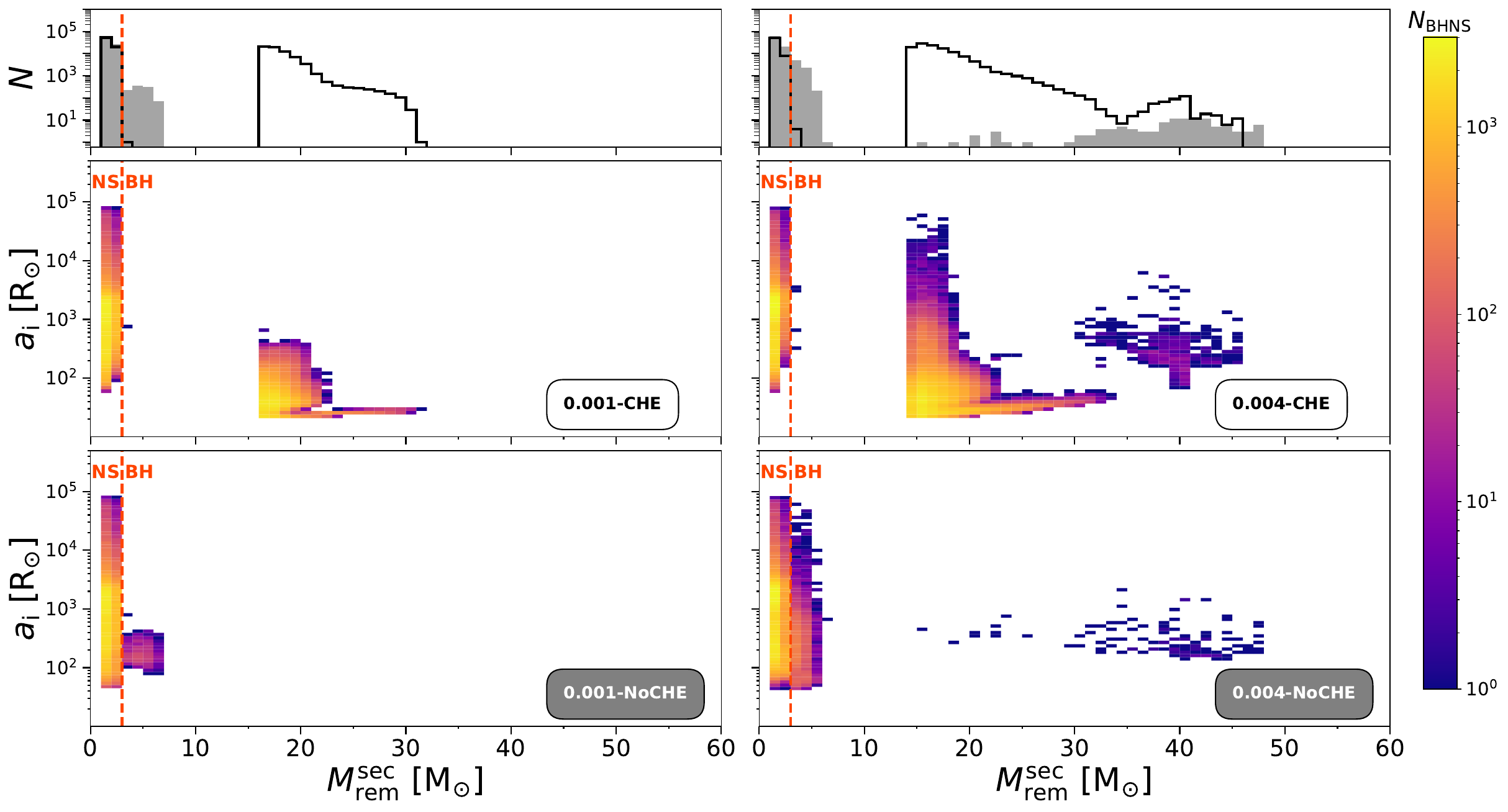}
    \caption{Same as Figure~\ref{fig:BHNS} but for the CHE20preMS model.}
        \label{fig:20BHNS}
\end{figure*}


\begin{figure*}
\centering
\includegraphics[width=0.95\textwidth]{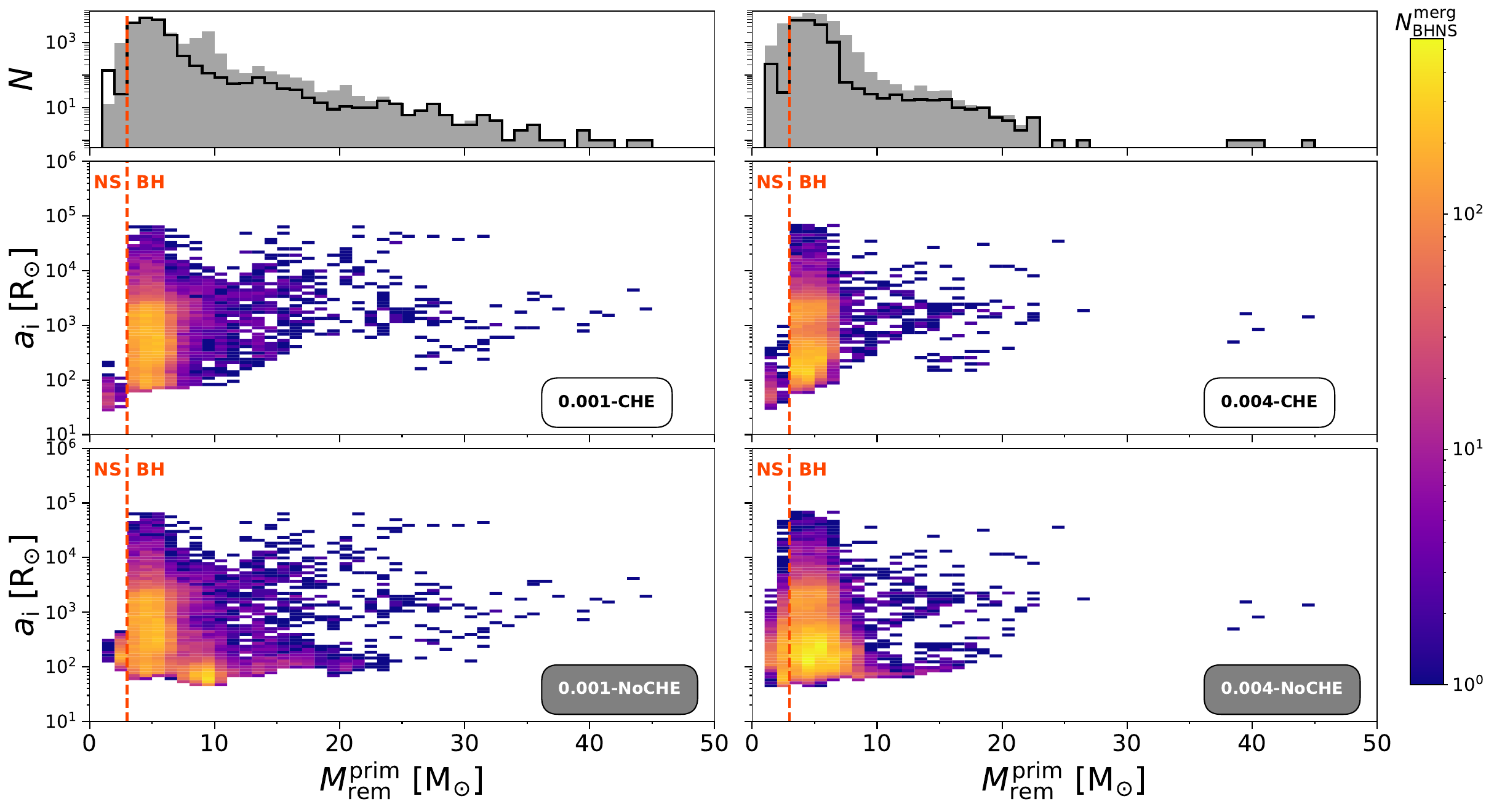}
\includegraphics[width=0.95\textwidth]{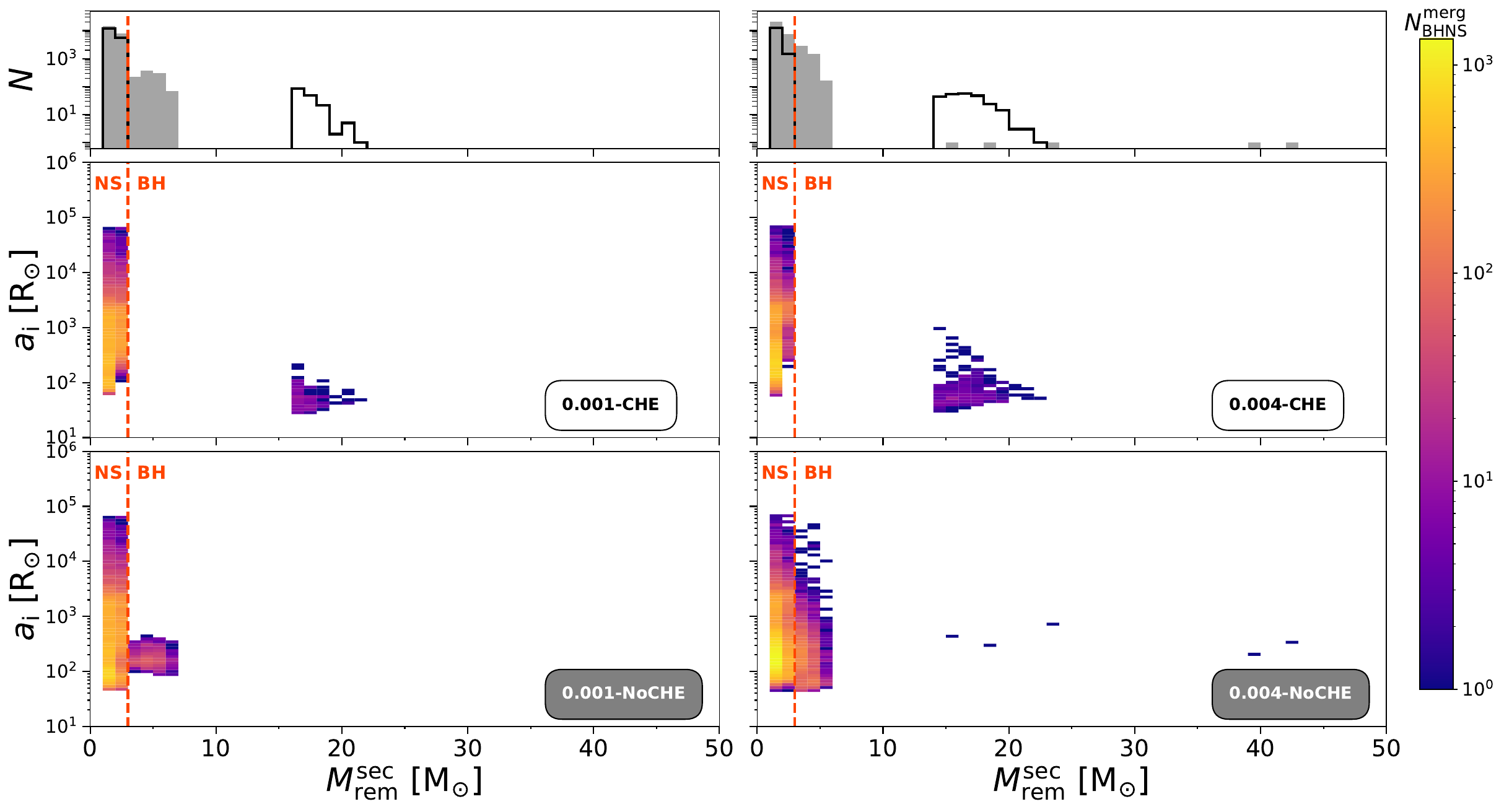}
    \caption{Same as Figure~\ref{fig:BHNS_mergers} but for the CHE20preMS model.}
        \label{fig:20BHNSmergers}
\end{figure*}

\begin{figure}
    \includegraphics[width=0.49\textwidth]{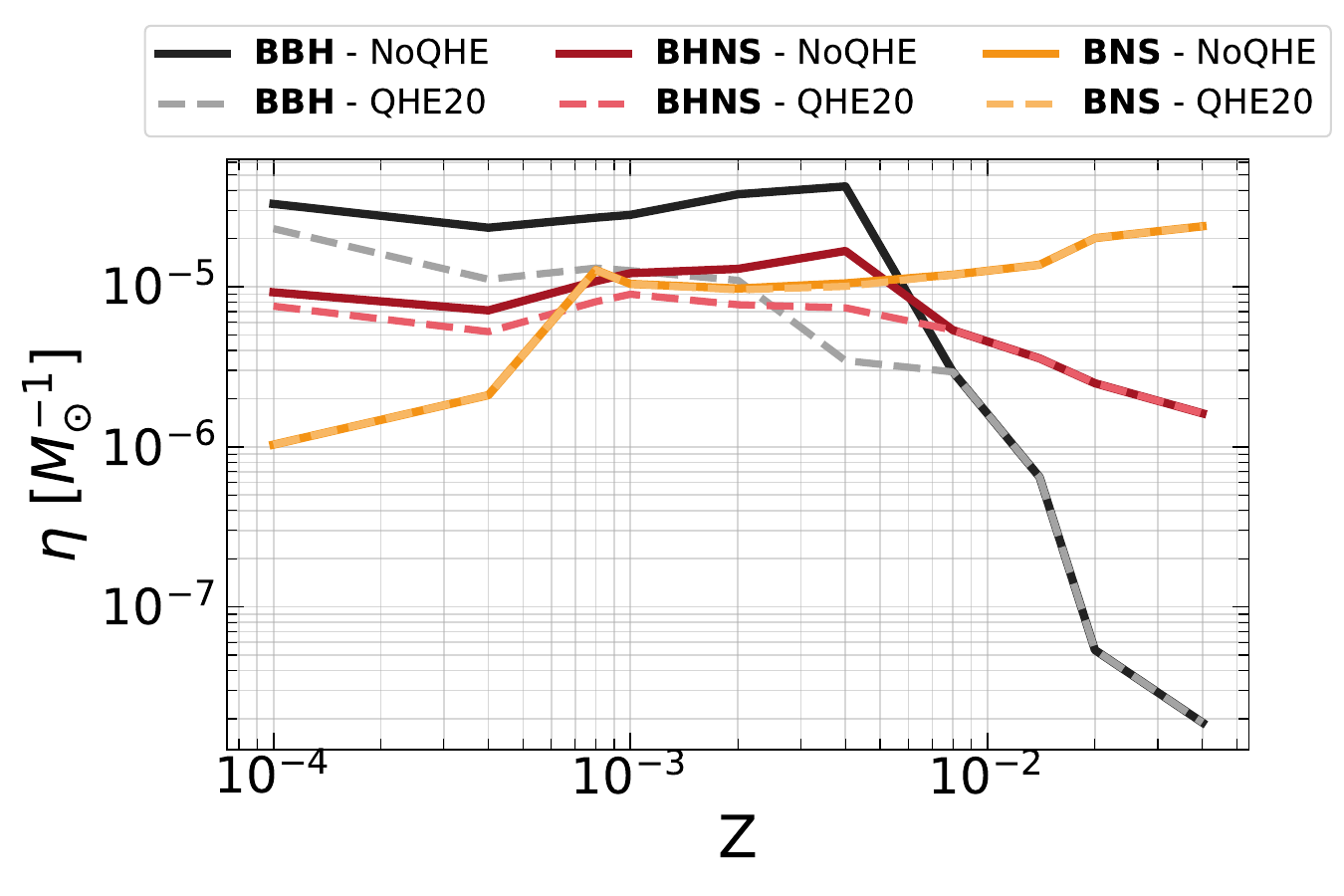}
    \caption{Same as Figure~\ref{fig:ME} but for the CHE20preMS model.}
    \label{fig:20ME}
\end{figure}


   \begin{figure}
   \centering
   \includegraphics[width=0.47\textwidth]{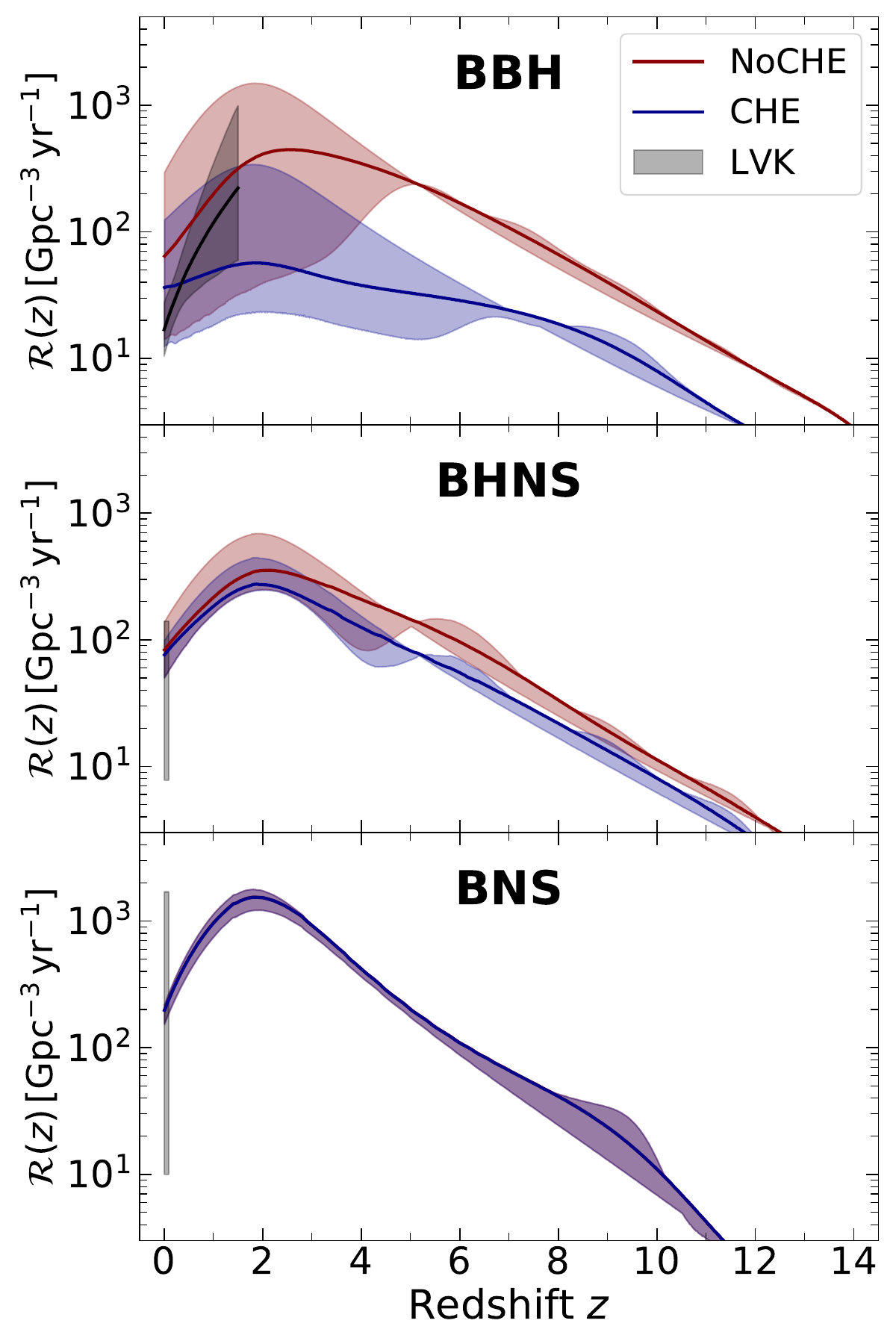}
      \caption{Same as Figure~\ref{fig:MRD} but for the CHE20preMS model.}
         \label{fig:20MRD}
   \end{figure}

\section{Merger rate density}\label{app:mrd}
We computed the redshift evolution of the merger rate density for all compact binary mergers in our models using {\sc cosmo}$\mathcal{R}${\sc ate} \citep{Santoliquido2020,Santoliquido2021}. {\sc cosmo}$\mathcal{R}${\sc ate} estimates the merger rate density $\mathcal{R}(z)$ of a given population of compact binary mergers at redshift $z$ as
\begin{equation}
\mathcal{R}(z) = \int_{z_{\text{max}}}^{z} \left[ \int_{Z_{\text{min}}}^{Z_{\text{max}}} \text{SFRD}(z',Z)\,\mathcal{F}(z',z,Z)\,\text{d}Z \, \right] \frac{\text{d}t(z')}{\text{d}z'}\,\text{d}z',
\end{equation}
where
\begin{align}
\text{SFRD}(z',Z)=\psi(z')\,p(z',Z),\\
\mathcal{F}(z',z,Z) = \frac{1}{M_{*}(Z)}\frac{\text{d}\mathcal{N}(z',z,Z)}{\text{d}t(z)},
\end{align}
and
\begin{equation}
\frac{\text{d}t(z')}{\text{d}z'} = [H_{\rm 0}\,(1+z')]^{-1}\,[(1+z')^{3}\,\Omega_{M}+\Omega_{\Lambda}]^{-1/2}.
\end{equation}
In the above equations, $\psi(z')$ is the cosmic star formation rate density at formation redshift $z'$, while $p(z',Z)$ is the log-normal distribution of metallicities $Z$ at formation redshift $z'$ computed with average $\mu(z')$ and metallicity spread $\sigma_{Z}$. We calculated the merger rate densities using $\mu(z)$ and $\psi(z)$ from \cite{Madau2017}, considering a metallicity spread of $\sigma_Z = 0.1$, $0.3$, and $0.7$, with $0.3$ adopted as the fiducial value. Our simulations produced the merger catalogs included in the $\text{d}\mathcal{N}(z',z,Z)/\text{d}t(z)$ term, which represents the rate of compact binary mergers that from from a stellar population with initial metallicity $Z$ at redshift $z'$, and that merge at redshift $z$. Finally, $H_{\rm 0}$ is the Hubble constant, while $\Omega_{M}$ and $\Omega_{\Lambda}$ are the matter and energy density parameters.

\end{appendix}

\end{document}